\documentclass[12pt
]{article}
\usepackage[height=8.5in,width=6.4in]{geometry} 
\usepackage{empheq}
\usepackage{ascmac}
\usepackage{amsmath,amsthm,amssymb}
\usepackage{type1cm}
\usepackage{hyperref}
\usepackage{tikz}
\usepackage{arydshln}
\usepackage{xcolor}

\hypersetup{
	colorlinks=false,
	citebordercolor=green,
	linkbordercolor=red,
	linktocpage = true,
	urlbordercolor=cyan,
}

\usetikzlibrary{intersections,calc,arrows.meta}

\renewcommand{\tilde}{\widetilde}

\allowdisplaybreaks

\author{}

\begin{document}

\begin{titlepage}
\begin{flushright} 
NITEP 259\\
\end{flushright}

\vskip .8cm

\begin{center}

{\LARGE \fontseries{mx}\selectfont
On bosonic vertex algebras associated with \\[1.5mm] 
3D reductions of Argyres-Douglas theories
\par
}

\vskip 1.2cm

Takahiro Nishinaka,$^{\clubsuit, 1,2,3}$ and Hikaru Sasaki,$^{\diamondsuit, 1}$

\vskip .8cm

{\it
$^1$ Department of Physics, Graduate School of Science\\
Osaka Metropolitan University, Osaka 558-8585, Japan
}

\vskip.4cm

{\it
$^2$ Nambu Yoichiro Institute of Theoretical and Experimental Physics (NITEP)\\
Osaka Metropolitan University, Osaka 558-8585, Japan
}

\vskip.4cm

{\it
$^3$ Osaka Central Advanced Mathematical Institute (OCAMI)\\
Osaka Metropolitan University, Osaka 558-8585, Japan
}

\end{center}

\vskip.7cm

\begin{abstract}
We study the bosonic VOA associated with the 3D $\mathcal{N}=4$ abelian
 linear quiver gauge theories arising from compactifying 4D $\mathcal{N}=2$
 Argyres-Douglas theories of $(A_1,A_{2n-1})$ and $(A_1,D_{2n})$
 types. These VOAs are obtained by cancelling the gauge anomaly of the
 H-twisted 3D theory on the half-space by Heisenberg algebras on the boundary. We
 particularly conjecture a complete set of strong generators of these
 bosonic VOAs, which contains more than the Virasoro stress
 tensor and those arising from Higgs branch operators. We also find that
 these bosonic VOAs contain copies of the $W_3$ vertex algebra at $c=-2$ as
 sub vertex algebras.
\end{abstract}
\end{titlepage}

\newpage

\setcounter{tocdepth}{2}

\tableofcontents

\section{Introduction}
\label{sec:introduction}

It was shown in \cite{Costello:2018fnz} that, from every 3D
$\mathcal{N}=4$ gauge theory on the half-space $\mathbb{R}_{\geq 0} \times
\mathbb{C}$, one can construct a vertex operator
algebra (VOA) on the boundary $\mathbb{C}$ via a bulk topological twist and
a holomorphic boundary condition. Here the topological twist
replaces the spacial rotation $\mathfrak{so}(3)$ with a
diagonal $\mathfrak{su}(2)$ sub-algebra of $\mathfrak{so}(3) \times
\mathfrak{so}(4)_R$, where $\mathfrak{so}(4)_R$ is the
R-symmetry algebra of the 3D $\mathcal{N}=4$ theory. Since
$\mathfrak{so}(4)_R \simeq
\mathfrak{su}(2)_H\times \mathfrak{su}(2)_C$, there are two
different twists, i.e.,
the H-twist and C-twist.
In this paper, we focus on the H-twist and therefore mix the spacial
rotation with $\mathfrak{su}(2)_H$ that acts trivially on the
Coulomb branch but non-trivially on the Higgs branch of the 3D theory. 

Since the three-dimensional spacetime $\mathbb{R}_{\geq 0}\times
\mathbb{C}$ has a boundary, one needs to specify a boundary
condition. The most familiar boundary condition for H-twisted theories is the $(2,2)$-preserving
one, which makes the theory completely
topological.
The authors of \cite{Costello:2018fnz} instead considered an appropriate
deformation of the holomorphic $(0,4)$-preserving boundary condition
so that 
the boundary theory on $\mathbb{C}$ is equipped with holomorphic operator product
expansions (OPEs) and therefore gives rise to a VOA.
Thus, one can construct a VOA from every 3D
$\mathcal{N}=4$ gauge theory. 
Since we consider the H-twist in the bulk, 
the resulting
VOA contains operators arising from the Higgs branch operators of the 3D
theory.
These VOAs and related topics have recently been studied extensively from various
viewpoints in
\cite{Costello:2018fnz, Costello:2020ndc, Garner:2022vds, Ballin:2022rto, Garner:2022rwe, Kuwabara, Yoshida:2023wyt, Beem:2023dub, Garner:2023zko, Ferrari:2023fez,
Coman:2023xcq, Dedushenko:2023cvd,  Creutzig:2024abs, Creutzig:2024ljv,ArabiArdehali:2024ysy, Gaiotto:2024ioj,
ArabiArdehali:2024vli, Gang:2024loa, Go:2025ixu, Ferrari:2025byw}.

It is known that a similar VOA can be constructed from 4D
$\mathcal{N}=2$ superconformal field theories (SCFTs)
\cite{Beem:2013sza}. Indeed, with translations twisted by $SU(2)_R$ symmetry,
the operator product expansions of every 4D $\mathcal{N}=2$ SCFT give
rise to a VOA that receives contributions from Higgs branch
operators. To distinguish them from the VOAs discussed in the previous paragraphs, we denote the VOAs associated with 4D $\mathcal{N}=2$ SCFTs by
$V^{(\text{4D})}$ and those associated with 3D $\mathcal{N}=4$ gauge
theories by $V^{(\text{3D})}$.

Suppose that a 4D $\mathcal{N}=2$ SCFT
$\mathcal{T}_\text{4D}$ is reduced down to a 3D $\mathcal{N}=4$ gauge theory
$\mathcal{T}_\text{3D}$ by $S^1$ compactification. 
Then one can construct two different VOAs; $V^{(\text{4D})}$ associated
 with $\mathcal{T}_\text{4D}$ and $V^{(\text{3D})}$ associated with
 its 3D reduction $\mathcal{T}_\text{3D}$. When $\mathcal{T}_\text{4D}$ is a Lagrangian
 theory, one can show that these two VOAs are identical
 \cite{Costello:2018fnz}. However, when $\mathcal{T}_\text{4D}$ does not admit a
 Lagrangian description, the relation between them is still to be
 understood \cite{Costello:2018fnz, Yoshida:2023wyt}.

In this paper, we study the relation between the VOA associated with
$\mathcal{T}_\text{4D}$ and that associated with its 3D reduction $\mathcal{T}_{3D}$,
focusing on the cases in which $\mathcal{T}_\text{4D}$ is an Argyres-Douglas (AD) theory of $(A_1,A_{2n-1})$ and $(A_1,D_{2n})$
types \cite{Cecotti:2010fi, Xie:2012hs}. While no $\mathcal{N}=2$ preserving Lagrangian for
these AD theories is known, there are conjectures on $V^{(\text{4D})}$
for them \cite{Beem:2013sza, Buican:2015ina,
Cordova:2015nma, Creutzig:2017qyf}. 
When
compactifying these theories on $S^1$, the resulting 3D $\mathcal{N}=4$
SCFTs $\mathcal{T}_{\text{3D}}$ are believed to be described by
abelian linear quiver gauge theories \cite{Xie:2012hs,
Xie:2021ewm} (See also \cite{Buican:2015hsa} for a consistency check).\footnote{Here, we are {\it not} considering the twisted
compactification discussed recently in \cite{Dedushenko:2023cvd, ArabiArdehali:2024ysy}. Understanding the
relation between the abelian gauge theories of \cite{Xie:2012hs,
Xie:2021ewm} and those obtained by twisted compactifications of AD
theories is important and left for future work.}
 The purpose of this paper is to reveal the structure
of $V^{(\text{3D})}$ associated with these quiver gauge theories and
then compare them with $V^{(\text{4D})}$ associated with their 4D
ancestors, i.e.,  AD theories
of $(A_1,A_{2n-1})$ and $(A_1, D_{2n})$ types. 

One subtlety here is that, to define $V^{(\text{3D})}$ for the above quiver
gauge theories, one needs to specify how to cancel
abelian gauge anomalies on the boundary. There are two known ways of anomaly
cancellation in the literature; one is to add Fermi multiplets \cite{Costello:2018fnz}
 and the other is to add the currents of Heisenberg vertex algebras
 \cite{Kuwabara}. 

In this paper, we add 
 Heisenberg algebras to cancel the boundary gauge anomaly.
The main reason for this is that it leads to a bosonic
 $V^{(\text{3D})}$. Indeed, when the anomaly is canceled by Heisenberg
 algebras, no fermionic operators are introduced in the construction of
 $V^{(\text{3D})}$. This is in contrast to when the anomaly is
 canceled by Fermi multiplets, in which case the resulting
 $V^{(\text{3D})}$ contains fermionic operators 
 \cite{Costello:2018fnz, Yoshida:2023wyt, Beem:2023dub}. Since
 $V^{(\text{4D})}$ associated
 with
 $(A_1,A_{2n-1})$ and $(A_1,D_{2n})$ is conjectured to be bosonic
 \cite{Beem:2013sza, Buican:2015ina, Cordova:2015nma,Buican:2015tda,
 Creutzig:2017qyf}, it would be interesting to study bosonic
 $V^{(\text{3D})}$ for the 3D reduction of $(A_1,A_{2n-1})$ and
 $(A_1,D_{2n})$ and then
 compare it with $V^{(\text{4D})}$ for the original 4D theories.

With this motivation,
we study the bosonic $V^{(\text{3D})}$ for the 3D reduction of the
 $(A_1,A_{2n-1})$ and $(A_1,D_{2n})$ theories in which 
 the anomaly is canceled by Heisenberg algebras.   In particular, we
 conjecture a complete set of strong generators of these
 VOAs by explicitly studying the OPEs arising from BRST reductions. Note
 that, while these BRST reductions have been studied in \cite{Kuwabara,Coman:2023xcq}, the
 resulting OPEs and strong generators are still to be understood.
Our results in this paper imply that the bosonic VOA
 contains more generators than the Virasoro stress tensor and those arising from
 Higgs branch operators. Indeed, it generically contains strong generators of
 dimensions $\frac{n}{2}+\ell$ for $\ell = 1,2,\cdots,\frac{n}{2}$ (or $\ell = 1,2,\cdots,\frac{n-1}{2}$) when associated with the 3D reduction
 of the $(A_1,A_{2n-1})$ theory (or the $(A_1,D_{2n})$ theory). Furthermore, we
 find that the bosonic
 $V^{(\text{3D})}$ for these 3D reductions contains copies of the $W_3$ vertex algebra at $c=-2$ as
 sub vertex algebras. Given
 these results, we also give a comparison of $V^{\text(4D)}$ for
 $(A_1,A_{2n-1})$ and $(A_1,D_{2n})$ with these bosonic $V^{(\text{3D})}$.

The organization of this paper is the following. In Sec.~\ref{sec:3D},
we briefly review the 3D reductions of the two series of 4D
Argyres-Douglas theories $(A_1,A_{2n-1})$ and $(A_1,D_{2n})$. In
Sec.~\ref{sec:review}, we describe the construction of the bosonic
$V^{(\text{3D})}$ for these 3D reductions, by applying the general
method proposed in \cite{Costello:2018fnz}. In particular, we describe
the BRST reduction that gives rise to these bosonic
$V^{(\text{3D})}$. In Sec.~\ref{sec:VOA1}, we study the BRST cohomology
for the 3D reduction of the $(A_1,A_{2n-1})$ theory, and conjecture a
complete set of the strong generators of the resulting VOA. In
Sec.~\ref{sec:VOA2}, we perform a similar analysis for 3D reduction of
the $(A_1,D_{2n})$ theory, and conjecture a complete set of strong
generators of the bosonic VOA associated with the 3D reduction. 
In
Sec.~\ref{sec:summary}, we give a summary and conclusions. In appendix
\ref{app:nilpotency}, we give explicit computations to see the
nilpotency of the BRST charge.
In appendix \ref{app:T}, we describe how to rewrite the canonical stress
tensor adding/extracting BRST exact terms. In appendix \ref{app:A1A5},
we list non-vanishing OPEs of the bosonic $V^{(\text{3D})}$ for the 3D
reduction of the $(A_1,A_5)$ theory. In appendix \ref{app:A1D6}, we list
non-vanishing OPEs of the bosonic $V^{(\text{3D})}$ for the 3D reduction
of $(A_1,D_6)$.

\section{3D reductions of AD theories}

\label{sec:3D}

In this section, we give a brief review of the 3D $\mathcal{N}=4$ abelian quiver gauge
theories obtained by compactifying the 4D $\mathcal{N}=2$ AD theories called
$(A_1,A_{2n-1})$ and $(A_1,D_{2n})$. To be more precise, we first consider the 3D mirror of
these 4D theories by applying the method of \cite{Xie:2012hs}, and then
take its mirror. 
We will focus on
the 3D reduction of $(A_1,A_{2n-1})$ theories in
Sec.~\ref{subsec:Higgs1}, and then move to the 3D reduction of
$(A_1,D_{2n})$ in Sec.~\ref{subsec:Higgs2}. The bosonic VOAs associated
with these 3D theories will be discussed in Sec.~\ref{sec:review},
\ref{sec:VOA1} and \ref{sec:VOA2}.

\subsection{$T_{[n-1,1]}^{[1^n]}(SU(n))$ as 3D reduction of $(A_1,A_{2n-1})$}
\label{subsec:Higgs1}

\begin{figure}
\centering
\begin{tikzpicture}[gauge/.style={circle,draw=black,inner sep=0pt,minimum size=10mm},flavor/.style={rectangle,draw=black,inner sep=0pt,minimum size=10mm},auto]
 \node[flavor] (0) at (-2,0) {\;$n$\;};
 \node[gauge] (1) at (0,0) {\;$1$\;} edge (0);
\end{tikzpicture}
\caption{The quiver diagram of the 3D mirror of $(A_1,A_{2n-1})$
 theory. The circle stands for a $U(1)$ vector multiplet, and the box
 stands for $n$ hypermultiplet coupled to it.}
\label{fig:mirror1}
\end{figure}
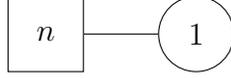

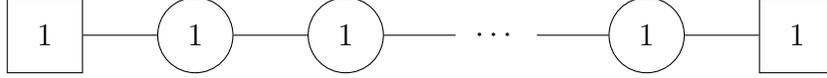
\begin{figure}
\centering
\begin{tikzpicture}[gauge/.style={circle,draw=black,inner sep=0pt,minimum size=10mm},flavor/.style={rectangle,draw=black,inner sep=0pt,minimum size=10mm},auto]
 \node[flavor] (0) at (-2,0) {\;$1$\;};
 \node[gauge] (1) at (0,0)  {\;$1$\;} edge (0);
 \node[gauge] (2) at (2,0)  {\;$1$\;} edge (1);
 \node (3) at (4,0) {\;$\cdots$\;} edge (2);
 \node[gauge] (4) at (6,0)  {\;$1$\;} edge (3);
 \node[flavor] (5) at (8,0)  {\;$1$\;} edge (4);
\end{tikzpicture}
\caption{The quiver diagram of the $T^{[1^n]}_{[n-1,1]}(SU(n))$
 theory. The diagram contains $(n-1)$ circles. Each circle stands for a $U(1)$ vector multiplet, and each box
 stands for a fundamental hypermultiplet. Each edge between two circles
 stand for a bifundamental hypermultiplet.  
 }
\label{fig:quiver_A1A2n-1}
\end{figure}

Let us first consider the $(A_1,A_{2n-1})$ theory. We assume the integer
$n$ is
larger than one because otherwise the theory is a free theory.
The 3D theory obtained by compactifying the $(A_1,A_{2n-1})$ theory on
$S^1$ is the $\mathcal{N}=4$ quiver gauge theory called
$T_\sigma^\rho(SU(n))$  with $\rho= [1^{n}]$ and $\sigma =
[n-1,1]$ being
partitions of $n$.\footnote{A series of 3D $\mathcal{N}=4$ gauge
theories called $T_\sigma^\rho(SU(n))$ was first introduced in
\cite{Gaiotto:2008ak}.} This can be derived 
as follows. First, the 3D mirror theory of $(A_1,A_{2n-1})$ theory is
identified in \cite{Xie:2012hs} as an $\mathcal{N}=4$ $U(1)$ gauge
theory coupled to $n$ hypermultiplets (Fig.~\ref{fig:mirror1}).
We take its mirror dual according to
\cite{deBoer:1996ck}, and then obtain the quiver gauge theory shown in
Fig.~\ref{fig:quiver_A1A2n-1}, which is called
$T^{[1^n]}_{[n-1,1]}(SU(n))$ theory.

We briefly discuss the Higgs branch chiral ring of
 the theory here.
 Let us denote by $q_i$ and
$\tilde{q}_i$ the two chiral multiplets sitting in the $i$-th hypermultiplet
for $i=1,2,\cdots,n$. Then $q_i$ and $\tilde{q}_i$ have
charge $(+1,-1)$ and $(-1,+1)$ under  $U(1)_i\times U(1)_{i+1}$, respectively.
Since the F-term condition implies that
\begin{align}
 q_1\tilde{q}_1 = q_2\tilde{q}_2 = \cdots = q_n\tilde{q}_n~, 
\end{align}
independent gauge invariant operators are composite operators built out of
\begin{align}
 e \coloneqq \prod_{i=1}^n q_i~,\qquad f \coloneqq \prod_{i=1}^n \tilde{q}_i~,\qquad u \coloneqq
 -\frac{1}{n}\sum_{i=1}^n q_i\tilde{q}_i~.
\label{eq:Higgs-gen1}
\end{align}
Therefore, $e,\,f$ and $u$ are the generators of the Higgs branch chiral ring.
Note that there is a chiral ring relation $ef = (-1)^nu^n$, which
implies that 
the Higgs branch of the theory is $\mathbb{C}^2/\mathbb{Z}_n$.

\subsection{$T^{[2,1^{n-1}]}_{[n-1,1^2]}(SU(n+1))$ as 3D reduction of $(A_1,D_{2n})$}
\label{subsec:Higgs2}

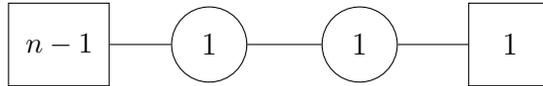
\begin{figure}
\centering
\begin{tikzpicture}[gauge/.style={circle,draw=black,inner sep=0pt,minimum size=10mm},flavor/.style={rectangle,draw=black,inner sep=0pt,minimum size=11mm},auto]
 \node[flavor] (0) at (-2,0) {\;\;\small $n-1$\;\;};
 \node[gauge] (1) at (0,0)  {\;$1$\;} edge (0);
 \node[gauge] (2) at (2,0)  {\;$1$\;} edge (1);
 \node[flavor] (3) at (4,0) {\;$1$\;} edge (2);
\end{tikzpicture}
\caption{The quiver diagram of the 3D mirror of $(A_1,D_{2n})$
 theory. The leftmost box stands for $n-1$ hypermultiplets }
\label{fig:mirror2}
\end{figure}

\begin{figure}[tbp]
         \centering
         \begin{tikzpicture}[gauge/.style={circle,draw=black,inner sep=0pt,minimum size=10mm},flavor/.style={rectangle,draw=black,inner sep=0pt,minimum size=10mm},auto]
 \node[flavor] (0) at (-2,0) {\;$1$\;};
 \node[gauge] (1) at (0,0) [shape=circle] {\;$1$\;} edge (0);
 \node[gauge] (2) at (2,0) [shape=circle] {\;$1$\;} edge (1);
 \node (3) at (4,0) {\;$\cdots$\;} edge (2);
 \node[gauge] (4) at (6,0) [shape=circle] {\;$1$\;} edge (3);
 \node[flavor] (5) at (8,0)  {\;$2$\;} edge (4);
        \end{tikzpicture}
        \caption{The quiver diagram of the
      $T^{[2,1^{n-1}]}_{[n-1,1^2]}(SU(n+1))$ theory. There are $(n-1)$
      $U(1)$ gauge groups. The difference from the
      $T^{[1^n]}_{[n-1,1]}(SU(n))$ is that the rightmost $U(1)$ gauge
      group is coupled to two fundamental hypermultiplets.
      }
     \label{fig:quiver_A1D2n}
    \end{figure}
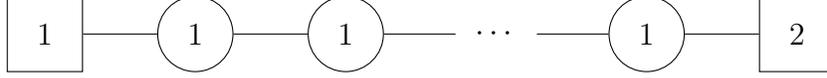

Let us next consider the $(A_1,D_{2n})$ theory. We assume $n>1$ so that
the theory is interacting. The 3D mirror of the theory is described by
the quiver diagram shown in Fig.~\ref{fig:mirror2}
\cite{Xie:2012hs}. Taking its mirror dual via the method of \cite{deBoer:1996ck}, we see that the $S^1$
compactification of the $(A_1,D_{2n})$ theory is described by the quiver
gauge theory shown in Fig.~\ref{fig:quiver_A1D2n}.

We review the Higgs branch chiral ring of the theory here.
Let
us denote by $q_i$ and $\tilde{q}_i$ the two chiral multiplets in the
$i$-th hypermultiplets for $i=1,\cdots,n-1$, and 
by $q_{n,1},\,q_{n_2},\,\tilde{q}_{n,1}$ and $\tilde{q}_{n,2}$ the four
chiral multiplets in the rightmost pair of hypermultiplets. We say $q_i$
and $\tilde{q}_i$ have charge $(+1,-1)$ and $(-1,+1)$ under
$U(1)_i\times U(1)_{i+1}$ for $i=1,\cdots,n-1$, respectively. Similarly,
$q_{n,j}$ and $\tilde{q}_{n,j}$ respectively have charge $1$ and $-1$ under $U(1)_n$
for $j=1,2$. Note that there is a flavor $SU(2)$ symmetry under which
$(q_{n,1},q_{n,2})$ and $(\tilde{q}_{n,1},\tilde{q}_{n,2})$ transform as doublets.
The Higgs branch chiral ring is composed of gauge invariant operators
built out of these chiral operators subject to the F-term conditions:
\begin{align}
 q_1\tilde{q}_1 = q_2\tilde{q}_2 = \cdots = q_{n-1}\tilde{q}_{n-1} = \sum_{j=1}^2q_{n,j}\tilde{q}_{n,j}~.
\end{align}
We see that this ring is generated by 
\begin{align}
 h &\coloneqq \frac{1}{2}\left(-q_{n,1}\tilde{q}_{n,1} +
 q_{n,2}\tilde{q}_{n,2}\right)~,\quad e \coloneqq
 q_{n,1}\tilde{q}_{n,2}~,\quad f \coloneqq q_{n,2}\tilde{q}_{n,1}~,
\label{eq:Higgs-gen2-1}
\\
 u &\coloneqq
 -\frac{1}{2n-1}\left(2\sum_{i=1}^{n-1}q_i\tilde{q}_i + \sum_{j=1}^2
 q_{n,j}\tilde{q}_{n,j}\right) 
~,
\\
x_+ &\coloneqq \left(\prod_{i=1}^{n-1}q_i\right)q_{n,1}~,\quad x_- \coloneqq
 \left(\prod_{i=1}^{n-1}q_i\right)q_{n,2}~,\quad
y_+ \coloneqq
 \left(\prod_{i=1}^{n-1}\tilde{q}_i\right)\tilde{q}_{n,2}~,\quad y_-
 \coloneqq \left(\prod_{i=1}^{n-1}\tilde{q}_i\right)\tilde{q}_{n,1}~,
\label{eq:Higgs-gen2-3}
\end{align}
subject to the ring relations $x_+y_- + x_-y_+ = 2u^n~,\, -x_+y_- +
x_-y_+ = 2u^{n-1}h~,\, x_+y_+ = u^{n-1}e$ and $x_-y_- = u^{n-1}f$.
We see from these relations that the Higgs branch is of
complex four dimensions.

\section{Construction of associated VOA}
\label{sec:review}

In this section, we describe the BRST construction of the VOA associated with the 3D
$\mathcal{N}=4$ theories discussed in the previous section. We first review the general
BRST procedure in Sec.~\ref{subsec:general-review}, and then apply it to
the
$T^{[1^n]}_{[n-1,1]}(SU(n))$ and $T^{[2,1^{n-1}]}_{[n-1,1^2]}(SU(n+1))$ theories
in Sec.~\ref{subsec:BRST-1} and Sec.~\ref{subsec:BRST-2},
respectively. The strong generators of the resulting VOAs will be
discussed in Sec.~\ref{sec:VOA1} and \ref{sec:VOA2}.

\subsection{Review of general construction}
\label{subsec:general-review}

We consider a 4D $\mathcal{N}=4$ gauge theory $\mathcal{T}$ with gauge group $G$ and
matter hypermultiplets in some representation of $G$. In this paper, we
only consider the case of abelian gauge group. We put the theory
on the interval $\mathbb{R}_{\geq 0} \times \mathbb{C}$ and consider the
H-twist in the bulk together with an a boundary condition for the
fields. The boundary conditions we impose here is an appropriate
deformation of the $\mathcal{N}=(0,4)$ boundary condition, which makes
the boundary theory holomorphic instead of topological. The resulting
OPEs on the boundary $\mathbb{C}$ then gives rise to a vertex algebra.

\subsubsection{Free hypermultiplet and vector multiplets}

As an example, let us consider the case that $\mathcal{T}$ is the theory
of a free hypermultiplet. While there are two natural $(0,4)$ boundary
conditions (i.e., the Neumann and Dirichlet boundary conditions), it turns out
that an appropriate deformation of the Neumann boundary condition is
consistent with the H-twist in the bulk, as shown in Appendix E of
 \cite{Gaiotto:2017euk}.\footnote{Here, by
Neumann (Dirichlet), we mean that the scalar fields in the
hypermultiplet have
 Neumann (Dirichlet) boundary conditions. The boundary conditions on the
 other fields are then fixed by supersymmetry.} 
 With this boundary condition, the H-twisted free
 hypermultiplet gives rise to a symplectic boson VOA on the boundary
\cite{Gaiotto:2017euk, Costello:2018fnz};
\begin{align}
 X(z)\, Y(0) \sim \frac{1}{z}~,
\label{eq:OPE-XY}
\end{align}
where $X$ and $Y$ are bosonic vertex operators of holomorphic dimension
$1/2$. When the original theory $\mathcal{T}$ is the theory of $n$
hypermultiplets, the same procedure leads to $n$ copies of the symplectic boson.

Let us next consider the case that $\mathcal{T}$ is the theory of a free
vector multiplet. In this case, an appropriate deformation of the Neumann boundary condition is
expected to be compatible with the H-twist \cite{Costello:2018fnz,
Gaiotto:2017euk}.\footnote{Here, the Neumann boundary condition means
that the Neumann boundary condition for the gauge fields. The boundary conditions for
the other fields in the vector multiplet are then fixed by supersymmetry.} The deformed boundary
condition leads to 
the kernel of $b_0$ in the $(b,c)$ ghost VOA of holomorphic dimension $(1,0)$:
\begin{align}
 b(z)\, c(0) \sim \frac{1}{z}~.
\label{eq:OPE-bc}
\end{align}
Note that taking the kernel of $b_0$ implies that this VOA does not contain $c(z)$ itself as an
operator while its derivatives are contained. This is the same
situation as the VOA associated with a free $\mathcal{N}=2$ vector
multiplet in four dimensions \cite{Beem:2013sza}.

When $\mathcal{T}$ is a general
$\mathcal{N}=4$ gauge theory obtained by gauging a $G$ symmetry acting
on a collection of hypermultiplets. Before gauging the $G$-symmetry, the
VOA is the tensor product of the $(b,c)$ VOA arising from the
vector multiplet and the symplectic
boson VOA arising from the hypermultiplets.
It is conjectured in \cite{Costello:2018fnz} that, when gauging the
$G$-symmetry,
the resulting VOA is the BRST reduction of this tensor product VOA (together with
an extra degree of freedom described below).

\subsubsection{Gauge theory and anomaly}

When considering the above gauging, the gauge anomaly needs to be
canceled. The absence of the gauge
anomaly is equivalent to the nilpotency of the BRST charge. 
The gauge anomaly is always canceled when the 3D theory $\mathcal{T}$ is
obtained by compactifying a 4D $\mathcal{N}=2$ SCFT with Lagrangian
description \cite{Costello:2018fnz, Yoshida:2023wyt}, but it can be
non-vanishing when $\mathcal{T}$
is not obtained from a 4D $\mathcal{N}=2$ Lagrangian theory.  In particular, when $\mathcal{T}$
is an abelian gauge theory, the gauge anomaly is always non-vanishing and therefore the
BRST charge is not nilpotent. 

The above fact implies that, for the BRST reduction associated with H-twisted $\mathcal{N}=4$ abelian
gauge theories on $\mathbb{R}_{\geq 0}\times \mathbb{C}$ (with the
deformed $(0,4)$ 
boundary condition imposed), one needs to add
an extra degree of freedom on the boundary to cancel the gauge anomaly. There are two
known ways to do it; the first one is to add a sufficient number of Heisenberg currents
\cite{Kuwabara}, and
the second one is to
add a sufficient
number of Fermi multiplets \cite{Costello:2018fnz}. For instance, when $\mathcal{T}$ is the
$U(1)$ gauge theory coupled to $N_f$ hypermultiplets, the first way means to
consider
\begin{align}
J_\text{BRST} =
c\left(\sum_{j=1}^{N_f}q_jX_jY_j + h\right)
\end{align}
as the BRST current, where $h$ is a Heisenberg current such that $h(z)h(0)\sim
\sum_{j=1}^{N_f}(q_j)^2/z$. In contrast, the second way of anomaly cancellation
replaces the above $h$ with a bilinear operator of extra fermions. 

In this paper, we focus on the first way of anomaly cancellation and
introduce Heisenberg currents. 
The main reason for this is that it leads
to a VOA without fermionic operators. Indeed, when the anomaly is
canceled by Heisenberg currents, the resulting VOA contains no fermionic
operator by definition. In contrast, when the anomaly is
canceled by Fermi multiplets, the resulting VOA always involves fermionic
generators \cite{Yoshida:2023wyt, Beem:2023dub}. As stated in
Sec.~\ref{sec:introduction}, we are interested in the relation to the
VOA associated with the 4D AD theories $\mathcal{T}_{\text{4D}}$ of $(A_1,A_{2n-1})$ and
$(A_1,D_{2n})$ types. These VOAs associated with $\mathcal{T}_\text{4D}$
are all known to be bosonic and contain no fermionic
operators \cite{Beem:2013sza, Buican:2015ina, Cordova:2015nma, Buican:2015tda, Creutzig:2017qyf}. In the next two sections, we will study the
VOAs associated with the 3D $\mathcal{N}=4$ theories
$\mathcal{T}_\text{3D}$ obtained by compactifying these AD
theories $\mathcal{T}_\text{4D}$ on $S^1$,
and see how they differ from the VOAs associated with the original AD
theories $\mathcal{T}_\text{4D}$.

\subsection{BRST reduction for $T^{[1^n]}_{[n-1,1]}(SU(n))$}
\label{subsec:BRST-1}

Let us focus on the 3D reduction of the $(A_1,A_{2n-1})$ theory, i.e.,
 $T_{[n-1,1]}^{[1^n]}(SU(n))$.
This theory contains $n$ hypermultiplets and $(n-1)$ $U(1)$ vector
 multiplets.
According to the general construction reviewed above, we associate a symplectic
boson $(X_i,Y_i)$ with the $i$-th hypermultiplet for
$i=1,\cdots,n$. 
The OPEs of these symplectic bosons are the $n$ copies of
\eqref{eq:OPE-XY}, i.e.~,
\begin{align}
 X_i(z)Y_j(0) \sim \frac{\delta_{ij}}{z}~.
\label{eq:OPE-XY-2}
\end{align}
We also associate a $(b,c)$-ghost, $(b_a,c^a)$, with
the $a$-th gauge node of $U(1)$ for $a=1,\cdots,n-1$. The OPEs of these
ghosts are written as
\begin{align}
 b_a(z)c^b(0) \sim\frac{\delta_a^b}{z}~.
\label{eq:OPE-bc-2}
\end{align}
 As reviewed in the previous sub-section, the naive BRST current $J_\text{BRST} =
\sum_{a=1}^{n-1}c^a(X_aY_a - X_{a+1}Y_{a+1})$ does not give rise to a
nilpotent BRST charge, which reflects the fact that the gauge
anomaly is not canceled. To cancel the anomaly, we introduce Heisenberg
currents $h_a$ with the OPE
 \begin{align}
  h_a(z)h_b(0) \sim \frac{C_{ab}}{z^2}~,
\label{eq:OPE-h}
\end{align}
where $a=1,\cdots,n-1$ and
\begin{align}
 C_{ab} \coloneqq \left\{
\begin{array}{l}
 2 \qquad \text{for} \qquad a=b\\
 -1 \qquad \text{for} \qquad |a-b| = 1\\
 0 \qquad \text{for the other cases}
\end{array}
\right.
\label{eq:C1}
\end{align}
is the Cartan matrix of $A_{n-1}$ \cite{Kuwabara, Yoshida:2023wyt}.
Using these Heisenberg currents, the BRST
current is defined by
\begin{align}
 J_\text{BRST} = \sum_{a=1}^{n-1}c^a\left(X_aY_a  - X_{a+1}Y_{a+1} +
 h_a\right)~.
\label{eq:BRST}
\end{align}
From the OPEs $X_i(z)Y_j(0) \sim \delta_{ij}/z$, $b_a(z)c^b(0) \sim
\delta_a^b/z$ and Eq.~\eqref{eq:OPE-h}, one can show that the BRST charge 
\begin{align}
Q_\text{BRST} \coloneqq \oint \frac{dz}{2\pi i}J_\text{BRST}(z)
\label{eq:BRSTcharge1}
\end{align}
satisfies $(Q_\text{BRST})^2 = 0$. Note that this nilpotency of $Q_\text{BRST}$ requires the
additional degrees of freedom $h_i$
introduced to cancel the gauge anomaly, which we review carefully in
Appendix \ref{app:nilpotency}.

The VOA associated with
$T^{[1^n]}_{[n-1,1]}(SU(n))$ is obtained by the BRST reduction of the
tensor product of the symplectic boson algebras generated by $X_i,Y_i$
and the Heisenberg algebras generated by $h_i$. In
Sec.~\ref{sec:VOA1}, we study this
BRST reduction carefully and conjecture the complete set of generators
of the resulting VOA.

\subsection{BRST reduction for $T^{[2,1^{n-1}]}_{[n-1,1^2]}(SU(n+1))$}
\label{subsec:BRST-2}

Let us next consider the 3D reduction of the $(A_1,D_{2n})$ theory, i.e.,
$T^{[2,1^{n-1}]}_{[n-1,1^2]}(SU(n+1))$. 
Note that the only difference from $T^{[1^n]}_{[n-1,1]}(SU(n))$ is that there
are two hypermultiplets at one end of the quiver for
$T^{[2,1^{n-1}]}_{[n-1,1^2]}(SU(n+1))$. Therefore, the VOA associated with
the latter is constructed in a quite similar way as in previous
sub-section. First, the $a$-th
$U(1)$ gauge group is associated with the $(b,c)$-ghost $(b_a,c^a)$
for $a=1,\cdots,n-1$, whose OPE is characterized by Eq.~\eqref{eq:OPE-bc-2}.
Second, the $i$-th hypermultiplet groups gives rise to a symplectic boson
$(X_i,Y_i)$ for $i=1,\cdots,n-1$, whose OPE is shown in Eq.~\eqref{eq:OPE-XY-2}.
Similarly, the rightmost pair of hypermultiplets
leads to a pair of $\beta\gamma$-system 
$(X_{n,j},\,Y_{n}{}^j)$ for $j=1,2$, whose 
(non-vanishing) OPEs
are characterized by
\begin{align}
X_{n,i}(z)Y_{n}{}^j(0) \sim \frac{\delta^j_i}{z}~,
\end{align}

The fact that there are two hypermultiplets at right end of the quiver
 affects the OPEs of the Heisenberg currents that we
need to introduce to cancel the gauge anomaly. Indeed, while the BRST current is
still written as \eqref{eq:BRST}, the Heisenberg currents now need to
have the OPEs of the form 
\begin{align}
 h_a(z)h_b(0)\sim \frac{\widetilde{C}_{ab}}{z^2}~,
\end{align}
where
\begin{align}
 \widetilde{C}_{ab} \coloneqq \left\{
\begin{array}{l}
2 \qquad \text{for}\qquad a=b \leq n-2\\
3 \qquad \text{for}\qquad a=b=n-1\\
-1 \qquad \text{for}\qquad |a-b|=1\\
0 \qquad \text{for the other cases}
\end{array}
\right.~. 
\label{eq:C2}
\end{align}
Note that this is {\it not} the Cartan matrix of a Lie algebra. The BRST
current is defined by
\begin{align}
 J_\text{BRST} = \sum_{a=1}^{n-2}c^a\left(X_aY_a -
 X_{a+1}Y_{a+1}+h_a\right) + c^{n-1}\left(X_{n-1}Y_{n-1} -
 X_{n,j}Y_{n}{}^j + h_{n-1}\right)~,
\label{eq:BRST2}
\end{align}
which leads to the nilpotent BRST charge $Q_\text{BRST} = \oint
\frac{dz}{2\pi i}J_\text{BRST}(z)$. The nilpotency of $Q_\text{BRST}$ is
carefully reviewed in Appendix \ref{app:nilpotency}.

The VOA associated with
$T^{[2,1^{n-1}]}_{[n-1,1^2]}(SU(n+1))$ is obtained by the BRST reduction of the
tensor product of the symplectic boson algebras generated by $X_i,Y_i$
and the Heisenberg algebras generated by $h_i$. In
Sec.~\ref{sec:VOA2}, we carefully study this
BRST reduction and conjecture the complete set of generators of the
resulting VOA.

\section{VOA for $T^{[1^n]}_{[n-1,1]}(SU(n))$ theory}
\label{sec:VOA1}

In this section, we study the strong generators of the bosonic VOA
associated with the 3D $\mathcal{N}=4$
theory $T^{[n-1,1]}_{[1^n]}(SU(n))$. As described in
Sec.~\ref{subsec:BRST-1}, 
 it is given by the BRST reduction of the tensor product
of the symplectic boson VOA generated by $(X_i,Y_i)$ for $i=1,\cdots,n$,
the $bc$-ghost VOA generated by $(b_a,\partial c^a)$ for
$a=1,\cdots,n-1$, and the Heisenberg VOA generated by $h_a$ for
$a=1,\cdots,n-1$. 
Here, the OPEs of the Heisenberg algebra are
characterized by Eq.~\eqref{eq:C1}. The relevant BRST current is shown in Eq.~\eqref{eq:BRST}.
 While it is generally quite hard to identify generators of a VOA obtained by
 this type of BRST reduction, we will conjecture a complete set of
 generators of this VOA in Sec.~\ref{subsec:conjecture1}.

To that end, in Sec.~\ref{subsubsec:cand1}, we begin with listing the trivial candidates for generators that can be
read off from the construction of the VOA. These candidates are
either those associated with Higgs branch operators or
the Virasoro stress tensor. In Sec.~\ref{subsubsec:cand2}, we list more non-trivial candidates for
generators, which are neither stress tensor nor those associated with
Higgs branch operators.
Finally, in Sec.~\ref{subsec:conjecture1}, we conjecture a complete set of generators of the VOA.
Our conjecture particularly implies that
the VOA generically contains $n$ copies of the $W_3$ algebra. Several
examples will be given in Sec.~\ref{subsec:example1}. Finally in
Sec.~\ref{subsec:comparison1}, we will compare the resulting VOA
with the VOA associated with 4D Argyres-Douglas theory of $(A_1,A_{2n-1})$.

\subsection{Generators read off from the construction}
\label{subsubsec:cand1}

Let us start with trivial candidates for strong generators of the VOA.
By construction, the generator of the Higgs branch chiral ring of the 3D
gauge theory $\mathcal{T}$ gives rise to a strong generator of the VOA
associated with $\mathcal{T}$. In addition, the VOA
always contains a Virasoro stress tensor, which is also a strong
generator. 

As discussed in Sec.~\ref{subsec:Higgs1}, the Higgs branch chiral
ring of
this theory is generated by $x,y$ and $u$ shown in \eqref{eq:Higgs-gen1}.
One expects that the VOA associated with
$T^{[1^n]}_{[n-1,1]}(SU(n))$ contains strong generators corresponding to
$x,y$ and $u$. 
In terms of $X_i$ and $Y_i$ defined in the previous
section, these generators are identified as
\begin{align}
\mathcal{X} = \prod_{i=1}^{n}X_i~,\qquad
 \mathcal{Y}=\prod_{i=1}^nY_i~,\qquad U =
 -\frac{1}{n}\sum_{i=1}^{n}X_iY_i~.
\label{eq:XYU}
\end{align}
In particular, $U$ generates a Heisenberg algebra of the form
\begin{align}
 U(z) U(0) \sim -\frac{1}{nz^2}~.
\label{eq:UU}
\end{align}
In addition to
these VOA generators corresponding to the generators of the Higgs
branch chiral ring, the stress tensor
\begin{align}
 T = T_\text{sb} + T_{bc} + T_h
\label{eq:totalT}
\end{align}
 is also a strong generator of the VOA, where 
\begin{align}
T_\text{sb} \coloneqq \frac{1}{2}\sum_{i=1}^n\left(X_i\partial Y_i -
 (\partial X_i)Y_i\right)~,\qquad T_{bc} \coloneqq
 \sum_{a=1}^{n-1}b_a\partial c_a~,\qquad T_h \coloneqq
 \frac{1}{2}\sum_{a,b=1}^{n-1}C^{ab}h_ah_b
\end{align}
with  $C^{ab}$ being the inverse of $C_{ab}$ shown in Eq.~\eqref{eq:C1}, i.e., $C^{ab}C_{bc} = \delta^a_c$.

\subsection{More non-trivial closed operators}
\label{subsubsec:cand2}

In addition to the above strong generators, we see that the following
are also BRST-closed operators:
\begin{align}
 T_i &\coloneqq \frac{1}{2}D_i \mathcal{U}_i~, \qquad W_i \coloneqq
 \sqrt{\frac{2}{27}}\left(\mathcal{D}_i\left(\mathcal{D}_i\mathcal{U}_i\right)
 + \frac{1}{2}\mathcal{D}_i\left(\mathcal{U}_i^2\right) +
 \frac{1}{2}\mathcal{U}_i\mathcal{D}_i\mathcal{U}_i +
 \mathcal{U}_i^3\right)~,
\label{eq:TiWi}
\end{align}
where $i=1,\cdots,n$, and\footnote{We define nested normal-ordered
products as
\begin{align}
 \mathcal{O}_1\mathcal{O}_2 \cdots \mathcal{O}_k \coloneqq
 \mathcal{O}_1(\mathcal{O}_1(\cdots(\mathcal{O}_{k-2}(\mathcal{O}_{k-1}\mathcal{O}_k)_0)_0\cdots)_0~,
\end{align}
where $(\mathcal{O}_1\mathcal{O}_2)_0$ is the normal-ordered product of
$\mathcal{O}_1$ and $\mathcal{O}_2$ in this ordering.
This means that $D_i \mathcal{O} = 2 \mathcal{U}_i \mathcal{O} +
X_iY_i\mathcal{O} = -2((X_iY_i)_0\mathcal{O})_0 + (X_i (Y_i \mathcal{O})_0)_0$.
}
\begin{align}
\mathcal{U}_i\coloneqq -X_iY_i~,\qquad  D_i\mathcal{O} \coloneqq 
2\,\mathcal{U}_i\mathcal{O}  + X_iY_i\mathcal{O}~, \qquad \mathcal{D}_i\mathcal{O} \coloneqq \mathcal{U}_i\mathcal{O} + X_iY_i\mathcal{O}~.
\label{eq:Di}
\end{align}
It follows from the results of \cite{Wang:1997ndk} that $T_i$ and $W_i$ form $n$ copies of $W_3$ algebra
\begin{align}
 T_i(z)T_{j}(0) &\sim \delta_{ij}\left(\frac{c}{2z^4} + \frac{2T_i(0)}{z^2} +
 \frac{\partial T_i(0)}{z}\right)~,
\label{eq:TiTj}
\\
T_i(z) W_j(0) &\sim \delta_{ij}\left(\frac{3W_i(w)}{z^2} +
 \frac{\partial W_i(0)}{z}\right)~,
\\
W_i(z)W_j(0) &\sim \delta_{ij}\Bigg(\frac{c}{3z^6} + \frac{2T_i(0)}{z^4}
 + \frac{\partial T_i(0)}{z^3} +
 \frac{1}{z^2}\left(\frac{32}{22+5c}\Lambda_i(0) + \frac{3}{10}
 \partial^2 T_i(0)\right)
\nonumber\\
&\qquad\qquad   + \frac{1}{z}\left(\frac{16}{22+5c}\partial \Lambda_i(0) +
 \frac{1}{15}\partial^3 T_i(0)\right)\Bigg)~,
\label{eq:WiWj}
\end{align}
where the central charge is $c=-2$, and $\Lambda_i$ is defined by
\begin{align}
\Lambda_i = (T_i)^2 -\frac{3}{10}\partial^2T_i~.
\end{align}
The action of the $W_3$ algebra on $\mathcal{X},\,\mathcal{Y}$ and $U$ is
characterized by
\begin{align}
T_i(z)\mathcal{X}(0)&\sim \frac{\mathcal{X}(0)}{z^2} +
 \frac{(D_i\mathcal{X})(0)}{z}~,
\label{eq:TiX}
\\
T_i(z)\mathcal{Y}(0)&\sim \frac{\mathcal{X}(0)}{z^2} -
 \frac{(D_i\mathcal{Y})(0)}{z}~,
\label{eq:TiY}
\\
T_i(z) U(0) &\sim 0~,
\\
W_i(z)\mathcal{X}(0) &\sim
 \sqrt{\frac{2}{3}}\left(\frac{\mathcal{X}(0)}{z^3} + \frac{3}{2}
 \frac{D_i \mathcal{X}(0)}{z^2} + 2 \frac{(T_i
 \mathcal{X})(0)}{z}\right)~,
\label{eq:WiX}
\\
W_i(z)\mathcal{Y}(0) &\sim
 \sqrt{\frac{2}{3}}\left(-\frac{\mathcal{Y}(0)}{z^3} + \frac{3}{2}
 \frac{D_i \mathcal{Y}(0)}{z^2} - 2\frac{(T_i
 \mathcal{Y})(0)}{z}\right)~,
\label{eq:WiY}
\\
W_i(z) U(0) &\sim 0~.
\end{align}
Note that $D_i$ defined in \eqref{eq:Di} appears in these OPEs.

We also see that the following operators give yet another set of BRST-closed operators:
\begin{align}
D_{i_1}D_{i_2}\cdots D_{i_\ell} \mathcal{X}~,\qquad D_{i_1}D_{i_2}\cdots D_{i_\ell}\mathcal{Y}~,
\end{align}
where $\ell$ is a positive integer less than or equal to $n$, and
$\{i_1,i_2,\cdots,i_\ell\}$ is a proper subset of $\{1,2,\cdots,n\}$ such
that $1\leq i_1<i_2<\cdots<i_\ell\leq n$. 
Note that $D_i$ and $D_j$ commute with each other when $i\neq j$.
 Note also that, while the numbers
of linearly independent $D_{i_1}D_{i_2}\cdots D_{i_\ell}\mathcal{X}$ and
$D_{i_1}D_{i_2}\cdots D_{i_\ell}\mathcal{Y}$ are both
\begin{align}
 \sum_{\ell=1}^n \binom{n}{\ell}  = 2^n-1~,
\end{align}
not all of them are strong generators of the VOA. We will conjecture
below that only some number of linear combinations of
such operators  for  $1\leq \ell\leq n/2$ are
strong generators.

\subsection{Conjecture on complete set of generators}
\label{subsec:conjecture1}

We have seen that the VOA associated with $T^{[1^n]}_{[n-1,1]}(SU(n))$
contains strong generators $\mathcal{X},\,\mathcal{Y},\, U$ and $T$ as
well as BRST closed operators $T_i,\,W_i,\,\mathcal{O}_{\mathcal{X},I}$
and $\mathcal{O}_{\mathcal{Y},I}$.

We now conjecture that one can take, as a complete set of
strong generators of the VOA associated with
$T^{[1^n]}_{[n-1,1]}(SU(n))$, the set $\mathtt{S}$ of the following
BRST-closed operators:
\begin{itemize}
 \item $U,\;\mathcal{X},\; \mathcal{Y},\; T_i,\; W_i$~,
\item for each integer $\ell$ such that $1\leq  \ell \leq n/2$,
\begin{itemize}
 \item  $N(n,\ell)$ different linear combinations of
      $D_{i_1}D_{i_2}\cdots D_{i_\ell}\mathcal{X}$~,
 \item  $N(n,\ell)$ different linear combinations of
      $D_{i_1}D_{i_2}\cdots D_{i_\ell}\mathcal{Y}$~,
\end{itemize}
\end{itemize}
where
\begin{align}
N(n,\ell)\coloneqq
\binom{n}{\ell } - \binom{n}{\ell-1}
=
 \frac{n-2\ell+1}{\ell}\binom{n}{\ell-1}~.
\label{eq:NnI}
\end{align}
Note here that, for $n=2,3$, there are accidental operator
relations involving the above generators. To be more specific, when $n=2$, $T_1+T_2$
is a composite operator of $U,\mathcal{X}$ and $\mathcal{Y}$, and $W_i$
are all composite operators of $U,\mathcal{X},\mathcal{Y}$ and
$T_1-T_2$. Similarly, when $n=3$, $W_1+W_2+W_3$ is a composite operator
of $U,\mathcal{X},\mathcal{Y}$ and $T_1+T_2+T_3$. See
Sec.\ref{subsec:example1} for more detail.

Note also that the stress tensor \eqref{eq:totalT} is written in terms
of $T_i$ and $U$ as
\begin{align}
  T= \sum_{i=1}^nT_i -\frac{n}{2}U^2~,
\label{eq:canonicalT}
\end{align}
up to BRST exact terms, as shown in Appendix \ref{app:rewritingT}.
Therefore we do not include it as a generator here. 
One can check
that this stress tensor has central charge $-2n+1$, i.e.,
\begin{align}
 T(z) T(0) \sim \frac{-2n+1}{2z^4} + \frac{2T}{z^2} + \frac{\partial T}{z}~.
\end{align}
The holomorphic dimension and the $U(1)$ charge of the generators are
summerized in Table~\ref{table:gen1}.

While the number
$N(n,\ell)$ that appears above might be
unfamiliar, we see that  it is always a positive integer when  $0\leq \ell \leq \frac{1}{2}(n-2)$. An interpretation
of this number as well as explicit expressions for the linear
combinations of
$D_{i_1}D_{i_2}\cdots D_{i_\ell}\mathcal{X}$ and those of
$D_{i_1}D_{i_2}\cdots D_{i_\ell}\mathcal{Y}$ included in
$\mathtt{S}$ will be
discussed in a separate paper by one of the authors \cite{Sasaki:2025}. In
the present paper, we instead show expressions for these linear combinations
explicitly in some examples in the next sub-section.

As a consistency check of the above conjecture,
we explicitly checked for $n=2,3,\cdots,7$ that the number of independent
generators of dimension up to $13/2$ is perfectly consistent
with Table~\ref{table:gen1}.\footnote{For this computation, we have used
the Mathematica package $\mathtt{OPEdefs}$.} 
Moreover, for $n=2$ and $n=3$, we explicitly computed the OPEs of
the generators
listed in Table~\ref{table:gen1}, and showed that these generators have
closed OPEs among themselves as shown in the
next sub-section.
These represent very strong
evidence for our conjecture.

\begin{table}
 \centering
\begin{tabular}{|c|c|c|}
\hline
 generator & dimension & $U(1)$ charge \\
\hline
$U$ & $1$ & $0$ \\[.5mm]
$\mathcal{X}$ & $n/2$ & $+1$ \\[.5mm]
$\mathcal{Y}$ & $n/2$ & $-1$ \\[.5mm]
$T_i$ & $2$ & $0$ \\[.5mm]
$W_i$ & $3$ & $0$ \\[.5mm]
linear combinations of $D_{i_1}D_{i_2}\cdots D_{i_\ell}\mathcal{X}$ & $n/2 + \ell$ &
	 $+1$ \\[.5mm]
linear combinations of $D_{i_1}D_{i_2}\cdots D_{i_\ell}\mathcal{Y}$ & $n/2 + \ell$
     &$-1$\\[.5mm]
\hline
\end{tabular}
\caption{The conjectural list of strong generators. Their dimensions and
 $U(1)$ charges are also shown. The index $i$ runs over
 $1,2,\cdots,n$, and $\ell$ runs over $1,2,\cdots, [\frac{n}{2}]$. The subscripts $i_1,\cdots,i_\ell$ are mutually different positive integers
 less than or equal to $n$. Only $N(n,\ell)$ linear combinations are
 counted as generators in each of the last two lows. For $n=2$, $T_1+T_2$ is a composite
 operator of $U,\mathcal{X}$ and $\mathcal{Y}$, and $W_1$ and $W_2$ are
 composite operators of $U, \mathcal{X},\mathcal{Y}$ and
 $T_1-T_2$. Similarly, for $n=3$, $W_1+W_2+W_3$ is a composite operator
 of $U,\mathcal{X},\mathcal{Y}$ and $T_1+T_2+T_3$.}
\label{table:gen1}
\end{table}

\subsection{Examples}
\label{subsec:example1}

In this section, we give the first two examples of the VOA associated with
$T^{[1^n]}_{[n-1,1]}(SU(n))$.

\subsubsection{$n=2$}

Let us first consider the case of $n=2$. Our conjecture stated in
Sec.~\ref{subsec:conjecture1} implies that the VOA is generated by
dimension-one operators $U,\, \mathcal{X}$ and $\mathcal{Y}$,
and operators generating $W_3$-algebras $T_i$ and $W_i$ for $i=1,2$,
together with  
$N(2,1) = 1$ linear combination of $D_1\mathcal{X}$ and
$D_2\mathcal{X}$, and a similar linear combination of 
$D_1\mathcal{Y}$ and $D_2\mathcal{Y}$.\footnote{Note that $N(2,1)=1$ as a special
case of $N(n,\ell)$ in \eqref{eq:NnI}.} 

As mentioned in Sec.~\ref{subsec:conjecture1}, however, there are
accidental operator relations in the case of $n=2$. The first
non-trivial relation is
\begin{align}
 T_1 + T_2 = 2U^2 -
 \frac{1}{2}(\mathcal{X}\mathcal{Y}+\mathcal{Y}\mathcal{X})~,
\label{eq:T1+T2}
\end{align}
which means that $T_1$ and $T_2$ are not independent generators when
$n=2$. Another set of operator relations special for $n=2$ is
\begin{align}
 W_i &= (-1)^{i+1}\left(\frac{2}{3}\right)^{\frac{3}{2}}\left(U\tau -
 \frac{1}{4}\mathcal{Y}(D_1-D_2)\mathcal{X} + \frac{1}{8}\partial \tau \right)
\nonumber\\
&\qquad + \sqrt{\frac{2}{3}}\left(-U\partial
 U + \frac{1}{4}\mathcal{X}\partial\mathcal{Y} -
 \frac{1}{4}\mathcal{Y}\partial\mathcal{X} +
 \frac{4}{3}U^3-U\mathcal{X}\mathcal{Y}+\frac{1}{3}\partial^2U\right)
\end{align}
for $i=1,2$, which implies that $W_1$ and $W_2$ are both composite operators of the
other generators when $n=2$.

Therefore, the VOA for $n=2$ has six independent generators,
which can be taken as
\begin{align}
U~,\quad \mathcal{X}~,\quad \mathcal{Y}~,\quad \tau \coloneqq
 T_1-T_2~,\quad \mathcal{X}_1 \equiv
 \frac{1}{2}(D_1-D_2)\mathcal{X}~,\quad \mathcal{Y}_1 \coloneqq
 \frac{1}{2}(D_1-D_2)\mathcal{Y}~.
\label{eq:gen1}
\end{align}
Note that $\mathcal{X}_1$ and $\mathcal{Y}_1$ are those listed in the
last and second last rows of Table~\ref{table:gen1}.
Note also that, since $T_1+T_2$ is a composite operator of the form \eqref{eq:T1+T2}, the canonical stress tensor \eqref{eq:canonicalT} is also
a composite operator of $U,\mathcal{X}$ and $\mathcal{Y}$, and expressed as
\begin{align}
 T = U^2 -\frac{1}{2}(\mathcal{X}\mathcal{Y} + \mathcal{Y}\mathcal{X})~.
\end{align}
As mentioned in Sec.~\ref{subsec:conjecture1}, the central charge of this
stress tensor is $-3$:
\begin{align}
 T(z) T(0) \sim \frac{-3}{2z^4} + \frac{2T}{z^2} +
 \frac{\partial T}{z}~.
\end{align}
It is straightforward to see that
$U,\mathcal{X},\mathcal{Y},\tau,\mathcal{X}_1$ and $\mathcal{Y}_1$ are
all Virasoro primary operators with respect to $T$. The holomorphic
dimension of $U,\mathcal{X}$ and $\mathcal{Y}$ are one, while that of
$\tau,\mathcal{X}_1$ and $\mathcal{Y}_1$ is two. From these, the OPEs of
the generators with $T$ are completely fixed.

Let us write down the OPEs among the generators shown in \eqref{eq:gen1}.
The dimension-one generator $U$ gives rise to the Heisenberg algebra of
the form
\begin{align}
 U(z)U(0) \sim \frac{-1}{2z^2}
\end{align}
It is also straightforward to see that
$\mathcal{X},\mathcal{Y},\tau,\mathcal{X}_1$ and $\mathcal{Y}_1$ are
primary operators with respect to $U$. The charges of
$\mathcal{X},\,\mathcal{X}_1$ are $+1$ so that
\begin{align}
 U(z)\mathcal{X}(0) \sim \frac{\mathcal{X}}{z}~,\qquad U(z)\mathcal{X}_1(0)
 \sim \frac{\mathcal{X}_1}{z}~,
\end{align}
and the charges of $\mathcal{Y}$ and $\mathcal{Y}_1$ are $-1$:
\begin{align}
 U(z) \mathcal{Y}(0) \sim -\frac{\mathcal{Y}}{z}~,\qquad U(z) \mathcal{Y}_1(0) \sim -\frac{\mathcal{Y}_1}{z}~,
\end{align}
while the charge of $\tau$ vanishes.
The non-vanishing OPEs with $\tau$ are written as
\begin{align}
\tau(z)\tau(0) & \sim \frac{-2}{z^4} + \frac{4U^2
 -2\mathcal{X}\mathcal{Y}-2\partial U}{z^2}  + \frac{4U\partial U
 -\mathcal{X}\partial \mathcal{Y} - \mathcal{Y}\partial\mathcal{X}}{z}~,
\label{eq:tautau}
\\
 \tau(z) \mathcal{X}(0) & \sim \frac{2\mathcal{X}_1}{z}~,
\\
 \tau(z)\mathcal{Y}(0) &\sim \frac{2\mathcal{Y}_1}{z}~,
\label{eq:tauY}
\\
 \tau(z) \mathcal{X}_{1} (0) &\sim \frac{2\mathcal{X}}{z^3} +\frac{U\mathcal{X} +\frac{1}{2}\partial \mathcal{X}}{z^2}
                - \frac{6 U\partial \mathcal{X} -\mathcal{X} \partial U
 - 2 UU\mathcal{X} +2 \mathcal{X} \mathcal{X} \mathcal{Y} -3\partial^2
 \mathcal{X}}{z}~,
\\
 \tau(z) \mathcal{Y}_{1} (0) &\sim -\frac{2\mathcal{Y}}{z^3} +\frac{U\mathcal{Y} -\frac{1}{2}\partial \mathcal{Y}}{z^2}
                - \frac{6 U\partial \mathcal{Y} -\mathcal{Y} \partial U
 + 2 UU\mathcal{Y} -2 \mathcal{Y} \mathcal{Y} \mathcal{X} +3\partial^2
 \mathcal{Y}}{z}~,
\end{align}
and the non-vanishing OPEs among
$\mathcal{X},\,\mathcal{Y},\,\mathcal{X}_1$ and $\mathcal{Y}_1$ are the following:
\begin{align}
 \mathcal{X}(z)\mathcal{Y}(0) & \sim \frac{1}{z^2} - \frac{2U}{z}~,
\\
\mathcal{X}(z)\mathcal{Y}_1(0) &\sim -\frac{\tau}{z}~,
\\
\mathcal{Y}(z)\mathcal{X}_1(0) &\sim \frac{\tau}{z}~,
\\
 \mathcal{X}_{1}(z) \mathcal{X}_{1}(0) &\sim
 \frac{\mathcal{X}^2}{2z^2} +\frac{\mathcal{X}\partial
 \mathcal{X}}{2z}~,
\\
 \mathcal{X}_{1}(z) \mathcal{Y}_{1}(0) &\sim \frac{1}{z^4}
 -\frac{2U}{z^3} -\frac{U^2 -\frac{3}{2}\mathcal{X} \mathcal{Y}
 +\frac{1}{2}\partial U}{z^2}
\\
                &\qquad +\frac{U \partial U -\frac{1}{2}\mathcal{X}
 \partial \mathcal{Y} +2 \mathcal{Y} \partial \mathcal{X} -2U^3
 +2U\mathcal{X} \mathcal{Y} -2\partial^2 U}{z}~,
\\
                \mathcal{Y}_{1}(z) \mathcal{Y}_{1}(0) &\sim \frac{\mathcal{Y}^2}{2z^2} +\frac{\mathcal{Y}\partial \mathcal{Y}}{2z}~.
\end{align}
The above is the complete set of non-vanishing OPEs among the
generators shown in \eqref{eq:gen1}, and all the other OPEs of the
generators vanish. We see that the conjectured set of generators
\eqref{eq:gen1} have closed OPEs.  Note that \eqref{eq:tautau}--\eqref{eq:tauY} are consistent with
\eqref{eq:TiTj}, \eqref{eq:TiX} and \eqref{eq:TiY}. 
Note also that the sub-algebra generated by $U,\mathcal{X}$ and
$\mathcal{Y}$ is the $\widehat{\mathfrak{su}(2)}_{-1}$  algebra.

\subsubsection{$n=3$}

We now turn to the example of $n=3$. Our general conjecture implies that
the VOA is generated by dimension-one $U$, dimension-$\frac{3}{2}$
$\mathcal{X}$ and $\mathcal{Y}$, three copies of $W_3$ generators
$(T_i,W_i)$ for $i=1,2,3$, and $N(3,1)=2$ linear combinations of
$D_i\mathcal{X}$ for $i=1,2,3$ as well as similar two linear
combinations of $D_i\mathcal{Y}$ for
$i=1,2,3$. However, as in the case of $n=2$, there is an accidental
operator relation
\begin{align}
 W_1 + W_2 + W_3 = \sqrt{\frac{2}{3}}\left(3UT + 3U\partial U
 -\mathcal{X}\mathcal{Y} -\frac{1}{2}\partial T - \frac{1}{2}\partial^2
 U\right)~.
\end{align}
This means that only two of $W_i$ are independent generators.

One can take the following as the complete set of independent
generators:
\begin{align}
 &U~,\quad \mathcal{X}~,\quad \mathcal{Y}~,
\label{eq:gen1-2-1}
\\
&T\coloneqq
 \sum_{i=1}^3T_i-\frac{3}{2}U^2~, \quad \tau_1\coloneqq T_1-T_2~,\quad \tau_2 \coloneqq
 \frac{1}{\sqrt{3}}(T_1+T_2-2T_3)~,
\\
 &\mathcal{\omega}_1 \coloneqq \frac{1}{2}\sqrt{\frac{3}{2}}(W_1-W_2), \quad \omega_2 \coloneqq \frac{1}{2\sqrt{2}}(W_1+W_2-2W_3)~,
\\
 &\mathcal{X}_1 \coloneqq
 \frac{1}{2}\left(D_1-D_2\right)\mathcal{X}~,\qquad \mathcal{X}_2
 \coloneqq \frac{1}{2\sqrt{3}}(D_1+D_2-2D_3)\mathcal{X}~,
\\
&\mathcal{Y}_1 \coloneqq \frac{1}{2}(D_1-D_2)\mathcal{Y}~,\qquad
 \mathcal{Y}_2 \coloneqq \frac{1}{2\sqrt{3}}(D_1+D_2-2D_3)\mathcal{Y}~.
\label{eq:gen1-2-2}
\end{align}
Note that $\mathcal{X}_i$ and $\mathcal{Y}_i$ for $i=1,2$ are those
listed in the last and second last rows of
Table~\ref{table:gen1}.\footnote{Note that there are $N(3,1) = 2$ such
generators for each of the last and second last rows of Table~\ref{table:gen1}.}
Note also that $U$ gives rise to the Heisenberg algebra \eqref{eq:UU}, and
$T$ gives us the Virasoro algebra at the central charge $-5$:
\begin{align}
 U(z)U(0) \sim -\frac{1}{3z^2}~,\qquad T(z)T(0) \sim -\frac{5}{2z^4} +
 \frac{2T}{z^2} + \frac{\partial T}{z}~.
\label{eq:self-UT2}
\end{align}
All the other generators turn out to be primary operators with respect
to $U$ and $T$. Their $U(1)$ charges and holomorphic dimensions can be
read off from Table~\ref{table:gen1}, which fix their OPEs with $T$ and
$U$. Note that, in contrast to the $n=2$ case, the holomorphic dimension
of $\mathcal{X}$ and $\mathcal{Y}$ is $3/2$, and the VOA has
only one dimension-one current $U$. One can check that the above twelve
generators form closed set of OPEs, which
we list in Appendix \ref{app:A1A5}.

\subsection{Comparison to the VOA associated with $(A_1,A_{2n-1})$}

\label{subsec:comparison1}

Let us now compare the bosonic VOA for $T^{[1^n]}_{n-1,1}(SU(n))$
with the VOA associated with the $(A_1,A_{2n-1})$ theory in the sense of
\cite{Beem:2013sza}.

It was conjectured in \cite{Creutzig:2017qyf} that the VOA associated with
$(A_1,A_{2n-1})$ is the logarithmic $\mathcal{B}_{n+1}$ algebra, which is
conjecturally isomorphic to $\mathcal{W}^{-\frac{n^2}{n+1}}(\mathfrak{sl}_n,f_\text{sub})$, i.e., the sub-regular quantum Drinfeld-Sokorov reduction
of the simple
affine $\mathfrak{sl}(n)$ algebra at level
$k=-\frac{n^2}{n+1}$. It is known that this vertex algebra
is generated by 
\begin{align}
 J~,\quad X~,\quad Y~,\quad T^{(2)}~,\quad T^{(3)}~,\quad \cdots
 ~,\quad T^{(n-1)}~,
\label{eq:gen3}
\end{align}
where $J$ is a dimension-one Heisenberg current, and the dimensions and
$U(1)$-charge of the above generators are listed in
Table~\ref{table:gen3}. See \cite{Genra:2018dca} for the explicit
construction of these strong generators.

\begin{table}
\centering
\begin{tabular}{|c|c|c|}\hline
generator & dimension & $U(1)$ charge
\\ \hline
$J$ & $1$ & $0$\\
$X$& $n/2$ & $+1$ \\
$Y$& $n/2$ & $-1$ \\
$T^{(i)}$ & $i$ & $0$ \\
\hline
\end{tabular}
\caption{List of strong generators of $\mathcal{W}^{-\frac{n^2}{n+1}}(\mathfrak{sl}_n,f_\text{sub})$, where
 $i=2,3,\cdots, n-1$. }
\label{table:gen3}
\end{table}

Comparing Table~\ref{table:gen3} with Table~\ref{table:gen1}, we see
that the charge and dimension of $U,\,\mathcal{X},\,\mathcal{Y}$ and
$\sum_{i=1}^nT_i$ in Table~\ref{table:gen1} are
identical to those of $J,\,X,\,Y$ and $T^{(2)}$ in
Table~\ref{table:gen3}, respectively. Similarly, $T^{(i)}$ for
$i=3,\cdots,n-1$ in Table~\ref{table:gen3} have the same charge and
dimension as $\sum_{i}(T_i)^k (W_i)^\ell$ for non-negative integers $k$ and
$\ell$ such that $2k+3\ell =i$. Therefore, it is tempting to identify these operators.
However, Table~\ref{table:gen1} clearly has generators which have no
counterpart in Table~\ref{table:gen3}.
This implies that the VOA associated with $T^{[1^n]}_{n-1,1}(SU(n))$ has more
generators than that associated with its 4D ancestor, i.e., $(A_1,A_{2n-1})$.

However, when we consider an appropriate $S_n$-quotient, the above discrepancy
between three and four dimensions becomes milder. To see this, let us
consider the following $S_n$-action on the symplectic bosons
$(X_i, Y_i)$ for $i=1,2,\cdots,n$ 
\begin{align}
 X_i \to X_{\sigma(i)}~,\qquad Y_i \to  Y_{\sigma(i)}~,
\end{align}
for $\sigma\in S_n$. This induces
an $S_n$-action on the VOA associated with
$T^{[1^n]}_{[n-1,1]}(SU(n))$. 
One can take the sub vertex algebra of
this VOA that contains only operators invariant under this
$S_n$. The resulting (sub-)vertex algebra turns out to
contain $\mathcal{W}^{-n+1}(\mathfrak{sl}_n,f_\text{sub})$ \cite{Kuwabara}, i.e., the sub-regular quantum Drinfeld-Sokorov reduction of
the affine $\mathfrak{sl}(n)$ algebra at level $k=-n+1$.\footnote{In terms of our
generators shown in Table~\ref{table:gen1}, this
$\mathcal{W}^{-n+1}(\mathfrak{sl}_n,f_\text{sub})$ is weakly
generated by $U,\,\mathcal{X}$ and $\mathcal{Y}$. The strong generators
are expected to be $U,\,\mathcal{X},\,\mathcal{Y}$ and
$\mathbb{Z}_n$-invariant sums of normal-ordered products of $T_i,\,W_i$.}

Note that $\mathcal{W}^{-n+1}(\mathfrak{sl}_n,f_\text{sub})$ is {\it not} precisely
identical to the VOA associated with $(A_1,A_{2n-1})$, i.e., $\mathcal{W}^{-\frac{n^2}{n+1}}(\mathfrak{sl}_n,f_\text{sub})$. Nevertheless,
these two algebras have the same number of generators with the same
dimensions and charges.
 The OPEs among the generators
are, however, different between $\mathcal{W}^{-n+1}(\mathfrak{sl}_n,f_\text{sub})$ and
$\mathcal{W}^{-\frac{n^2}{n+1}}(\mathfrak{sl}_n,f_\text{sub})$.

\section{VOA for $T^{[2,1^{n-1}]}_{[n-1,1^2]}(SU(n+1))$}

\label{sec:VOA2}

We now turn to the $T^{[2,1^{n-1}]}_{[n-1,1^2]}(SU(n+1))$ theory, and
discuss the strong generators of the bosonic VOA associated with it. 
As described in Sec.~\ref{subsec:BRST-2}, it is given by the following BRST
reduction. We  first consider the tensor product of the symplectic
boson VOA generated by $(X_i,Y_i)$ for $i=1,\cdots,n-1$ and $(X_n^{(j)},
Y_n^{(j)})$ for $j=1,2$,
the $bc$-ghost VOA generated by $(b_a, \partial c_a)$ for
$a=1,\cdots,n-1$, and the Heisenberg VOA generated by $h_a$ for
$a=1,\cdots,n-1$. Here, the OPEs of the Heisenberg algebra are
characterized by \eqref{eq:C2}. We then consider the BRST reduction of this tensor
product VOA with respect to the nilpotent BRST charge
\eqref{eq:BRSTcharge1}.
In this section, we conjecture a complete set of
 generators of the VOA obtained by this BRST reduction.

To that end, in Sec.~\ref{subsec:cand3}, we list generators read off
from the construction of the VOA, i.e., the Virasoro stress tensor and
those associated with Higgs
branch operators. In Sec.~\ref{subsec:cand4}, we list other candidates for
generators which are neither stress tensor nor those associated with
Higgs branch operators. 
We then conjecture a complete set of generators of the VOA in Sec.~\ref{subsec:conjecture2}.
Our conjecture implies that
the VOA generically contains $(n-1)$ copies of the $W_3$ algebra. Several
examples will be given in Sec.~\ref{subsec:example2}. Finally in
Sec.~\ref{subsec:comparison2}, we will compare the resulting VOA
with the VOA associated with 4D Argyres-Douglas theories called $(A_1,D_{2n})$.

\subsection{Generators read off from the construction}
\label{subsec:cand3}

The Higgs branch chiral ring of $T^{[2,1^{n-1}]}_{[n-1,1^2]}(SU(n+1))$ is
generated by $e,f,h,u,x_{\pm}$ and $y_{\pm}$ shown in
Eqs.~\eqref{eq:Higgs-gen2-1}--\eqref{eq:Higgs-gen2-3}.
The associated VOA includes a strong generator for each
of them, which is easily identified as follows:
\begin{align}
E &= X_{n,1}Y_{n}{}^2~,\qquad F = X_{n,2}Y_{n}{}^1 ~,\qquad H =
 \frac{1}{2}\left(-X_{n,1}Y_{n}{}^1 + X_{n,2}Y_{n}{}^2\right)~,
\label{eq:H}
\\
U &= -\frac{2}{2n-1}\left(\sum_{i=1}^{n-1}X_iY_i +
 \frac{1}{2}\left(X_{n,1}Y_{n}{}^1 + X_{n,2}Y_{n}{}^2\right)\right)~,
\label{eq:U}
\\
\mathcal{X}_{+} &= \left(\prod_{i=1}^{n-1}X_i\right)X_{n,1}~,\qquad
 \mathcal{X}_{-} = \left(\prod_{i=1}^{n-1}X_i\right)X_{n,2}~,
\\
 \mathcal{Y}_{+} &= \left(\prod_{i=1}^{n-1}Y_i\right)Y_{n}{}^2~,\qquad \mathcal{Y}_{-} = \left(\prod_{i=1}^{n-1}Y_i\right)Y_{n}{}^1~.
\end{align}
Among them, $E,F$ and $H$ generate the $\widehat{\mathfrak{su}(2)}_{-1}$
vertex sub-algebra:
\begin{align}
 H(z)H(0) \sim -\frac{1}{2z^2}~,\qquad H(z)E(0) \sim \frac{E}{z}~,\qquad
 H(z)F(0) \sim -\frac{F}{z} ~,\qquad E(z)F(0)\sim -\frac{1}{z^2} + \frac{2H}{z}~,
\end{align}
and $U$ generates the Heisenberg vertex sub-algebra:
\begin{align}
 U(z) U(0) \sim -\frac{2}{(2n-1)z^2}~.
\end{align}
    In addition, 
the stress tensor $T$
is written
as 
\begin{align}
 T &= T_\text{sb} + T_{bc} + T_{h}~,
\label{eq:T2}
\\
T_\text{sb} &= \frac{1}{2}\sum_{i=1}^{n-1}\left(X_i \partial Y_i -
 (\partial X_i)Y_i\right) + \frac{1}{2}\sum_{j=1}^2\left(
 X_{n,j}\partial Y_{n}{}^j - (\partial X_{n,j})Y_{n}{}^j\right)~,
\label{eq:Tsb2}
\\
T_{bc} &= -\sum_{a=1}^{n-1}b_a\partial c_a~,\qquad T_h =
 \frac{1}{2}\sum_{i,j=1}^{n-1}\tilde{C}^{ij}h_ih_j~, 
\label{eq:Tbch2}
\end{align}
where $\tilde{C}^{ij}$ is the inverse of $\tilde{C}_{ij}$ shown in \eqref{eq:C2}.
While these operators turn out to generate the whole VOA in the case of
    $n=2$, there are more generators when $n\geq 3$. We will describe
    them below.

\subsection{More non-trivial generators}
\label{subsec:cand4}

In addition to the above strong generators, the VOA associated with
$T^{[2,1^{n-1}]}_{n-1,1^2}(SU(n+1))$ also contains the following BRST
closed operators. The first series of closed operators are those given
by Eq.~\eqref{eq:TiWi} for $i=1,\cdots,n-1$, as well as their cousins
charged under the $SU(2)$ symmetry:
\begin{align}
 T_{n}^{(j)} \coloneqq \frac{1}{2} D_{n}^{(j)}\mathcal{U}_n^{(j)}~,\qquad W_n^{(j)} \coloneqq
 \sqrt{\frac{2}{27}}\left(\mathcal{D}_n^{(j)}\left(\mathcal{D}_n^{(j)}\mathcal{U}_n^{(j)}\right)
 + \frac{1}{2}\mathcal{D}_{n}^{(j)}(\mathcal{U}_n^{(j)})^2 +
 \frac{1}{2}\mathcal{U}_n^{(j)}\mathcal{D}_n^{(j)}\mathcal{U}_n^{(j)} +
 \left(\mathcal{U}_n^{(j)}\right)^3\right)~,
\label{eq:TiWij}
\end{align}
where
\begin{align}
 \mathcal{U}_n^{(j)} \coloneqq - X_{n,j}Y_n{}^{j}~,\qquad
 \mathcal{D}_n^{(j)}\mathcal{O} \coloneqq \mathcal{U}_n^{(j)}\mathcal{O} +
 X_{n,j}Y_n{}^{j}\mathcal{O}~,\qquad
 D_{n}^{(j)}\mathcal{O}\coloneqq
2\, \mathcal{U}_n^{(j)}\mathcal{O} + X_{n,j}Y_n{}^j\mathcal{O}
 ~,
\label{eq:Dij}
\end{align}
where we do {\it not} take the sum over $j$ on the RHS.
Since $(T_i,W_i)$ for $i=1,\cdots,n-1$ and $(T_n^{(j)}, W_n^{(j)})$ for
$j=1,2$ separately form $W_3$ algebra at $c=-2$, the VOA associated with
$T^{[2,1^{n-1}]}_{n-1,1^2}(SU(n+1))$ contains $(n+1)$ copies of this
$W_3$ algebra. Note also that the stress tensor \eqref{eq:T2} can be
expressed as
\begin{align}
 T = \sum_{i=1}^{n-1}T_i + \sum_{j=1}^2 T_n^{(j)} -H^2 - \frac{2n-1}{4}U^2~,
\label{eq:TTi2}
\end{align}
up to BRST exact terms, as shown in Appendix~\ref{app:rewritingT2}.
We see that this $T$ generates the Virasoro vertex sub-algebra at
central charge $-2n$:
\begin{align}
 T(z)T(0) \sim -\frac{n}{z^4} + \frac{2T}{z^2} + \frac{\partial T}{z}~.
\end{align}

There are yet another class of BRST-closed operators that are not
generated by those described above:\footnote{In addition,
$D_{i_1}D_{i_2}\cdots D_{i_\ell} D_{n}^{(j)}\mathcal{X}_\pm$ and
$D_{i_1}D_{i_2}\cdots D_{i_\ell} D_{n}^{(j)}\mathcal{Y}_\pm$ are also
BRST-closed, but we do not need them to give our conjecture in the
next sub-section.}
\begin{align}
  &D_{n}^{(j)}E~,\qquad
 D_n^{(j)}F~, \qquad D_{i_1}D_{i_2}\cdots D_{i_\ell} \mathcal{X}_\pm ~,\qquad
 D_{i_1}D_{i_2}\cdots D_{i_\ell}\mathcal{Y}_\pm
\end{align}
where $\ell$ is a positive integer less than or equal to $n-1$, and
$\{i_1,i_2,\cdots,i_\ell\}$ is a proper subset of $\{1,2,\cdots,n-1\}$.
As in the previous sub-section, not all of these BRST-closed operators
are strong generators of the VOA. We will give a conjecture on
a complete set of strong generators below.

\subsection{Conjecture on complete set of generators}
\label{subsec:conjecture2}

\begin{table}[h]
 \centering
\begin{tabular}{|c|c|c|c|}
\hline
 generator & dimension & $U(1)$ charge & $SU(2)$ charge \\
\hline
$U$ & $1$ & $0$ & $0$ \\[.5mm]
\hdashline
$H$ & $1$ & $0$ & $0$ \\[.5mm]
$E$ & $1$ & $0$ & $+1$ \\[.5mm]
$F$ & $1$ & $0$ & $-1$\\[.5mm]
\hdashline
$\mathcal{X}_\pm$ & $n/2$ & $+1$ & $\pm 1/2$ \\[.5mm]
\hdashline
$\mathcal{Y}_\pm$ & $n/2$ & $-1$ & $\pm 1/2$ \\[.5mm]
\hdashline
$T_i$ & $2$ & $0$ & $0$ \\[.5mm]
\hdashline
$W_i,$ & $3$ & $0$ & $0$\\[.5mm]
\hdashline
$\frac{1}{2}(T_n^{(1)} - T_n^{(2)})$ & $2$ & $0$ & $0$ \\[.5mm]
$\frac{1}{2}(D_n^{(1)} + D_n^{(2)})E$ & $2$ & $0$ & $+1$\\[.5mm]
$\frac{1}{2}(D_n^{(1)} + D_n^{(2)})F$ & $2$ & $0$ & $-1$\\[.5mm]
\hdashline
linear combinations of $D_{i_1}D_{i_2}\cdots D_{i_\ell}\mathcal{X}_\pm$ & $n/2 + \ell$ &
	 $+1$ & $\pm 1/2$\\[.5mm]
\hdashline
linear combinations of $D_{i_1}D_{i_2}\cdots D_{i_\ell}\mathcal{Y}_\pm$ & $n/2 + \ell$
     &$-1$ & $\pm 1/2$\\[.5mm]
\hline
\end{tabular}
\caption{The conjectural list of strong generators. Their dimensions and
 $U(1)$ and $SU(2)$ charges are also shown. The dashed lines separate different $SU(2)$
 multiplets. The index $i$ runs over $1,2,\cdots,n-1$. The positive integer $\ell$ is less than or
 equal to $\frac{n-1}{2}$, and $i_1,\cdots,i_\ell$ are mutually
 different positive integers less than or equal to $n-1$. Only
 $N(n-1,\ell)$ linear combinations are counted as generators in each of
 the
 last two lows. For $n=2$, generators of dimension larger than one are
 accidentaly composite. Similarly, for $n=3$, $W_1+W_2$ is a
 composite operator of other generators.}
\label{table:gen2}
\end{table}

We have seen that the VOA associated with $T^{[2,1^{n-1}]}_{[n-1,1^2]}(SU(n+1))$
contains strong generators $U,\,H,\,E,\,F,\mathcal{X}_\pm,\,\mathcal{Y}_\pm$ and $T$ as
well as BRST closed operators $T_i,\, T_n^{(j)}\,W_i,\,W_n^{(j)}\,
D_n^{(j)}E,\,D_n^{(j)}F$, $D_{i_1}D_{i_2}\cdots
D_{i_\ell}\mathcal{X}_\pm$ and $D_{i_1}D_{i_2}\cdots D_{i_\ell}\mathcal{Y}_\pm$.

Now, we conjecture that one can take, as a complete set of
strong generators, the set $\mathtt{S}$ of the following operators:
\begin{itemize}
 \item $U,\; H,\; E,\; F,\; \mathcal{X}_\pm,\; \mathcal{Y}_\pm$~,
\item $T_i,\;  W_i$ for $i=1,2,\cdots, n-1$~,
\item $\frac{1}{2}(T_n^{(1)} - T_n^{(2)})$,\; $\frac{1}{2}(D_n^{(1)} +
      D_n^{(2)})E$ and
      $\frac{1}{2}(D_n^{(1)}+D_n^{(2)})F$ \;(These form an
      $\mathfrak{su}(2)$ triplet as $H,E$ and $F$.)
\item for each integer $\ell$ such that $1\leq  \ell \leq (n-1)/2$,
\begin{itemize}
 \item $N(n-1,\ell)$ different linear combinations of
      $D_{i_1}D_{i_2}\cdots D_{i_\ell}\mathcal{X}_\pm$~,
 \item $N(n-1,\ell)$ different linear combinations of
      $D_{i_1}D_{i_2}\cdots D_{i_\ell}\mathcal{Y}_\pm$~,
\end{itemize}
\end{itemize}
where $I\subset\{1,2,3,\cdots,n-1\}$, and $N(n,\ell)$ is an integer 
given by \eqref{eq:NnI}. Note also that, since the stress tensor is written in
terms of $T_i$ as in Eq.~\eqref{eq:TTi2}, it is not included in
$\mathtt{S}$. we do not include it as a generator here. The holomorphic
dimension, $U(1)$ charge and $SU(2)$ charge of these generators are
summerized in Table~\ref{table:gen2}.

Note that neither of $\mathcal{F}_{E,\{1\}} -
\mathcal{F}_{E,\{2\}}$ or $\mathcal{F}_{F,\{1\}} -
\mathcal{F}_{F,\{2\}}$ is included in $\mathtt{S}$, because
they are written as a composite operators of the other
generators. Similarly, $\mathcal{O}_{\mathcal{X}_\pm, I, J}$ and
$\mathcal{O}_{\mathcal{Y}_\pm, I, J}$ for $J\neq \emptyset$ are
realized as composite operators of the operators included in $\mathtt{S}$.

As we will see in Sec.~\ref{subsec:example2}, for lower values of $n$, some of the above strong generators are
accidentally composite operators of lower-dimensional generators.
Indeed, for $n=2$, operators of dimension larger than one are all
composite. Similarly, for $n=3$, $W_1+W_2$ is a composite operator of
generators of lower dimensions.

As a consistency check of the above conjecture,
we checked for $n=2,3,4$ and $5$ that the number of independent
generators of dimension up to five is perfectly consistent with Table~\ref{table:gen2}.\footnote{For this computation, we used
the Mathematica package $\mathtt{OPEdefs}$.} Moreover, for $n=2$ and $n=3$, we explicitly computed the OPEs of
the generators
listed in Table~\ref{table:gen2}, and showed that these generators have closed OPEs among themselves as shown in the
next sub-section. These represent very strong
evidence for our conjecture.

\subsection{Examples}
\label{subsec:example2}

In this section, we give some examples of the VOA associated
with $T^{[2,1^{n-1}]}_{[n-1,1^2]}(SU(n+1))$.

\subsubsection{$n=2$}

We begin with the case of $n=2$. According to our general conjecture
stated in the previous sub-section, the VOA associated with
$T^{[2,1]}_{[1^3]}(SU(3))$ is generated by 
\begin{align}
 U~,\qquad H~,\qquad E~,\qquad F~,\qquad \mathcal{X}_\pm~,\qquad
 \mathcal{Y}_{\pm}~,
\label{eq:UHEFXY}
\end{align}
together with\footnote{Note that there is no generator corresponding
to the last or second last row of Table~\ref{table:gen2} in the case
of $n=2$.}
\begin{align}
 T_1~,\qquad W_1~,\qquad \frac{1}{2}(T_2^{(1)}-T_2^{(2)})~,\qquad
 \frac{1}{2}(D_2^{(1)} + D_2^{(2)})E~,\qquad
 \frac{1}{2}(D_2^{(1)} + D_2^{(2)})F~.
\label{eq:TWTF}
\end{align}
However, it turns out that all the operators listed in \eqref{eq:TWTF}
are composite operators of the generators listed in \eqref{eq:UHEFXY},
accidentally in the case of $n=2$. Indeed, we find the following
operator relations:
\begin{align}
                     T_1 &= \frac{9}{8}U^2 -\frac{1}{2}H^2 -\frac{1}{4}\left(EF+FE\right)-\frac{1}{4}\left(\mathcal{X}_+ \mathcal{Y}_- +\mathcal{Y}_- \mathcal{X}_+\right)
                    -\frac{1}{4} \left(\mathcal{X}_{-} \mathcal{Y}_{+}
 +\mathcal{Y}_{+} \mathcal{X}_{-}\right)~,
\\
W_1 &= \frac{1}{\sqrt{6}}\Bigg(\frac{9}{4}U^3 + \frac{U''}{4}
 -\frac{1}{2}U\left(EF+FE+2H^2\right) -
 \frac{3}{4}U\left(\mathcal{X}_+\mathcal{Y}_-
 +\mathcal{Y}_-\mathcal{X}_++\mathcal{X}_-\mathcal{Y}_+ +
 \mathcal{Y}_+\mathcal{X}_-\right)
\nonumber\\
&\qquad + \frac{1}{12}\left(EF+FE+2H^2\right)'
 -\frac{1}{2}\left(\mathcal{X}_+'\mathcal{Y}_-
 -\mathcal{X}_+\mathcal{Y}_-' +
 \mathcal{X}_-'\mathcal{Y}_+-\mathcal{X}_-\mathcal{Y}_+'\right)
\nonumber\\
&\qquad - \frac{1}{3}\Big(E\mathcal{X}_-\mathcal{Y}_- +
 F\mathcal{X}_+\mathcal{Y}_+ -H\left(\mathcal{X}_+\mathcal{Y}_- + \mathcal{X}_-\mathcal{Y}_+\right)\Big)\Bigg)~,
\\
                    T_{2}^{(j)} &= \frac{3}{2}(-1)^{j-1}UH +H^2
 +\frac{1}{4}\left(EF +FE\right)
\nonumber\\
&\qquad +\frac{(-1)^j}{4}\left(\mathcal{X}_{+} \mathcal{Y}_{-} +\mathcal{Y}_{-} \mathcal{X}_{+}\right)
                    +\frac{(-1)^{j-1}}{4} \left(\mathcal{X}_{-}
 \mathcal{Y}_{+} + \mathcal{Y}_{+} \mathcal{X}_{-}\right)~,
\\
                    D_2^{(1)}E &= \frac{3}{2} UE + HE +
 \mathcal{X}_{+}\mathcal{Y}_{+}~,\qquad  D_2^{(2)}E = \frac{3}{2} UE - HE +
 \mathcal{Y}_{+}\mathcal{X}_{+}~,
\\
D_2^{(1)}F &= \frac{3}{2}UF + HF +
 \mathcal{Y}_-\mathcal{X}_-~,\qquad D_2^{(2)}F = \frac{3}{2}UF-HF+\mathcal{X}_-\mathcal{Y}_-~,
\end{align}
where $j=1,2$.

The remaining generators \eqref{eq:UHEFXY} are all of dimension one, and
form the $\widehat{\mathfrak{sl}(3)}_{-1}$ Kac-Moody algebra. To be
more explicit, in terms of
\begin{align}
 J_1 &\coloneqq -\frac{E+F}{2}~,\qquad J_2 \coloneqq
 \frac{i(E-F)}{2}~,\qquad J_3 \coloneqq H~,\qquad J_4 \coloneqq
 -\frac{\mathcal{X}_+ - \mathcal{Y}_m}{2}~,
\\
J_5 &\coloneqq \frac{i(\mathcal{X}_+ + \mathcal{Y}_m)}{2}~,\qquad J_6
 \coloneqq -\frac{\mathcal{X}_- - \mathcal{Y}_+}{2}~,\qquad J_7 \coloneqq
 \frac{i(\mathcal{X}_- + \mathcal{Y}_+)}{2}~,\qquad J_8 \coloneqq \frac{\sqrt{3}}{2}U~,
\end{align}
the OPEs among them are summerized as
\begin{align}
 J_\ell(z) J_m(0) \sim -\frac{1}{2z^2} + \frac{1}{z}\sum_{p=1}^8 i f_{\ell m p}
 J_p(0)~,
\end{align}
where $f_{\ell m p}$ are structure constants of
$\mathfrak{sl}(3)$.\footnote{Our convention for the structure constant is such that $\sum_{m, p =1}^8f_{\ell
m p}f_{q m p} = 3 \delta_{\ell q}$.}
Hence, our general conjecture implies that the VOA associated
with $T^{[2,1]}_{[1^3]}(SU(3))$ is the $\widehat{\mathfrak{sl}(3)}_{-1}$ Kac-Moody
algebra. Note that the stress tensor \eqref{eq:TTi2} is precisely
identical to the Sugawara stress tensor $T= \frac{1}{-1+h^\vee}\sum_{\ell=1}^8J_\ell J_\ell$ with
$h^\vee=3$ for $\mathfrak{sl}(3)$.

\subsubsection{$n=3$}
\label{subsubsec:A1D6}

Let us move to the case of $n=3$. Our conjecture stated in
Sec.~\ref{subsec:conjecture2} implies that the VOA associated with
$T^{[2,1^2]}_{[2,1^2]}(SU(4))$ is generated by the following generators;
$U,\, H,\,E$ and $F$ of dimension
one, $\mathcal{X}_\pm$ and $\mathcal{Y}_\pm$ of dimension $3/2$,
$T_1,\, T_2,\, \frac{1}{2}(T_3^{(1)}-T_3^{(2)}),\, \frac{1}{2}(D_3^{(1)} +
D_3^{(2)})E$ and $\frac{1}{2}(D_3^{(1)} + D_3^{(2)})F$ of dimension two, $W_1,\, W_2$ of  dimension three, and
$N(2,1)=1$ linear combination of
$D_1\mathcal{X}_\pm$ and $D_{2}\mathcal{X}_\pm$
as well as a similar linear combination of 
$D_1\mathcal{Y}_\pm$ and
$D_2\mathcal{Y}_\pm$. See Table \ref{table:gen2} for their
$U(1)$ and $SU(2)$ charges.

In contrast to the $n=2$ case, most of the above operators are
independent generators of the VOA. The only
exception is $W_1+ W_2$, which turns out to be a composite operator of
the other generators of the form
\begin{align}
W_1 + W_2 &= 
                    \frac{5U T}{\sqrt{6}}-\frac{H \tilde{H}}{3 \sqrt{6}}-\frac{1}{3} \sqrt{\frac{2}{3}} E \tilde{F} +\frac{5}{4} \sqrt{\frac{3}{2}} U \partial U
                    +\frac{5 U \partial H}{2\sqrt{6}}+\frac{H \partial H}{\sqrt{6}}+\frac{E \partial F}{2\sqrt{6}}+\frac{F \partial E}{2 \sqrt{6}}\\
                    &\quad +\frac{25 U U U }{24 \sqrt{6}} -\frac{5U H H }{2 \sqrt{6}}-\frac{5 U E F }{2\sqrt{6}}-\frac{ \mathcal{X}_{+} \mathcal{Y}_{-} }{\sqrt{6}}
                    -\frac{\mathcal{X}_{-} \mathcal{Y}_{+}}{\sqrt{6}}-\frac{5 \partial^2 U}{6\sqrt{6}}-\frac{\partial T}{\sqrt{6}}+\frac{\partial \tilde{H}}{6 \sqrt{6}}~.
\end{align}
Therefore, one can take the strong generators of this VOA as
\begin{align}
 &U~,\quad H~,\quad E~,\quad F~,\quad \mathcal{X}_\pm~,\quad
 \mathcal{Y}_\pm~,\quad T~,\quad \tau \coloneqq T_1-T_2~,
\label{eq:A1D6-gen1}
\\
 &\widetilde{H} \coloneqq
 T_3^{(1)}-T_3^{(2)}~,\quad \widetilde{E}\coloneqq
 \frac{1}{2}\left(
\mathcal{F}_{E,\{1\}} + \mathcal{F}_{E,\{2\}}
\right)
~,\quad
 \widetilde{F}\coloneqq
 \frac{1}{2}\left (
\mathcal{F}_{F,\{1\}}+\mathcal{F}_{F,\{2\}}
\right)
~,
\\
&\widetilde{\mathcal{X}}_\pm \coloneqq
 \frac{1}{2}\left(D_1-D_2\right)\mathcal{X}_\pm~,\quad
 \widetilde{\mathcal{Y}}_\pm\coloneqq
 \frac{1}{2}\left(D_1-D_2\right)\mathcal{Y}_\pm~,\quad \mathcal{W}\coloneqq \frac{1}{2}\sqrt{\frac{3}{2}}(W_1-W_2)~,
\label{eq:A1D6-gen4}
\end{align}
where $T$ is the stress tensor given by Eq.~\eqref{eq:TTi2} generating the Virasoro vertex (sub-)algebra at $c=-6$:
\begin{align}
 T(z)T(0) \sim -\frac{6}{2z^4} + \frac{2T}{z^2} + \frac{\partial T}{z}~.
\end{align}
Note that $\widetilde{\mathcal{X}}_\pm$ and
$\widetilde{\mathcal{Y}}_\pm$ are those listed in the last and second
last rows of Table~\ref{table:gen2}.
All the other generators are Virasoro primary
operators, 
which
completely fixes
their OPE with the stress tensor $T$. When focusing on dimension-one
Virasoro primaries, $U$ generates the Heisenberg vertex (sub-)algebra:
\begin{align}
 U(z)U(0) \sim -\frac{2}{5z^2}~,
\label{eq:A1D6-currents1}
\end{align}
and $H,E$ and $F$ generate the $\widehat{\mathfrak{su}(2)}_{-1}$ vertex
(sub-)algebra:
\begin{align}
 H(z)H(0)\sim -\frac{1}{2z^2}~,\qquad H(z)E(0)\sim \frac{E}{z}~,\qquad
 H(z)F(0)\sim -\frac{F}{z}~,\qquad  E(z)F(0)\sim -\frac{1}{z^2} +
 \frac{2H}{z}~.
\end{align} 
All the other OPEs among $U,H,E$ and $F$ vanish. Similarly, the generators except
for $T,U,H,E,F$ are primary operator with respect to this Heisenberg and
$\widehat{\mathfrak{su}(2)}_{-1}$ algebras. This almost completely fixes their
OPEs with $U,H,E$ and $F$ in terms of their $U(1)$ and $SU(2)$ charges
that can be read off from Table \ref{table:gen2}, except for the
relative normalization of the $\mathfrak{su}(2)$ triplet $\widetilde{H},
\, \widetilde{E},\,\widetilde{F}$ and the $\mathfrak{su}(2)$ doublets $\mathcal{X}_\pm,\,\mathcal{Y}_\pm,\,\widetilde{\mathcal{X}}_\pm,\,\widetilde{\mathcal{Y}}_\pm$; our normalization is
such that
\begin{align}
 E(z)\widetilde{H}(0) &\sim -\frac{2\widetilde{E}}{z}~,\qquad F(z)
 \widetilde{H}(0) \sim \frac{2\widetilde{F}}{z}~,
\\
E(z)\widetilde{F}(0) &\sim \frac{\tilde{H}}{z}~,\;\;\;\;\qquad F(z)\widetilde{E}(0) \sim -\frac{\tilde{H}}{z}~,
\\
E(z)\mathcal{X}_-(0) &\sim -\frac{\mathcal{X}_+}{z}~,\qquad
 F(z)\mathcal{X}_+(0) \sim - \frac{\mathcal{X}_-}{z}~,
\\
 E(z)\mathcal{Y}_-(0) &\sim +\frac{\mathcal{Y}_+}{z}~,\qquad
 F(z)\mathcal{Y}_+(0) \sim + \frac{\mathcal{Y}_-}{z}~,
\\
E(z)\widetilde{\mathcal{X}}_-(0) &\sim -\frac{\widetilde{\mathcal{X}}_+}{z}~,\qquad
 F(z)\widetilde{\mathcal{X}}_+(0) \sim - \frac{\widetilde{\mathcal{X}}_-}{z}~,
\\
 E(z)\widetilde{\mathcal{Y}}_-(0) &\sim +\frac{\widetilde{\mathcal{Y}}_+}{z}~,\qquad
 F(z)\widetilde{\mathcal{Y}}_+(0) \sim +
 \frac{\widetilde{\mathcal{Y}}_-}{z}~.
\label{eq:A1D6-currents8}
\end{align}
 The other OPEs of the
generators are not fixed by dimension or charges, which we describe in
appendix \ref{app:A1D6} in detail.

\subsection{Comparison to the VOA associated with $(A_1,D_{2n})$}
\label{subsec:comparison2}

Let us now compare the bosonic VOA for $T^{[2,1^{n-1}]}_{[n-1,1^2]}(SU(n+1))$
with the VOA associated with the $(A_1,D_{2n})$ theory in the sense of
\cite{Beem:2013sza}.

\begin{table}
\centering
\begin{tabular}{|c|c|c|c|}\hline
generator & dimension & $U(1)$ charge & $SU(2)$ charge
\\ \hline
$J$ & $1$ & $0$ & $0$ \\
\hdashline
$\mathsf{J}_0$ & $1$ & $0$ & $0$ \\
$\mathsf{J}_\pm$ & $1$ & $0$ & $\pm1$\\
\hdashline
$X_\pm$& $n/2$ & $+1$ & $\pm 1/2$\\
\hdashline
 $Y_\pm$& $n/2$ & $-1$ & $\pm 1/2$\\
\hdashline
$T^{(i)}$ & $i$ & $0$ & $0$\\
\hline
\end{tabular}
\caption{List of strong generators of
 $\mathcal{W}^{-\frac{n^2-1}{n}}(\mathfrak{sl}_{n+1},f)$, where $f$ is
 a nilpotent element of $\mathfrak{sl}_{n+1}$ corresponding to the
 partition $[n-1,1^2]$. The dashed lines separate different $SU(2)$ multiplets.
 The index $i$ runs over $2,3,\cdots, n-1$. }
\label{table:gen4}
\end{table}

It was conjectured in \cite{Creutzig:2017qyf} that the VOA associated with
$(A_1,D_{2n})$ is conjectured to be
$\mathcal{W}^{-\frac{n^2-1}{n}}(\mathfrak{sl}_{n+1},f)$, where $f$ is a nilpotent
element of $\mathfrak{sl}_{n+1}$ corresponding to the partition $[n-1,1^2]$.
This vertex algebra
is strongly generated by 
\begin{align}
 J,~\quad \mathsf{J}_0~\quad \mathsf{J}_+~,\quad \mathsf{J}_-,~,\quad X_\pm ~,\quad Y_\pm~,\quad T^{(2)}~,\quad T^{(3)}~,\quad \cdots
 ~,\quad T^{(n-1)}~,
\label{eq:gen3}
\end{align}
where $J$ is a dimension-one Heisenberg current,
$(\mathsf{J}_{0},\,\mathsf{J}_\pm)$ are affine $\mathfrak{sl}_2$
currents, $X_\pm$ and $Y_\pm$ are operators of dimension $n/2$, and
$T^{(i)}$ for $i=2,\cdots,n-1$ are operators of dimension $i$. The
$U(1)$ and $SU(2)$ charges of these generators are shown in Table~\ref{table:gen4}.

Comparing Table~\ref{table:gen4} with Table~\ref{table:gen2}, we see
that the charge and dimension of
$U,\,H,\,E,\,F$, $\mathcal{X}_\pm,\,\mathcal{Y}_\pm$ and $\sum_i T_i$ in Table~\ref{table:gen2} are
identical to those of $J,\,
\mathsf{J}_0,\,\mathsf{J}_+,\,\mathsf{J}_-,\,X_\pm,\,Y_\pm$  and
$T^{(2)}$ in
Table~\ref{table:gen4}, respectively. Similarly, the charge and
dimension of $T^{(i)}$ for $i=3,4,\cdots, n-1$ are identical to
$\sum_{i} (T_i)^{k} (W_i)^{\ell}$  such that $2k + 3\ell =i$. It is
therefore tempting to identify these operators. However, we see that
the other generators in Table~\ref{table:gen2} have no counterpart in
Table~\ref{table:gen4}, which implies that the VOA associated with
$T^{[2,1^{n-1}]}_{[n-1,1^2]}(SU(n+1)))$ has more generators than that
associated with $(A_1,D_{2n})$.

As in the case of $(A_1,A_{2n-1})$, this discrepancy becomes milder when
we consider an $S_{n-1}$ action induced by
\begin{align}
 X_i \to X_{\sigma(i)}~,\qquad Y_i \to Y_{\sigma(i)}~,
\end{align}
where $\sigma \in S_{n-1}$. Indeed, taking the $S_{n-1}$-invariant
vertex sub-algebra, one can remove $T_i$ and $W_i$ shown in
Table~\ref{table:gen2}. However
\begin{align}
 \frac{1}{2}(T_n^{(1)} - T_n^{(2)})~,\qquad
 \frac{1}{2}(D_n^{(1)}+D_n^{(2)})E~,\qquad \frac{1}{2}(D_n^{(1)} + D_n^{(2)})F
\end{align}
survive this $S_{n-1}$-quotient. 

Furthermore, even though the generators $U,\, H,\,
E,\,F,\,\mathcal{X}_\pm,\,\mathcal{Y}_\pm$ and $\sum_{i}T_i$ in
Table~\ref{table:gen2} have the same charge and dimension as
$J,\,\mathsf{J}_0,\,\mathsf{J}_+,\,\mathsf{J}_-,\,X_\pm,\,Y_\pm$ and
$T^{(2)}$ in Table~\ref{table:gen4}, their OPEs are different. For
instance, the level of the affine $\mathfrak{su}(2)$ currents $H,E$ and
$F$ of our VOA is $-1$ as explained in
Sec.~\ref{subsec:cand3}, while the affine $\mathfrak{su}(2)$ currents
$\mathsf{J}_-,\,\mathsf{J}_+$ and $\mathsf{J}_-$ have fractional level
$-\frac{n^2-1}{n}$. Therefore, the VOA associated with
$T^{[2,1^{n-1}]}_{[n-1,1^2]}(SU(n+1))$ and that associated with
$(A_1,D_{2n})$ have different OPEs even for generators of the same charges and
dimensions.

Let us briefly comment on the special feature of $n=2$. As seen in
Sec.~\ref{subsec:example2}, our bosonic VOA is accidentally simplified
to be $\widehat{\mathfrak{su}(3)}_{-1}$. Interestingly, the VOA associated with the 4D
ancestor, i.e. $(A_1,D_4)$, is also an affine $\mathfrak{su}(3)$ vertex
algebra, but with a different value of the level, $-3/2$. Although the levels
are different, they share the same number of generators of the same
dimensions.

\section{Summary and Conclusions}
\label{sec:summary}

In this paper, we have studied bosonic VOAs associated with 3D reduction
of 4D Argyres-Douglas theories of $(A_1,A_{2n-1})$ and $(A_1,D_{2n})$
types. These 3D reductions are 3D $\mathcal{N}=4$ abelian linear quiver
gauge theories called $T^{[1^n]}_{[n-1,1]}(SU(n))$ and
$T^{[2,1^{n-1}]}_{[n-1,1^2]}(SU(n+1))$. 
The bosonic VOA we consider is obtained by putting the 3D theory on
$\mathbb{R}_{\geq 0} \times \mathbb{C}$, considering the bulk H-twist  with a holomorphic boundary condition \cite{Costello:2018fnz}, and canceling the gauge anomaly
by Heisenberg algebras on the boundary instead of Fermi
multiplets. Technically, these VOAs are obtained by a certain BRST
reduction of the tensor product of symplectic bosons, $bc$-ghosts and the Heisenberg algebras.

In
particular, we have conjectured a complete set of strong generators of the
bosonic VOAs for $T^{[1^n]}_{n-1,1}(SU(n))$ and
$T^{[2,1^{n-1}]}_{[n-1,1^2]}(SU(n+1))$. The conjectured generators for
$T^{[1^n]}_{[n-1,1]}(SU(n))$ are shown in Table~\ref{table:gen1}, and
those for $T^{[2,1n^{n-1}]}_{[n-1,1^2]}(SU(n+1))$ are shown in
Table~\ref{table:gen2}.  We have explicitly expressed these strong generators as
representatives of the BRST cohomologies, except for the last two lines
of Table~\ref{table:gen1} and Table~\ref{table:gen2}. Note that, for lower values of $n$, some of
these generators are composite operators of lower-dimensional
generators, as described.

As a consistency check, we have shown for $n=2$ and $3$ that our
conjectured generators for $T^{[1^n]}_{[n-1,1]}(SU(n))$ and $T^{[2,1^{n-1}]}_{[n-1,1^2]}(SU(n+1))$
have closed OPEs among themselves. These OPEs are explicitly shown in
Sec.~\ref{subsec:example1} and \ref{subsec:example2}, and
Appendix~\ref{app:A1A5} and \ref{app:A1D6}. Moreover, for
$T^{[1^n]}_{[n-1,1]}(SU(n))$ with $n=2,3,\cdots, 7$, we have checked that the
number of strong generators of dimension less than or equal to $13/2$ is
perfectly consistent with Table~\ref{table:gen1}. We have also done a
similar check for $T^{[2,1^{n-1}]}_{[n-1,1^2]}(SU(n+1))$ with
$n=2,3,\cdots,5$ up to dimension five. These represent very strong evidence for our conjecture.

We have also compared our conjectured generators for
$T^{[1^n]}_{[n-1,1]}(SU(n))$ and $T^{[2,1^{n-1}]}_{n-1,1^2}(SU(n+1))$
with the generators of the VOAs associated 
 with their 4D ancestors, $(A_1,A_{2n-1})$ and
$(A_1,D_{2n})$ respectively. We see that the bosonic VOAs for the 3D theories
generally have a larger number of generators than those for their 4D ancestors. While taking a sub-algebra
invariant under a symmetric group $S_n$ or $S_{n-1}$ makes this
discrepancy milder, OPEs of the generators are generally different.

As a future direction, it would be interesting to study the associated
variety \cite{Arakawa} of the bosonic VOAs we studied in this
paper. Since the gauge anomaly is canceled by Heisenberg algebras
instead of Fermi multiplets, the associated variety is a continuous family of
symplectic singularities \cite{Kuwabara}. This is in physics expected to be the family
of the Higgs branch of the 3D $\mathcal{N}=4$ gauge theory deformed by
(an $\mathcal{N}=4$ partner of) FI parameters.

Another future direction will be to search for a deformation of the
BRST reduction so that the bosonic VOA associated with
$T^{[1^n]}_{[n-1,1]}(SU(n))$ and $T^{[2,1^{n-1}]}_{n-1,1^2}(SU(n+1))$
coincide with the VOAs associated with their 4D ancestors. One
possibility might be to take into account non-vanishing FI parameters
for the $U(1)$ gauge groups. Indeed, it was shown in
\cite{Buican:2015hsa} that the $q\to 1$ limit of the Schur index of the
$(A_1,A_{2n-1})$ and $(A_1,D_{2n})$ theories respectively coincide with the $S^3$
partition function of $T^{[1^n]}_{[n-1,1]}(SU(n))$ and
$T^{[2,1^{n-1}]}_{n-1,1^2}(SU(n+1))$, only when complex FI
parameters are turned on in three dimensions. It would be interesting to
see if these complex FI parameters could modify the BRST construction of the
bosonic VOAs studied in this paper.

\bigskip
\bigskip

\begin{center}
\noindent{\bf Acknowledgments}
\end{center}

We are grateful to Ryo Hamachika and Yutaka Yoshida for illuminating discussions. Most of our
computations have been carried out with the Mathematica package OPEdefs provided by K. Thielemans
\cite{Thielemans:1991uw, Thielemans:1994er} to whom the authors are grateful. The authors' research
	is partially supported by JSPS KAKENHI Grant Number JP21H04993.
	In addition, T.~N’s research is partially supported by JSPS KAKENHI
	Grant Numbers JP18K13547, 23K03394 and 23K03393.

\appendix

\section{Nilpotency of the BRST charge}
\label{app:nilpotency}

\begin{figure}
\centering
\begin{tikzpicture}[gauge/.style={circle,draw=black,inner sep=0pt,minimum size=10mm},flavor/.style={rectangle,draw=black,inner sep=0pt,minimum size=10mm},auto]
 \node[flavor] (0) at (-2,0) {\;$1$\;};
 \node[gauge] (1) at (0,0)  {\;$1$\;} edge (0);
 \node[gauge] (2) at (2,0)  {\;$1$\;} edge (1);
 \node (3) at (4,0) {\;$\cdots$\;} edge (2);
 \node[gauge] (4) at (6,0)  {\;$1$\;} edge (3);
 \node[flavor] (5) at (8,0)  {\;$N_f$\;} edge (4);
\end{tikzpicture}
\caption{The quiver diagram of a 3D $\mathcal{N}=4$ abelian quiver gauge
 theory. There are $(n-1)$ $U(1)$ vector multiplets and $n-2$
 bifundamental hypermultiplets. At the left end of the quiver, there is
 a hypermultiplet coupled to the leftmost $U(1)$, while $N_f$
 hypermultiplets are coupled to the rightmost $U(1)$.
 }
\label{fig:quiver5}
\end{figure}
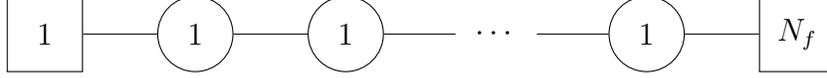

In this appendix, we explicitly check that the BRST currents given in
\eqref{eq:BRST} and \eqref{eq:BRST2} give rise to a nilpotent BRST
charge by $Q_\text{BRST} = \oint \frac{dz}{2\pi i}J_\text{BRST}(z)$.
To discuss \eqref{eq:BRST} for $T^{[1^n]}_{n-1,1}SU(n)$ and
\eqref{eq:BRST2} for $T^{[2,1^{n-1}]}_{[n-1,1^2]}(SU(n+1))$
simultaneously, we here consider the 3D $\mathcal{N}=4$ quiver gauge
theory described by the diagram in Fig.~\ref{fig:quiver5}. We see that
the $N_f=1$ case corresponds to $T^{[1^n]}_{[n-1,1]}(SU(n))$ and the
$N_f=2$ case leads to $T^{[2,1^{n-1}]}_{[n-1,1^2]}(SU(n+1))$.

The BRST current for general $N_f$ is defined by
\begin{align}
 J_\text{BRST} = \sum_{a=1}^{n-1}c^a\left(J^\text{sb}_a + J^h_a\right)~,
\label{eq:BRST3}
\end{align}
where
\begin{align}
 J^\text{sb}_a = \sum_{i=1}^{n-1}Q^{i}_a
 X_iY_i + \sum_{j=1}^{N_f}Q_{a,j}^{n}X_n^{(j)}Y_n^{(j)}~,\qquad
 J^{h}_a=\sum_{i=1}^{n-1}\tilde{Q}_{a}^i h_i~.
\end{align}
Here $X_n^{(j)}$ and $Y_n^{(j)}$ for $j=1,2,\cdots,N_f$ are the
symplectic bosons associated with the $N_f$ hypermulitplets at the right
end of the quiver, and $Q_a^i,\,Q_{a,j}^{n}$ and $\tilde{Q}^{i}_a$ are
coefficients to be determined so that $Q_\text{BRST}^2 = 0$.

By a straightforward computation, we see that 
\begin{align}
J_\text{BRST}(z)J_\text{BRST}(0) &\sim
 \sum_{a,b=1}^{n-1}\left [\left(\sum_{i,j=1}^{n-1}\widehat{C}_{ij}\tilde{Q}^{i}_a\tilde{Q}^{j}_b
 - \sum_{i=1}^{n-1}Q^{i}_a Q^{i}_b -
 \sum_{j=1}^{N_f}Q_{a,j}^{n} Q_{b,j}^{n}\right)\frac{(\partial
 c^a)c^b}{z}\right]~,
\label{eq:JJ}
\end{align}
where $\widehat{C}_{ij}$ is the matrix characterizing the Heisenberg
algebra
\begin{align}
 h_i(z) h_j(0) \sim \frac{\widehat{C}_{ij}}{z}.
\end{align}
We see from \eqref{eq:JJ} that $Q_\text{BRST}^2 = 0$ if and only if
\begin{align}
  \sum_{i=1}^{n-1}Q_a^iQ_b^{i} + \sum_{j=1}^{N_f}Q_{a,j}^{n}Q_{b,j}^{n} = \sum_{i,j=1}^{n-1}\widehat{C}_{ij}\tilde{Q}_a^{i}\tilde{Q}_b^{j}~.
\end{align}
Note that this constraint can be satisfied by
\begin{align}
 Q_a^{i} = \delta_a^{i} - \delta_a^{i-1}~,\qquad Q_{a,j}^{n} =
 -\delta^{n-1}_a~,\qquad \tilde{Q}^{i}_a = \delta^{i}_a~,
\label{eq:const1}
\end{align}
and
\begin{align}
 \widehat C_{ij}  = \left\{
\begin{array}{l} 
2 \qquad \text{if}\quad i = j = 1,2,\cdots,n-2\\
1 + N_f \qquad \text{if} \quad i = j = n-1\\
-1 \qquad \text{if} \quad |i-j|=1\\
0 \qquad \text{otherwise}\\
\end{array}
\right.~.
\label{eq:const2}
\end{align}
One can check that our BRST currents \eqref{eq:BRST} and
\eqref{eq:BRST2} are precisely identical to \eqref{eq:BRST3} with
\eqref{eq:const1} and \eqref{eq:const2} imposed, respectively for
$N_f=1$ and $N_f=2$.

\section{Rewriting the Virasoro stress tensor}
\label{app:T}

In this appendix, we show that the Virasoro stress tensor is written in
terms of symplectic bosons up to BRST exact terms.
In Sec.~\ref{app:rewritingT}, we show that the stress tensor defined in
\eqref{eq:totalT} for the VOA associated with
$T^{[1^n]}_{[n-1,1]}(SU(n))$ is rewritten as \eqref{eq:canonicalT} up to
BRST exact terms. Similarly, in Sec.~\ref{app:rewritingT2}, we show that
\eqref{eq:T2} defined for the VOA associated with
$T^{[2,1^{n-1}]}_{[n-1,1^2]}(SU(n+1))$ is identical to \eqref{eq:TTi2}
up to BRST exact terms.

\subsection{$T^{[1^n]}_{[n-1,1]}(SU(n))$ theory}

\label{app:rewritingT}
Here, we show that the stress tensor \eqref{eq:totalT} is written as
        \eqref{eq:canonicalT} up to BRST-exact terms.
First note that
the following relations hold:
        \begin{align}
            h_i = \mathcal{U}_i - \mathcal{U}_{i+1} +\{Q_\text{BRST},\, b_i\}~,
\qquad
            \big[Q_\text{BRST},\,
\mathcal{U}_{i}
	 -\mathcal{U}_{i+1}
\big] = \sum_{j=1}^{n-1}C_{ij} \partial c_j~,
        \end{align}
where $\mathcal{U}_i$ is defined in \eqref{eq:Di}.
        Using these relations, the expression for $h_i$ can be transformed as follows\footnote{"$A \equiv B$" means that $A$ is equal to $B$ up to $Q$-exact terms.}
        \begin{align}
            \frac{1}{2}\sum_{i,j=1}^{n-1}C^{ij}h_i h_j &=
	 \frac{1}{2}\sum_{i,j=1}^{n-1}C^{ij}\left(\mathcal{U}_i -
	 \mathcal{U}_{i+1} +\{Q_\text{BRST},\, b_i\} \right)\left(\mathcal{U}_j -
	 \mathcal{U}_{j+1} +\{Q_\text{BRST},\, b_j\}\right) \notag\\
            &\equiv \frac{1}{2}\sum_{i,j=1}^{n-1}C^{ij}\left(\mathcal{U}_{i}\mathcal{U}_{j}-\mathcal{U}_{i}\mathcal{U}_{j+1}-\mathcal{U}_{i+1}\mathcal{U}_{j}+\mathcal{U}_{i+1}\mathcal{U}_{j+1}\right) +\sum_{i=1}^{n-1} b_i \partial c_i~,
        \end{align}
where the symbol ``$\equiv$'' in the last line means equivalence up to
$Q_\text{BRST}$-exact terms, and $C^{ij}$ is the matrix element of the
inverse Cartan matrix of $\mathfrak{su}(n)$.
        Thus, the terms 
in the stress tensor involving 
$h_i~,\ b_i~,\ c_i$ can be written in terms of 
$\mathcal{U}_i$ as 
        \begin{align}
            \label{relation:h,b,c,U_i}
            \frac{1}{2}\sum_{i,j=1}^{n-1}C^{ij}h_i h_j -\sum_{i=1}^{n-1}b_i \partial c_i \equiv
            \frac{1}{2}\sum_{i,j=1}^{n-1}C^{ij}\left(\mathcal{U}_{i}\mathcal{U}_{j}-\mathcal{U}_{i}\mathcal{U}_{j+1}-\mathcal{U}_{i+1}\mathcal{U}_{j}+\mathcal{U}_{i+1}\mathcal{U}_{j+1}\right)~.
        \end{align}

Note that the inverse Cartan matrix element $C^{ij}$ 
is written as
\cite{Wei:2017}
        \begin{align}
            C^{ij} = \min \left(i,j\right) -\frac{i j}{n} \quad (i,j
	 =1,\dots,n-1)~.
\label{eq:Wei}
        \end{align}
        For convenience, the range of $i,j$ may be extended to
	$i,j=0,\dots,n$ from $i,j = 1,\dots,n-1$, with
$C^{0j}=C^{i0}=C^{nj}=C^{in}=0$.
Using this extended matrix elements, one can write 
the right hand side of \eqref{relation:h,b,c,U_i}
as 
        \begin{align}
            \frac{1}{2}\sum_{i,j=1}^{n}\left(C^{ij}-C^{i-1,j}-C^{i,j-1}+C^{i-1,j-1}\right)\mathcal{U}_{i}\mathcal{U}_{j}~.
        \end{align}
Using \eqref{eq:Wei}, the expression in the above bracket can be
        simplied as
        \begin{align}
            C^{ij} -C^{i-1,j} -C^{i,j-1}+C^{i-1,j-1} = \delta_{ij} -\frac{1}{n}~,
        \end{align}
and therefore we find
        \begin{align}
            \frac{1}{2}\sum_{i,j=1}^{n}\left(C^{ij}-C^{i-1,j}-C^{i,j-1}+C^{i-1,j-1}\right)\mathcal{U}_{i}\mathcal{U}_{j}
            &= \frac{1}{2} \sum_{i=1}^{n} (\mathcal{U}_{i})^2
 -\frac{n}{2} U^2~,
        \end{align}
where  $U$ is defined in
       \eqref{eq:XYU} and therefore $U =
       \frac{1}{n}\sum_{i=1}^n\mathcal{U}_i$.
       In other words, 
\eqref{relation:h,b,c,U_i} 
can be expressed as
        \begin{align}
            \frac{1}{2}\sum_{i,j=1}^{n-1}C^{ij}h_i h_j -\sum_{i=1}^{n-1}b_i \partial c_i \equiv
            \frac{1}{2} \sum_{i=1}^{n} (\mathcal{U}_{i})^2
	 -\frac{n}{2} U^2~,
\label{eq:app-T1}
        \end{align}
up to exact terms.

Similarly,  the terms 
in the stress tensor that involve
$X_i$ and $Y_i$ can be rewritten as
        \begin{align}
            \frac{1}{2}\sum_{i=1}^{n} \left(X_i \partial Y_i -\partial
	 X_i Y_i\right) = \frac{1}{2}\sum_{i=1}^{n}\mathcal{D}_i
	 \mathcal{U}_i~,
\label{eq:app-T2}
        \end{align}
where $\mathcal{D}_i$ is defined in \eqref{eq:Di}.
        
Combining \eqref{eq:app-T1} and \eqref{eq:app-T2}, we see that the total
stress tensor is written as
        \begin{align}
            T &\equiv  \frac{1}{2}\sum_{i=1}^{n}(\mathcal{D}_i + \mathcal{U}_i)\mathcal{U}_i - \frac{n}{2}U^2= \sum_{i=1}^{n}T_i - \frac{n}{2}U^2
        \end{align}
where $T_i$ is defined in \eqref{eq:TiWi}. This implies the equivalence
of \eqref{eq:totalT} and \eqref{eq:canonicalT} up to exact terms.

\subsection{$T^{[2,1^{n-1}]}_{[n-1,1^2]}(SU(n+1))$ theory}
\label{app:rewritingT2}

We here show that \eqref{eq:T2} can be written as \eqref{eq:TTi2} up to
BRST exact terms.

To that end, we first show that the matrix $\tilde{C}$ whose elements are given by
\eqref{eq:C2} is invertible, by explicitly constructing
the inverse of $\tilde{C}$ using the Sherman-Morrison formula
\cite{Golub}. Let us define a column vector $u\in
\mathbb{R}^{n-1}$ by
\begin{align}
 u = \left(
\begin{array}{c} 
0\\
\vdots\\
0\\
1\\
\end{array}\right)~,
\end{align}
and then compute $1 + u^TC^{-1}u$, where $C$ is the Cartan matrix of $\mathfrak{su}(n)$. Using
\eqref{eq:Wei}, we see that 
\begin{align}
 1 + u^TC^{-1}u = 1 + C^{n-1,n-1} = \frac{2n-1}{n}\neq 0~.
\end{align}
Then the inverse of $\tilde{C} = C + uu^T$ is explicitly given by the
Sherman-Morrison formula:
\begin{align}
 \tilde{C}^{-1} = (C+ uu^T)^{-1} = C^{-1} - \frac{C^{-1}u u^T
 C^{-1}}{1+u^T C^{-1}u}~.
\label{eq:S-M formula}
\end{align} 
From \eqref{eq:Wei} and \eqref{eq:S-M formula}, we see that the matrix
element of $\widetilde{C}^{-1}$ is written as
        \begin{align}
            \tilde{C}^{ij} 
            &= \min \left(i,j\right) -\frac{2 ij}{2n-1}~.
\label{eq:tildeC-ij}
        \end{align}

Below, we will show the equivalence of \eqref{eq:T2} and \eqref{eq:TTi2}
up to exact terms using \eqref{eq:tildeC-ij}.
Note first that $\mathcal{U}_i$ defined in \eqref{eq:Di} and
        $\mathcal{U}_i^{(j)}$ defined in \eqref{eq:Dij} satisfy the
        following relations:
        \begin{align}
            h_i &= \mathcal{U}_i -\mathcal{U}_{i+1} + \{Q_\text{BRST},\, b_i\}~,\\
            h_{n-1} &= \mathcal{U}_{n-1}
	 -\sum_{j=1}^{2}\mathcal{U}_{n}^{(j)} + \{Q_\text{BRST},\,
	 b_{n-1}\}
~,\\
            \Big[Q_\text{BRST},\;\mathcal{U}_{i} -\mathcal{U}_{i+1}\Big]
	 &= \sum_{k=1}^{n-1}\tilde{C}_{ik} \partial c_k~,\\
            \Big[Q_\text{BRST},\;\mathcal{U}_{n-1} -\sum_{j=1}^{2} \mathcal{U}_{n}^{(j)}\Big] &= \sum_{k=1}^{n-1}\tilde{C}_{n-1,k}\partial c_k~, 
        \end{align}
where $i=1,2,\cdots,n-2$.
Using these 
relations, we find that the sum of the two terms in \eqref{eq:Tbch2} is
written up to exact terms as
        \begin{align}
            \label{relation:h,b,c,U_i,U_n}
T_{bc} + T_h
&\equiv 
            \frac{1}{2} \sum_{i,j=1}^{n-2}\tilde{C}^{ij}\left(\mathcal{U}_{i} -\mathcal{U}_{i+1}\right) \left(\mathcal{U}_{j} -\mathcal{U}_{j+1}\right) \notag \\
            &\quad +\frac{1}{2} \sum_{j=1}^{n-2} \tilde{C}^{n-1,j}\left(\mathcal{U}_{n-1} -\sum_{l=1}^{2} \mathcal{U}_{n}^{(l)}\right)\left(\mathcal{U}_{j} -\mathcal{U}_{j+1}\right) \notag \\
            &\quad +\frac{1}{2} \sum_{i=1}^{n-2} \tilde{C}^{i,n-1} \left(\mathcal{U}_{i} -\mathcal{U}_{i+1} \right) \left(\mathcal{U}_{n-1} -\sum_{l=1}^{2}\mathcal{U}_{n}^{(l)}\right) \notag \\
            &\quad +\frac{1}{2} \tilde{C}^{n-1,n-1}
	 \left(\mathcal{U}_{n-1} -\sum_{l=1}^{2} \mathcal{U}_{n}^{(l)}
	 \right)\left(\mathcal{U}_{n-1} -\sum_{m=1}^{2}
	 \mathcal{U}_{n}^{(m)} \right)~,
        \end{align}
where the symbol ``$\equiv$'' stands for equivalence up to
        $Q_\text{BRST}$-exact terms.
        For convenience, let us define 
$\tilde{C}^{0j} = \tilde{C}^{i0}=0~$. Then each term on the RHS have
        simpler expressions as follows. The first term on the RHS of
        \eqref{relation:h,b,c,U_i,U_n} is expressed as
        \begin{align}
            &\frac{1}{2} \sum_{i,j=1}^{n-2}\tilde{C}^{ij}\left(\mathcal{U}_{i} -\mathcal{U}_{i+1}\right) \left(\mathcal{U}_{j} -\mathcal{U}_{j+1}\right) \notag \\
            &= \frac{1}{2} \sum_{i,j=1}^{n-2}\left(\tilde{C}^{ij}-\tilde{C}^{i,j-1}-\tilde{C}^{i-1,j}+\tilde{C}^{i-1,j-1}\right)\mathcal{U}_{i} \mathcal{U}_{j} +\frac{1}{2} \sum_{i=1}^{n-2} \left(\tilde{C}^{i-1,n-2} -\tilde{C}^{i,n-2}\right) \mathcal{U}_{i} \mathcal{U}_{n-1} \notag \\
            &\qquad +\frac{1}{2} \sum_{j=1}^{n-2}
	 \left(\tilde{C}^{n-2,j-1} -\tilde{C}^{n-2,j}\right)
	 \mathcal{U}_{n-1} \mathcal{U}_{j} +\frac{1}{2}
	 \tilde{C}^{n-2,n-2} \mathcal{U}_{n-1} \mathcal{U}_{n-1}~,
\label{eq:first}
\end{align}
and the second term can be rewritten as
\begin{align}
             &\frac{1}{2} \sum_{i=1}^{n-2} \tilde{C}^{i,n-1} \left(\mathcal{U}_{i} -\mathcal{U}_{i+1}\right)\left(\mathcal{U}_{n-1} -\sum_{m=1}^{2} \mathcal{U}_{n}^{(m)}\right)\notag \\
            &= \frac{1}{2} \sum_{i=1}^{n-2} \left(\tilde{C}^{i,n-1} -\tilde{C}^{i-1,n-1} \right) \mathcal{U}_i \mathcal{U}_{n-1} -\frac{1}{2} \tilde{C}^{n-2,n-1}\mathcal{U}_{n-1} \mathcal{U}_{n-1} \notag \\
            &\quad -\frac{1}{2}\sum_{i=1}^{n-2} \left(\tilde{C}^{i,n-1}
 -\tilde{C}^{i-1,n-1}\right)\mathcal{U}_{i}
 \left(\sum_{m=1}^{2}\mathcal{U}_{n}^{(m)} \right) +\frac{1}{2}
 \tilde{C}^{n-2,n-1} \mathcal{U}_{n-1}
 \left(\sum_{m=1}^{2}\mathcal{U}_{n}^{(m)}\right) ~.
\label{eq:second}
\end{align}
Similalry, the third term on the RHS of \eqref{relation:h,b,c,U_i,U_n}
is expressed as
\begin{align}
            &\frac{1}{2} \sum_{j=1}^{n-2} \tilde{C}^{n-1,j} \left(\mathcal{U}_{n-1} -\sum_{l=1}^{2}\mathcal{U}_{n}^{(l)} \right) \left(\mathcal{U}_{j} -\mathcal{U}_{j+1}\right) \notag \\
            &= \frac{1}{2} \sum_{j=1}^{n-2} \left(\tilde{C}^{n-1,j} -\tilde{C}^{n-1,j-1}\right) \mathcal{U}_{n-1} \mathcal{U}_{j} -\frac{1}{2} \tilde{C}^{n-1,n-2} \mathcal{U}_{n-1} \mathcal{U}_{n-1} \notag \\
            &\quad -\frac{1}{2}\sum_{j=1}^{n-2}\left(\tilde{C}^{n-1,j}
 -\tilde{C}^{n-1,j-1}\right)
 \left(\sum_{l=1}^{2}\mathcal{U}_{n}^{(l)}\right) \mathcal{U}_{j}
 +\frac{1}{2}\tilde{C}^{n-1,n-2}
 \left(\sum_{l=1}^{2}\mathcal{U}_{n}^{(l)} \right) \mathcal{U}_{n-1}~,
\label{eq:third}
\end{align}
and the fourth term can be rewritten as
\begin{align}
            &\frac{1}{2} \tilde{C}^{n-1,n-1} \left(\mathcal{U}_{n-1} -\sum_{l=1}^{2} \mathcal{U}_{n}^{(l)} \right)\left(\mathcal{U}_{n-1} -\sum_{m=1}^{2} \mathcal{U}_{n}^{(m)} \right) \notag \\
            &= \frac{1}{2} \tilde{C}^{n-1,n-1} \mathcal{U}_{n-1} \mathcal{U}_{n-1} -\frac{1}{2} \tilde{C}^{n-1,n-1} \mathcal{U}_{n-1} \left(\sum_{m=1}^{2} \mathcal{U}_{n}^{(m)}\right) \notag \\
            &\quad -\frac{1}{2} \tilde{C}^{n-1,n-1} \left(\sum_{l=1}^{2}
 \mathcal{U}_{n}^{(l)}\right) \mathcal{U}_{n-1} +\frac{1}{2}
 \tilde{C}^{n-1,n-1} \left(\sum_{l=1}^{2} \mathcal{U}_{n}^{(l)}\right)
 \left(\sum_{m=1}^{2} \mathcal{U}_{n}^{(m)}\right)~.
\label{eq:fourth}
        \end{align}
Moreover, 
the explicit expression \eqref{eq:tildeC-ij} for $\tilde{C}^{ij}$
implies that
        \begin{align}
            \tilde{C}^{ij}-\tilde{C}^{i,j-1}-\tilde{C}^{i-1,j}+\tilde{C}^{i-1,j-1} &= \delta_{ij} -\frac{2}{2n-1}~,\\
            \tilde{C}^{i-1,n-2} -\tilde{C}^{i,n-2} = \tilde{C}^{n-2,j-1} -\tilde{C}^{n-2,j} &= -\frac{3}{2n-1}~,\\
            \tilde{C}^{n-2,n-2} &= \frac{3n-6}{2n-1} ~,\\
            \tilde{C}^{i,n-1} -\tilde{C}^{i-1,n-1} = \tilde{C}^{n-1,j} -\tilde{C}^{n-1,j-1} &= \frac{1}{2n-1}~,\\
            \tilde{C}^{n-2,n-1} = \tilde{C}^{n-1,n-2} &= \frac{n-2}{2n-1} ~,\\
            \tilde{C}^{n-1,n-1} &= \frac{n-1}{2n-1}~.
        \end{align}
Substituting these into \eqref{eq:first}--\eqref{eq:fourth} and plugging
the results into \eqref{relation:h,b,c,U_i,U_n}, we find that        
        \begin{align}
           T_{bc} + T_h &\equiv \frac{1}{2}\sum_{i=1}^{n-1}
	 \left(\mathcal{U}_i\right)^2  +\frac{1}{2}\sum_{j=1}^{2}
	 \left(\mathcal{U}_{n}^{(j)}\right)^2 -\frac{2n-1}{4}U^2 -H^2~,
\label{eq:TTT1}
        \end{align}
where we used 
\begin{align}
 U = \frac{2}{2n-1}\left(\sum_{i=1}^{n-1}\mathcal{U}_i +
 \frac{1}{2}\sum_{j=1}^2\mathcal{U}_n^{(j)}\right)~,\qquad  H = \frac{1}{2}\left(\mathcal{U}_n^{(1)}-\mathcal{U}_n^{(2)}\right)~,
\end{align}
that follow from \eqref{eq:H} and \eqref{eq:U}.

In addition, we find that \eqref{eq:Tsb2} is expressed as
\begin{align}
T_\text{sb} &= \frac{1}{2}\sum_{i=1}^{n-1}\mathcal{D}_i\mathcal{U}_i +
 \frac{1}{2}\sum_{j=1}^2\mathcal{D}_n^{(j)}\mathcal{U}_n^{(j)}~,
\label{eq:TTT2}
\end{align}
where $\mathcal{D}_i$ is defined in \eqref{eq:Di} and
$\mathcal{D}_n^{(j)}$ is defined in \eqref{eq:Dij}.

Combining \eqref{eq:TTT1} and \eqref{eq:TTT2}, we finally find 
        Thus, using $T_i~,\ T_n^{(j)}~,\ U~,\ H$, the stress tensor $T$ can be expressed by
        \begin{align}
            T = T_\text{sb} + T_{bc} + T_h  \equiv \sum_{i=1}^{n-1} T_i +\sum_{j=1}^{2}T_{n}^{(j)} -\frac{2n-1}{4}U^2 -H^2~,
        \end{align}
where $T_i = \frac{1}{2}(\mathcal{D}_i + \mathcal{U}_i)\mathcal{U}_i$
and $T_n^{(j)} = \frac{1}{2}(\mathcal{D}_n^{(j)} +
\mathcal{U}_n^{(j)})\mathcal{U}_n^{(j)}$ as seen from \eqref{eq:TiWi}
and \eqref{eq:TiWij}. Thus, we see that \eqref{eq:T2} is identical to
\eqref{eq:TTi2} up to BRST exact terms.

\section{OPEs of the bosonic VOA for  $T^{[1^3]}_{[2,1]}(SU(3))$}
\label{app:A1A5}

Here, we list all non-vanishing OPEs among the conjectured set of
generators \eqref{eq:gen1-2-1}--\eqref{eq:gen1-2-2} of the VOA
associated with $T^{[1^3]}_{[2,1]}(SU(3))$. First of all, the self-OPEs of
Heisenberg
current $U$ and the stress tensor $T$ are shown in
\eqref{eq:self-UT2}. In addition, from the $U(1)$ charge and the
holomorphic dimension of the generators, one can read off the following OPEs:
\begin{align}
 T(z)U(0) &\sim \frac{U}{z^2}+ \frac{\partial U}{z}~, \qquad T(z)\tau_a(0) \sim
 \frac{2\tau_a}{z^2} + \frac{\partial \tau_a}{z}~,
\\
T(z)\omega_a(0) &\sim \frac{3\omega_a}{z^2} + \frac{\partial
 \omega_a}{z}~, \qquad T(z)\mathcal{X}(0) \sim
 \frac{\frac{3}{2}\mathcal{X}}{z^2} + \frac{\partial \mathcal{X}}{z}~,
 \qquad T(z)\mathcal{Y}(0) \sim \frac{\frac{3}{2}\mathcal{Y}}{z^2} +
 \frac{\partial \mathcal{Y}'}{z}~,
\\
T(z)\mathcal{X}_a(0) &\sim
 \frac{\frac{5}{2}\mathcal{X}_a}{z^2} + \frac{\partial \mathcal{X}_a}{z}~,
 \qquad T(z)\mathcal{Y}_a(0) \sim \frac{\frac{5}{2}\mathcal{Y}_a}{z^2} +
 \frac{\partial \mathcal{Y}_a}{z}~,
\\
 U(z)\mathcal{X}(0) &\sim \frac{\mathcal{X}}{z}~,\qquad U(z)
 \mathcal{Y}(0) \sim -\frac{\mathcal{Y}}{z}~,\qquad U(z) \mathcal{X}_a(0) \sim \frac{\mathcal{X}_a}{z}~,\qquad 
U(z) \mathcal{Y}_a(0) \sim -\frac{\mathcal{Y}_a}{z}~,
\end{align}
where $a=1,2$. Note that all the other OPEs with $T$ or $U$ vanish.

The OPEs among $\tau_a, \omega_a$ can be read off from the OPEs among
$T_i$ and $W_i$ shown in
\eqref{eq:TiTj}--\eqref{eq:WiWj}. The non-vanishing OPEs turn out to be
the following:
\begin{align}
            \tau_{a}(z) \tau_{a}(0) &\sim -\frac{2}{z^4} +\frac{2U^2 +\frac{4}{3}T +(-1)^{a+1}\frac{2}{\sqrt{3}}\tau_2}{z^2}
            +\frac{2U \partial U +\frac{2}{3} \partial T +(-1)^{a+1}\frac{1}{\sqrt{3}}\partial \tau_{2}}{z},\\
            \tau_{1}(z) \tau_{2}(0) &\sim
 \frac{\frac{2}{\sqrt{3}}\tau_{2}}{z^2}
 +\frac{\frac{1}{\sqrt{3}}\partial \tau_1}{z}~,
\\
            \tau_{a}(z) \omega_a(0) &\sim \frac{3UT +3U\partial U
 -\mathcal{X} \mathcal{Y} +(-1)^{a+1}\sqrt{3} \omega_{2} -\frac{1}{2}\partial
 T -\frac{1}{2}\partial^2 U}{z^2} 
\\
            &\quad + \frac{U \partial T-\frac{1}{3}\mathcal{X} \partial
 \mathcal{Y} -\frac{1}{3} \mathcal{Y}\partial \mathcal{X} +T \partial U
 +(-1)^{a+1}\frac{1}{\sqrt{3}}\partial \omega_{2}-\frac{7}{12}\partial^3
 U}{z}~,
\\
            \tau_{1}(z) \omega_2(0) &\sim \frac{\sqrt{3}
 \omega_{1}}{z^2}+ \frac{\frac{1}{\sqrt{3}}\partial
 \omega_{1}}{z}~,
\\ 
            \tau_{2}(z) \omega_{1}(0) &\sim \frac{\sqrt{3}\omega_1}{z^2} +\frac{\frac{1}{\sqrt{3}}\partial \omega_1}{z},\\
            \omega_{a}(z) \omega_{a}(0) &\sim -\frac{1}{2z^6}
 +\frac{\frac{3}{4}U^2 +\frac{1}{2}T +(-1)^{a+1}\frac{\sqrt{3}}{4}
 \tau_2}{z^4} +\frac{\frac{3}{4}U \partial U +\frac{1}{4}\partial T
 +(-1)^{a+1}\frac{\sqrt{3}}{8}\partial \tau_2}{z^3}
\nonumber \\
            &\quad +\frac{1}{z^2}\left(
                \frac{2}{3}UUT
 +(-1)^{a+1}\frac{UU\tau_{2}}{\sqrt{3}}-\frac{3}{8}(\partial U)^2 -\frac{1}{24}U\partial^2 U 
                +\frac{1}{2}U^4 \right.\\
                &\qquad\quad \left.+\frac{2}{9}T^2 +(-1)^{a+1}\frac{2 T\tau_{2}}{3\sqrt{3}} +\frac{1}{2} (\tau_{1})^2+\frac{1}{6} (\tau_{2})^2 
                -\frac{\partial^2 T}{8}+(-1)^{a}\frac{1}{16} \sqrt{3}\partial^2 \tau_{2}
             \right)
\nonumber \\
            &\quad +\frac{1}{z} \left(
                \frac{1}{3}UU\partial T+(-1)^{a+1}\frac{UU\partial \tau_{2}}{2 \sqrt{3}}+\frac{2}{3}UT\partial U +(-1)^{a+1}\frac{U\tau_{2}\partial U}{\sqrt{3}}\right.                
\nonumber\\
&\qquad\quad  -\frac{11}{12}\partial U \partial^2 U+UUU\partial U  -\frac{5}{12}U\partial^3 U
                +\frac{2}{9} T\partial T +(-1)^{a+1}\frac{T\partial
 \tau_{2}}{3 \sqrt{3}}
\nonumber\\
&\qquad \quad \left.+\frac{1}{2}\tau_{1} \partial \tau_{1} +(-1)^{a+1}\frac{\tau_{2}\partial T}{3 \sqrt{3}}
                +\frac{1}{6} \tau_{2} \partial \tau_{2} -\frac{5}{36} \partial^3 T 
                +(-1)^a\frac{5 \partial^3 \tau_{2}}{24 \sqrt{3}}
            \right)~,
\\
\omega_1(z)\omega_2(0) &\sim \frac{\sqrt{3}\tau_1}{4z^4} +\frac{\sqrt{3}
 \partial \tau_1}{8 z^3} +\frac{1}{z^2} \left\{\frac{U U \tau_1
 }{\sqrt{3}}+\frac{2 T \tau_1}{3 \sqrt{3}}+\frac{1}{3} \tau_1 \tau_2
 -\frac{1}{16} \sqrt{3} \partial \tau_1'\right\} 
\nonumber\\
            &\quad +\frac{1}{z} \left\{\frac{U U \partial \tau_1}{2\sqrt{3}}+\frac{U \tau_1 \partial U}{\sqrt{3}}
            +\frac{T \partial \tau_1}{3\sqrt{3}}+\frac{\tau_1 \partial T}{3\sqrt{3}}
            +\frac{1}{6} \tau_1 \partial \tau_2 +\frac{1}{6}
            \tau_2 \partial \tau_1 -\frac{5 \partial^3 \tau_1}{24\sqrt{3}} \right\},
\end{align}
where $a=1,2$. 
Similarly, the OPEs of $\tau_a$ with
$\mathcal{X},\mathcal{Y},\mathcal{X}_a,\mathcal{Y}_a$ are written as
\begin{align}
            \tau_a(z) \mathcal{X}(0) &\sim \frac{2\mathcal{X}_a}{z}~,\qquad\tau_a(z) \mathcal{Y}(0) \sim -\frac{2\mathcal{Y}_a}{z}~,
\\[2mm]
            \tau_a(z) \mathcal{X}_a(0) &\sim \frac{2\mathcal{X}}{z^3}
 +\frac{U\mathcal{X} +(-1)^{a+1}\frac{1}{\sqrt{3}} \mathcal{X}_{2}
 +\frac{1}{3}\partial \mathcal{X}}{z^2}+
 \frac{1}{z}\bigg((-1)^a\,2\sqrt{3}U\mathcal{X}_2 -U\partial \mathcal{X}
 +\mathcal{X}T
\nonumber\\
&\qquad \quad +(-1)^{a+1}\sqrt{3}\mathcal{X}\tau_2 +
 \frac{5}{2}\mathcal{X}\partial U +(-1)^{a+1}\frac{4}{\sqrt{3}}\partial
 \mathcal{X}_2 + \frac{4}{3}\partial^2\mathcal{X}\bigg)~,
\\[2mm]
            \tau_a(z) \mathcal{Y}_a(0) &\sim -\frac{2\mathcal{Y}}{z^3}
 +\frac{U\mathcal{Y} +(-1)^{a+1}\frac{1}{\sqrt{3}} \mathcal{Y}_{2}
 -\frac{1}{3}\partial \mathcal{Y}}{z^2} +
 \frac{1}{z}\bigg((-1)^{a+1}2\sqrt{3}U\mathcal{Y}_2  - U\partial
 \mathcal{Y} -\mathcal{Y}T
\nonumber\\
            &\qquad\quad   +(-1)^{a}\sqrt{3}\mathcal{Y} \tau_2
 +\frac{5}{2}\mathcal{Y}\partial U +(-1)^{a+1}\frac{4}{\sqrt{3}} \partial
 \mathcal{Y}_{2} -\frac{4}{3}\partial^2 \mathcal{Y}\bigg)~,
\\[2mm]
            \tau_{1}(z) \mathcal{X}_{2}(0) &\sim
 \frac{\frac{1}{\sqrt{3}}\mathcal{X}_1}{z^2} +\frac{-2\sqrt{3} U
 \mathcal{X}_1 +\sqrt{3} \mathcal{X} \tau_1 +\frac{4}{\sqrt{3}}\partial
 \mathcal{X}_1}{z}~,
\\
            \tau_{2}(z) \mathcal{X}_{1}(0) &\sim
 \frac{\frac{1}{\sqrt{3}}\mathcal{X}_1}{z^2} +\frac{-2\sqrt{3}U
 \mathcal{X}_1 +\sqrt{3} \mathcal{X} \tau_1 +\frac{4}{\sqrt{3}}\partial
 \mathcal{X}_1}{z}~,
\\
            \tau_{1}(z) \mathcal{Y}_{2}(0) &\sim
 \frac{\frac{1}{\sqrt{3}}\mathcal{Y}_1}{z^2} +\frac{2\sqrt{3}
 U\mathcal{Y}_1 -\sqrt{3} \mathcal{Y} \tau_1 +\frac{4}{\sqrt{3}}
 \partial \mathcal{Y}_1}{z}~,
\\ 
            \tau_{2}(z) \mathcal{Y}_{1}(0) &\sim \frac{\frac{1}{\sqrt{3}}\mathcal{Y}_{1}}{z^2} +\frac{2\sqrt{3} U \mathcal{Y}_{1} -\sqrt{3} \mathcal{Y} \tau_1 +\frac{4}{\sqrt{3}} \partial \mathcal{Y}_{1}}{z}~,
\end{align}
where $a=1,2$. The OPEs of $\omega_a$ with
$\mathcal{X},\mathcal{Y},\mathcal{X}_a,\mathcal{Y}_a$ are written as
\begin{align}
             \omega_a (z) \mathcal{X}(0) &\sim \frac{\frac{3}{2}
 \mathcal{X}_a}{z^2} +\frac{\mathcal{X}\tau_a +2 \partial
 \mathcal{X}_a}{z}~,\qquad \omega_a(z) \mathcal{Y}(0) \sim
 \frac{\frac{3}{2}\mathcal{Y}_a}{z^2} +\frac{-\mathcal{Y}\tau_a
 +2\partial \mathcal{Y}_a}{z}~,
\\[3mm]
            \omega_a(z) \mathcal{X}_a(0) &\sim
 \frac{\frac{3}{2}\mathcal{X}}{z^4} +\frac{\frac{3}{2}U \mathcal{X}
 +(-1)^{a+1}\frac{\sqrt{3}}{2} \mathcal{X}_2 +\frac{1}{2}\partial
 \mathcal{X}}{z^3}
\nonumber\\
            &\quad \frac{ -\frac{3U\partial \mathcal{X}}{4}  +\frac{23\mathcal{X}\partial U}{8}  
            +\frac{UU\mathcal{X}}{2} +(-1)^a\, \frac{3\sqrt{3}U\mathcal{X}_{2}}{2}   +\frac{13\mathcal{X}T}{12} 
            +(-1)^{a+1}\frac{11 \mathcal{X}\tau_{2}}{4
 \sqrt{3}}+\frac{19\partial^2 \mathcal{X}}{12}+(-1)^{a+1}\frac{4
 \partial \mathcal{X}_{2}}{\sqrt{3}}}{z^2}
\nonumber \\[2mm]
            &\quad +\frac{U\mathcal{X} \partial U +\frac{1}{2} U\partial^2 \mathcal{X} +\frac{1}{2} \mathcal{X}\partial^2 U 
            +\frac{1}{3} \mathcal{X}\partial T +\frac{\mathcal{X}
 \partial \tau_{2}}{2 \sqrt{3}}+ \tau_a \mathcal{X}_a +\frac{1}{6}
 \partial^3 \mathcal{X} +\frac{\partial^2 \mathcal{X}_{2}}{2
 \sqrt{3}}}{z}~,
\\
            \omega_{1}(z) \mathcal{X}_{2}(0) &\sim
 \frac{\frac{\sqrt{3}}{2} \mathcal{X}_{1}}{z^3}
 +\frac{-\frac{3\sqrt{3}}{2}U \mathcal{X}_{1} +\frac{11}{4\sqrt{3}}
 \mathcal{X} \tau_1 +\frac{4}{\sqrt{3}} \partial \mathcal{X}_1}{z^2} 
\nonumber\\
            &\quad +\frac{\sqrt{3} U\mathcal{X} \tau_{1} -\sqrt{3} \mathcal{X}_{1} \partial U 
            -3 \sqrt{3} UU \mathcal{X}_{1} +\frac{11 \mathcal{X} \partial \tau_{1}}{4 \sqrt{3}}+\frac{4 \tau_{1} \partial \mathcal{X}}{\sqrt{3}}
            +\sqrt{3} \mathcal{X} \omega_{1} -\tau_{2}
 \mathcal{X}_{1} -\frac{2\partial \mathcal{X}_{1}}{\sqrt{3}} }{z}~,
\\
            \omega_{2}(z) \mathcal{X}_{1}(0) &\sim \frac{\frac{\sqrt{3}}{2}\mathcal{X}_{1}}{z^3} +\frac{-\frac{3\sqrt{3}}{2} U \mathcal{X}_{1} +\frac{11}{4\sqrt{3}} U \tau_1 +\frac{4}{\sqrt{3}}\partial \mathcal{X}_{1}}{z^2}
            +\frac{\frac{1}{2\sqrt{3}}U \partial \tau_1 +\tau_2
 \mathcal{X}_{1} +\frac{1}{2\sqrt{3}}\partial^2 \mathcal{X}_1}{z}~,
\\
            \omega_a(z) \mathcal{Y}_a(0) &\sim
 \frac{\frac{3}{2}\mathcal{Y}}{z^4} +\frac{-\frac{3}{2}U \mathcal{Y}
 +(-1)^a\,\frac{\sqrt{3}}{2}\mathcal{Y}_{2} +\frac{1}{2}\partial
 \mathcal{Y}}{z^3} 
\nonumber\\
            &\quad +\frac{\frac{3 U\partial \mathcal{Y}}{4} -\frac{23\mathcal{Y} \partial U}{8}  +\frac{UU\mathcal{Y}}{2}  +(-1)^a\,\frac{3\sqrt{3}U \mathcal{Y}_{2}}{2}   
            +\frac{13\mathcal{Y}T}{12} +(-1)^{a+1}\frac{11 \mathcal{Y}
 \tau_{2}}{4 \sqrt{3}}+\frac{19\partial^2 \mathcal{Y}}{12}+(-1)^a\frac{4
 \partial \mathcal{Y}_{2}}{\sqrt{3}}}{z^2} 
\nonumber\\
            &\quad +\frac{U\mathcal{Y} \partial U -\frac{1}{2} U \partial^2 \mathcal{Y} -\frac{1}{2} \mathcal{Y} \partial^2 U +\frac{1}{3} \mathcal{Y} \partial T 
            +(-1)^{a+1}\frac{\mathcal{Y} \partial \tau_{2} }{2
 \sqrt{3}}- \tau_a \mathcal{Y}_a +\frac{1}{6}\partial^3
 \mathcal{Y}+(-1)^a\frac{\partial^2 \mathcal{Y}_{2}}{2 \sqrt{3}}}{z}~,
\\
            \omega_{1}(z) \mathcal{Y}_{2}(0) &\sim
 \frac{-\frac{\sqrt{3}}{2} \mathcal{Y}_{1}}{z^3}
 +\frac{-\frac{3\sqrt{3}}{2} U \mathcal{Y}_{1}
 +\frac{11}{4\sqrt{3}}\mathcal{Y} \tau_1 -\frac{4}{\sqrt{3}} \partial
 \mathcal{Y}_1}{z^2} 
\nonumber\\
            &\quad +\frac{-\sqrt{3} U \mathcal{Y} \tau_{1} -\sqrt{3} \mathcal{Y}_{1} \partial U +3 \sqrt{3} UU \mathcal{Y}_{1} +\frac{11 \mathcal{Y} \partial \tau_{1}}{4 \sqrt{3}}
            +\frac{4 \tau_{1} \partial \mathcal{Y}
 }{\sqrt{3}}-\sqrt{3}\mathcal{Y}\omega_{1} +\tau_{2}
 \mathcal{Y}_{1} +\frac{2\partial^2 \mathcal{Y}_{1}}{\sqrt{3}} }{z}~,
\\
            \omega_{2}(z) \mathcal{Y}_{1}(0) &\sim \frac{-\frac{\sqrt{3}}{2} \mathcal{Y}_{1}}{z^3} +\frac{-\frac{3\sqrt{3}}{2} U\mathcal{Y}_{1} +\frac{11}{4\sqrt{3}} \mathcal{Y} \tau_1 -\frac{4}{\sqrt{3}} \partial \mathcal{Y}_{1}}{z^2}
            +\frac{\frac{1}{2 \sqrt{3}}\mathcal{Y} \partial \tau_1
 -\tau_2 \mathcal{Y}_{1} -\frac{1}{2\sqrt{3}} \partial^2
 \mathcal{Y}_{1}}{z}~,
\end{align}
where $a=1,2$.
Finally, the non-vanishing OPEs among
$\mathcal{X},\mathcal{Y},\mathcal{X}_a$ and $\mathcal{Y}_a$ are
summerized as follows:
\begin{align}
            \mathcal{X}(z) \mathcal{Y}(0) &\sim \frac{1}{z^3}
 +\frac{-3U}{z^2} +\frac{3 U^2 -T -\frac{3}{2}\partial U}{z}~,
\\
            \mathcal{X}(z) \mathcal{Y}_a(0) &\sim -\frac{\tau_a}{z^2}
 +\frac{3U\tau_a -3\omega_a -\frac{1}{4}\partial \tau_a}{z}~,
\\
            \mathcal{Y}(z) \mathcal{X}_a(0) &\sim -\frac{\tau_a}{z^2} +\frac{-3 U\tau_a +3 \omega_a -\frac{1}{4}\partial \tau_a}{z},\\
            \mathcal{X}_a(z) \mathcal{X}_a(0) &\sim
 \frac{\frac{1}{2}\mathcal{X} \mathcal{X}}{z^2}
 +\frac{\frac{1}{2}\mathcal{X} \partial \mathcal{X}}{z}~,
\\
 \mathcal{Y}_a(z) \mathcal{Y}_a(0) &\sim
 \frac{\frac{1}{2}\mathcal{Y} \mathcal{Y}}{z^2}
 +\frac{\frac{1}{2}\mathcal{Y} \partial \mathcal{Y}}{z}~,
\\
            \mathcal{X}_{1}(z) \mathcal{Y}_{1}(0) &\sim \frac{1}{z^5} +\frac{-3U}{z^4} +\frac{\frac{5}{2}U^2 -\frac{4}{3}T +\frac{(-1)^a}{2\sqrt{3}}\tau_{2} -\frac{3}{2}\partial U}{z^3} \\
            &\quad +\frac{-2 UT+\frac{(-1)^{a+1}}{2} \sqrt{3}U\tau_{2}
 -\frac{7}{2}U \partial U +\frac{3}{2} U^3 +2 \mathcal{X} \mathcal{Y} +\frac{\partial^2 U}{2}+\frac{\partial T}{3}+(-1)^{a}\frac{\partial \tau_{2}}{4 \sqrt{3}}
            -\sqrt{3} \omega_{2}}{z^2} 
\nonumber\\
            &\quad + \frac{1}{z} \left(-U \partial T +\frac{(-1)^{a+1}}{8} \sqrt{3}U \partial \tau_{2} -T\partial U +\frac{3}{2} UUT+(-1)^{a+1}6\sqrt{3} UU\tau_{2} \right.
\nonumber\\
            &\qquad +(-1)^a\frac{21}{2} \sqrt{3}U\omega_{2}
            +\frac{45}{8} \partial U \partial U +\frac{9}{4} UU\partial U +\frac{9}{2} U\partial^2 U 
            +(-1)^{a+1}\frac{5 \mathcal{X} \mathcal{Y}_{2}}{2 \sqrt{3}}
\nonumber\\
&\qquad+(-1)^{a+1}\frac{2 \mathcal{Y}\mathcal{X}_{2}}{\sqrt{3}}+2\mathcal{Y} \partial \mathcal{X} -\frac{1}{2} T^2
            +(-1)^{a+1}\frac{3}{2} (\tau_{1})^2 +(-1)^a\frac{3}{2} (\tau_{2})^2
 +\frac{23}{24} \partial^3 U 
\\
            &\qquad \left.-\frac{3\partial^2 T}{4}+\frac{(-1)^a}{4} \sqrt{3}
 \partial^2 \tau_{2}+(-1)^a\frac{\partial \omega_{2}}{\sqrt{3}} \right)~,
\\
            \mathcal{X}_{1}(z) \mathcal{Y}_{2}(0) &\sim \frac{-\frac{1}{2\sqrt{3}} \tau_{1}}{z^3} +\frac{\frac{\sqrt{3}}{2} U\tau_1 -\sqrt{3}\omega_{1} -\frac{1}{4\sqrt{3}}\partial \tau_1}{z^2} \nonumber\\
            &\quad + \frac{\frac{1}{8} \sqrt{3} U\partial \tau_{1} -\frac{3}{2}\sqrt{3} UU\tau_{1} +3 \sqrt{3} U \omega_{1} +\frac{\mathcal{X} \mathcal{Y}_{1}}{4\sqrt{3}}
            -\frac{\mathcal{Y}\mathcal{X}_{1}}{4 \sqrt{3}}-\frac{1}{2} \sqrt{3} T\tau_{1} +\frac{1}{8} \sqrt{3} \partial^2 \tau_{1}-\frac{\partial \omega_{1}}{\sqrt{3}}}{z},\\
            \mathcal{X}_{2}(z) \mathcal{Y}_{1}(0) &\sim \frac{-\frac{1}{2\sqrt{3}}\tau_1}{z^3} +\frac{\frac{\sqrt{3}}{2} U \tau_1 -\sqrt{3} \omega_{1} -\frac{1}{4 \sqrt{3}}\partial \tau_1}{z^2} \nonumber\\
            &\quad + \frac{\frac{1}{8} \sqrt{3} U\partial \tau_{1} -\frac{3}{2} \sqrt{3} UU\tau_{1} +3 \sqrt{3} U \omega_{1} +\frac{\mathcal{X} \mathcal{Y}_{1}}{4\sqrt{3}}
            -\frac{\mathcal{Y}\mathcal{X}_{1}}{4 \sqrt{3}}-\frac{1}{2} \sqrt{3} T\tau_{1}+\frac{1}{8} \sqrt{3}\partial^2 \tau_{1}-\frac{\partial \omega_{1}}{\sqrt{3}}}{z}~,
\end{align}
where $a=1,2$.

\section{OPEs of the bosonic VOA for $T^{[2,1^2]}_{[2,1^2]}(SU(4))$}
\label{app:A1D6}

In this appendix, we list non-vanishing OPEs among the conjectured
set of generators \eqref{eq:A1D6-gen1}--\eqref{eq:A1D6-gen4} of the VOA associated with
$T^{[2,1^2]}_{[2,1^2]}(SU(4))$.

Note that the OPEs of the generators with $T,U,H,E$ and $F$
are completely fixed by their dimension and $U(1)$ and $SU(2)$ charges,
as explained at the end of Sec.~\ref{subsubsec:A1D6}. In particular, the
OPEs with the stress tensor $T$ are written as
\begin{align}
 T(z)T(0) &\sim -\frac{6}{2z^4} + \frac{2T}{z^2} + \frac{\partial
 T}{z}~,\qquad T(z)U(0) \sim \frac{U}{z^2} + \frac{\partial
 U}{z}~,\qquad T(z)H(0) \sim \frac{H}{z^2} + \frac{\partial H}{z}~,
\\
T(z)E(0) &\sim \frac{E}{z^2} + \frac{\partial E}{z}~,\qquad T(z)F(0)\sim
 \frac{F}{z^2} + \frac{\partial F}{z}~,\qquad T(z)\tau(0)\sim
 \frac{2\tau}{z^2} + \frac{\partial \tau}{z}~,
\\
T(z)\widetilde{H}(0) &\sim \frac{2\widetilde{H}}{z^2} + \frac{\partial
 \widetilde{H}}{z}~, \qquad T(z) \widetilde{E}(0) \sim
 \frac{2\widetilde{E}}{z^2} + \frac{\partial \widetilde{E}}{z}~,\qquad
 T(z)\widetilde{F}(0) \sim \frac{2\widetilde{F}}{z^2} + \frac{\partial\widetilde{F}}{z}~,
\\
T(z)\mathcal{W}(0) &\sim
 \frac{3\mathcal{W}}{z^2} + \frac{\partial\mathcal{W}}{z}~,\qquad
 T(z)\mathcal{X}_\pm(0) \sim \frac{\frac{3}{2}\mathcal{X}_\pm}{z^2} +
 \frac{\partial \mathcal{X}_\pm}{z}~,\qquad T(z)\mathcal{Y}_\pm(0) \sim \frac{\frac{3}{2}\mathcal{Y}_\pm}{z^2} +
 \frac{\partial \mathcal{Y}_\pm}{z}~,
\\
T(z)\widetilde{X}_\pm(0) &\sim
 \frac{\frac{5}{2}\widetilde{\mathcal{X}}_\pm}{z^2} +
 \frac{\partial\widetilde{\mathcal{X}}_\pm}{z}~,\qquad T(z)
 \widetilde{\mathcal{Y}}_\pm (0) \sim
 \frac{\frac{5}{2}\widetilde{\mathcal{Y}}_\pm}{z^2} + \frac{\partial\widetilde{\mathcal{Y}}_\pm}{z}~.
\end{align}
Similarly, the non-vanishing OPEs with $U,H,E$ and $F$ are expressed as
Eqs.~\eqref{eq:A1D6-currents1}--\eqref{eq:A1D6-currents8}, and
\begin{align}
U(z)\mathcal{X}_\pm(0) &\sim \frac{\mathcal{X}_\pm}{z}~,\qquad
 U(z)\mathcal{Y}_\pm(0) \sim - \frac{\mathcal{Y}_\pm}{z}~,\qquad
 U(z)\widetilde{\mathcal{X}}_\pm(0) \sim
 \frac{\widetilde{\mathcal{X}}_\pm}{z}~,\qquad
 U(z)\widetilde{\mathcal{Y}}_\pm(0) \sim -\frac{\widetilde{\mathcal{Y}}_\pm}{z}~,
\\
 H(z)\mathcal{X}_\pm(0) &\sim \pm\frac{\mathcal{X}_\pm}{2z}~,\qquad
 H(z)\mathcal{Y}_\pm (0) \sim \pm \frac{\mathcal{Y}_\pm}{2z}~,\qquad
 H(z)\widetilde{\mathcal{X}}\pm(0)\sim
 \pm\frac{\widetilde{\mathcal{X}}_\pm}{2z}~,\qquad
 H(z)\widetilde{\mathcal{Y}}_\pm(0) \sim \pm
 \frac{\widetilde{\mathcal{Y}}_\pm}{2z}~,
\\
H(z)\widetilde{E}(0) &\sim \frac{\widetilde{E}}{z}~,\qquad
 H(z)\widetilde{F}(0) \sim - \frac{\widetilde{F}}{z}~.
\end{align}
   
Below, we list the non-vanishing OPEs among $\tau,
\widetilde{H},\widetilde{E},\widetilde{F},\mathcal{W},\mathcal{X}_\pm,\mathcal{Y}_\pm,\widetilde{\mathcal{X}}_\pm$
and $\widetilde{\mathcal{Y}}_\pm$, which are not fixed purely by their
dimension or charges. First, the non-vanishing OPEs with $\tau$ are
the following:
\begin{align}
                   \tau(z) \tau(0) &\sim -\frac{2}{z^4} +\frac{\frac{5}{2}U^2 -2H^2 -2 EF+2T +2 \partial H}{z^2}
                    +\frac{\frac{5}{2}U \partial U -2 H \partial H -E \partial F- F \partial E +\partial T}{z},\\
                    \tau(z) \mathcal{W}(0) &\sim \frac{1}{z^2}\left\{
                        \frac{15}{4} U T-\frac{1}{4} H \widetilde{H}-\frac{1}{2} E \widetilde{F} +\frac{45}{16} U \partial U +\frac{15}{8} U \partial H
                        +\frac{3}{4} H \partial H +\frac{3}{8} E
 \partial F +\frac{3}{8} F \partial E \right.
\nonumber\\
                        &\quad\qquad \left. +\frac{25}{32}U^3 -\frac{15}{8} U H^2 -\frac{15}{8} U E F 
                        -\frac{3}{4} \mathcal{X}_{+} \mathcal{Y}_{-} -\frac{3}{4} \mathcal{X}_{-} \mathcal{Y}_{+} -\frac{5 \partial^2 U}{8}-\frac{3\partial T}{4}+\frac{\partial \widetilde{H}}{8}
                    \right\} 
\nonumber\\
                    &\quad +\frac{1}{z}\left\{
                    \frac{5}{4} U \partial T +\frac{5}{4} T \partial U -\frac{1}{4} \widetilde{H} \partial H -\frac{1}{4} \widetilde{E} \partial F 
                    -\frac{1}{4}\widetilde{F} \partial E
 +\frac{5}{8}\partial U \partial H +\frac{25}{32} U U \partial U \right.
\nonumber\\
                    &\quad \qquad -\frac{5}{4} U H \partial H -\frac{5}{8} U E \partial F
                    -\frac{5}{8} U F \partial E -\frac{5}{8} H H
 \partial U -\frac{5}{8} E F \partial U -\frac{1}{4} \mathcal{X}_{+}
 \partial \mathcal{Y}_{-} 
\nonumber\\
                    &\quad\qquad \left.-\frac{1}{4} \mathcal{X}_{-} \partial
 \mathcal{Y}_{+} -\frac{1}{4}\mathcal{Y}_{+} \partial \mathcal{X}_{-} -\frac{1}{4}\mathcal{Y}_{-} \partial \mathcal{X}_{+} -\frac{35}{48} \partial^3 U
                    \right\},
\\
\tau(z)\mathcal{X}_\pm(0) &\sim \frac{2\widetilde{X}_\pm}{z}~,
\\
 \tau(z) \mathcal{Y}_\pm(0) &\sim -\frac{2\widetilde{\mathcal{Y}}_\pm}{z}~,
\\
                    \tau(z) \widetilde{\mathcal{X}}_{+}(0) &\sim \frac{2 \mathcal{X}_{+}}{z^3}
                    +\frac{1}{z^2}\left\{
                        \frac{5}{4} U \mathcal{X}_{+} -\frac{1}{2}H \mathcal{X}_{+}
                        +\frac{1}{2} E \mathcal{X}_{-} +\frac{\partial \mathcal{X}_{+}}{2}
                     \right\}
\nonumber\\
                    &\quad +\frac{1}{z} \left\{
                        -\frac{5}{2} U \partial \mathcal{X}_{+} + H \partial \mathcal{X}_{+} -E \partial \mathcal{X}_{-} 
                        +\frac{15}{4} \mathcal{X}_{+} \partial U +\frac{1}{2} \mathcal{X}_{+} \partial H +2 \mathcal{X}_{-} \partial E 
                        -\frac{5}{8} U U \mathcal{X}_{+}\right.
\nonumber\\
                        &\quad \left.+\frac{5}{2} U H \mathcal{X}_{+} -\frac{5}{2} U E \mathcal{X}_{-} -\frac{5}{2} H H \mathcal{X}_{+}-\frac{5}{2} E F \mathcal{X}_{+} +2 \mathcal{X}_{+} T +\partial^2 \mathcal{X}_{+}
                    \right\}~,
\\
                    \tau(z) \widetilde{\mathcal{X}}_{-}(0) &\sim \frac{2 \mathcal{X}_{-}}{z^3}
                    +\frac{1}{z^2} \left\{
                        \frac{5}{4} U \mathcal{X}_{-} +\frac{1}{2} H \mathcal{X}_{-} +\frac{1}{2} F \mathcal{X}_{+} +\frac{\partial \mathcal{X}_{-}}{2}
                    \right\}
\nonumber\\
                    &\quad +\frac{1}{z}\left\{
                        -\frac{5}{2} U \partial \mathcal{X}_{-} -H \partial \mathcal{X}_{-} -F \partial \mathcal{X}_{+} +2 \mathcal{X}_{+} \partial F +\frac{15}{4} \mathcal{X}_{-} \partial U
                        +\frac{9}{2}\mathcal{X}_{-} \partial H
 -\frac{5}{8} U U \mathcal{X}_{-}\right.
\nonumber\\
                        &\quad \left. -\frac{5}{2} U H \mathcal{X}_{-} -\frac{5}{2} U F \mathcal{X}_{+}
                        -\frac{5}{2} H H \mathcal{X}_{-} -\frac{5}{2} E F \mathcal{X}_{-} +2 \mathcal{X}_{-} T -\frac{\partial^2 \mathcal{X}_{-}}{4}
                    \right\}~,
\\
                    \tau(z) \widetilde{\mathcal{Y}}_{+}(0) &\sim \frac{-2 \mathcal{Y}_{+}}{z^3}
                    +\frac{1}{z^2}\left\{
                        \frac{5}{4} U \mathcal{Y}_{+}+\frac{1}{2}H \mathcal{Y}_{+} +\frac{1}{2} E \mathcal{Y}_{-} -\frac{\partial \mathcal{Y}_{+}}{2}
                    \right\}
\nonumber\\
                    &\quad +\frac{1}{z} \left\{
                        -\frac{5}{2} U \partial \mathcal{Y}_{+} -H \partial \mathcal{Y}_{+} -E \partial \mathcal{Y}_{-} +\frac{15}{4} \mathcal{Y}_{+} \partial U -\frac{1}{2}\mathcal{Y}_{+} \partial H
                        +2 \mathcal{Y}_{-} \partial E +\frac{5}{8} U U
 \mathcal{Y}_{+}\right.
\nonumber\\
                        &\quad\left. +\frac{5}{2} U H \mathcal{Y}_{+}
                        +\frac{5}{2} U E \mathcal{Y}_{-}+\frac{5}{2}H H \mathcal{Y}_{+} +\frac{5}{2} E F \mathcal{Y}_{+} -2\mathcal{Y}_{+} T -\partial^2 \mathcal{Y}_{+}
                    \right\}~,
\\
                    \tau(z) \widetilde{\mathcal{Y}}_{-}(0) &\sim \frac{-2 \mathcal{Y}_{-}}{z^3}
                    +\frac{1}{z^2} \left\{
                        \frac{5}{4} U \mathcal{Y}_{-} -\frac{1}{2} H \mathcal{Y}_{-} +\frac{1}{2} F \mathcal{Y}_{+} -\frac{\partial \mathcal{Y}_{-}}{2}
                    \right\}
\nonumber\\
                    &\quad +\frac{1}{z} \left\{
                        -\frac{5}{2} U \partial \mathcal{Y}_{-} +H \partial \mathcal{Y}_{-} -F \partial \mathcal{Y}_{+} +2 \mathcal{Y}_{+} \partial F +\frac{15}{4} \mathcal{Y}_{-} \partial U -\frac{9}{2} \mathcal{Y}_{-} \partial H 
                        +\frac{5}{8} U U \mathcal{Y}_{-}\right.
\nonumber\\
                        &\quad \left. -\frac{5}{2} U H \mathcal{Y}_{-} +\frac{5}{2} U F \mathcal{Y}_{+} +\frac{5}{2} H H \mathcal{Y}_{-} +\frac{5}{2} E F \mathcal{Y}_{-} -2 \mathcal{Y}_{-} T +\frac{\partial^2 \mathcal{Y}_{-}}{4}
                    \right\}~.
\end{align}
Similarly, the non-vanishing OPEs with $\mathcal{W}$ are the following:
\begin{align}
 \mathcal{W}(z)\mathcal{W}(0) &\sim  -\frac{1}{2z^6}
                    +\frac{1}{z^4}\left\{
                        \frac{15}{16} U^2 -\frac{3}{4} H^2 -\frac{3}{4} E F +\frac{3 \partial H}{4}+\frac{3T}{4}
                    \right\}\\
                    &\qquad +\frac{1}{z^3}\left\{
                        \frac{15}{16} U \partial U -\frac{3}{4}H \partial H -\frac{3}{8} E \partial F -\frac{3}{8} F \partial E +\frac{3 \partial T}{8}
                    \right\}\\
                    &\qquad +\frac{1}{z^2}\left\{
                        T \partial H +\frac{5}{4}U U T - H H T - E F T
 -\frac{15}{32} (\partial U)^2 + \frac{3}{4} (\partial H)^2
                        -\frac{5}{8} \partial E \partial F \right.
\nonumber\\
                        &\qquad\quad +\frac{5}{4} U U \partial H -\frac{1}{2}H H \partial H +\frac{5}{4}H E \partial F
                        -\frac{5}{4}H F \partial E -2E F \partial
 H+\frac{5}{32}U \partial^2 U 
\nonumber\\
                        &\qquad\quad -\frac{13}{8}H \partial^2 H
                        +\frac{7}{8}E \partial^2 F-\frac{1}{8}F
 \partial^2 E +\frac{25}{32}U^4 -\frac{5}{4} U U H^2 -\frac{5}{4} U U E F
\nonumber\\
                        &\qquad\quad \left.  +\frac{1}{2}H H E F +\frac{1}{2} E E F^2
                        +\frac{1}{2} T^2+\frac{1}{2} \tau^2 +\frac{1}{8}\widetilde{H}^2 -\frac{19}{48} \partial^3 H-\frac{3\partial^2 T}{16}
                    \right\}
\nonumber\\
                    &\qquad +\frac{1}{z}\left\{
                        \frac{1}{2} \partial H \partial T +\frac{5}{8}U U \partial T +\frac{5}{4}U T \partial U -\frac{1}{2}H H \partial T -H T \partial H -\frac{1}{2} E F \partial T
                        \right.
\nonumber\\
                        &\qquad \quad -\frac{1}{2}E T \partial F -\frac{1}{2}F T \partial E -\frac{15}{16}\partial U \partial^2 U -\frac{3}{4} \partial H \partial^2 H
                        +\frac{5}{8} \partial E \partial^2 F
 +\frac{1}{8} \partial F \partial^2 E 
\nonumber\\
&\qquad \quad +\frac{5}{4} U \partial U \partial H -\frac{1}{2}H
 (\partial H)^2 -\frac{3}{8}E \partial H \partial F -\frac{13}{8}F
 \partial H \partial E +\frac{25}{16}U U U \partial U
\nonumber\\
                        &\qquad \quad  -\frac{5}{4}U U H \partial H
                        -\frac{5}{8}U U E \partial F -\frac{5}{8}U U F
 \partial E -\frac{5}{4}U H H \partial U -\frac{5}{4}U E F \partial U
\nonumber\\
                        &\qquad \quad  +\frac{1}{4} H H E \partial F
                        +\frac{1}{4}H H F \partial E +\frac{1}{2}H E F
 \partial H +\frac{1}{2}E E F \partial F +\frac{1}{2}E F F \partial E
\nonumber\\
                        &\qquad \quad  +\frac{3}{8} H E \partial^2 F
                        -\frac{3}{8}H F \partial^2 E -\frac{25}{48}U
 \partial^3 U +\frac{1}{3}E \partial^3 F +\frac{1}{3}F \partial^3 E
 +\frac{1}{2}T \partial T
\nonumber\\
                        &\qquad \quad \left. +\frac{1}{2}\tau \partial \tau +\frac{1}{8}\widetilde{H} \partial \widetilde{H} -\frac{1}{12}\partial^4 H-\frac{5}{24} \partial^3 T
                    \right\}~,
\\
                    \mathcal{W}(z) \mathcal{X}_{\pm}(0) &\sim \frac{3
 \widetilde{\mathcal{X}}_{\pm}}{2z^2} +\frac{\mathcal{X}_{\pm} \tau +2
 \partial \widetilde{\mathcal{X}}_{\pm}}{z}~,
\\
     \mathcal{W}(z) \mathcal{Y}_{\pm}(0) &\sim \frac{3
 \widetilde{\mathcal{Y}}_{\pm}}{2z^2} +\frac{-\mathcal{Y}_{\pm} \tau +2
 \partial \widetilde{\mathcal{Y}}_{\pm}}{z}~,
\\
                    \mathcal{W}(z) \widetilde{\mathcal{X}}_{+}(0) &\sim \frac{3 \mathcal{X}_{+}}{2z^4} +\frac{1}{z^3} \left\{
                        \frac{15}{8} U \mathcal{X}_{+} -\frac{3}{4}H \mathcal{X}_{+} +\frac{3}{4}E \mathcal{X}_{-} +\frac{3 \partial \mathcal{X}_{+}}{4}
                    \right\}
\nonumber\\
                    &\qquad +\frac{1}{z^2}\left\{
                        -\frac{15}{8} U \partial \mathcal{X}_{+}+\frac{3}{4}H \partial \mathcal{X}_{+} -\frac{3}{4}E \partial \mathcal{X}_{-}
                        +\frac{65}{16}\mathcal{X}_{+} \partial U
 +\frac{3}{8} \mathcal{X}_{+} \partial H 
 \right.
\nonumber\\
                        &\qquad\qquad \left. +2 \mathcal{X}_{-} \partial
 E + \frac{5}{32}U U \mathcal{X}_{+}+\frac{15}{8}U H \mathcal{X}_{+}
 -\frac{15}{8} U E \mathcal{X}_{-} -\frac{19}{8} H H \mathcal{X}_{+}
 \right.
\nonumber\\
&\qquad \qquad\left.-\frac{19}{8} E F \mathcal{X}_{+} +2 \mathcal{X}_{+} T +\frac{5 \partial^2 \mathcal{X}_{+}}{4}
                    \right\}
\nonumber\\
                    &\qquad +\frac{1}{z} \left\{
                        -\frac{3}{2} E \mathcal{X}_{+} \widetilde{F} +\frac{3}{2} F \mathcal{X}_{+} \widetilde{E} +\frac{1}{2} \partial H \partial \mathcal{X}_{+} -\frac{5}{4} \partial E \partial \mathcal{X}_{-}
                        +\frac{5}{4} U \mathcal{X}_{+} \partial U
  \right.
\nonumber\\
                        &\qquad\qquad +\frac{1}{2}H \mathcal{X}_{-}
 \partial E - \frac{1}{4} E \mathcal{X}_{+} \partial F -\frac{1}{2} E \mathcal{X}_{-} \partial H
                        +\frac{1}{4} F \mathcal{X}_{+} \partial E +\frac{5}{8} U \partial^2 \mathcal{X}_{+}
\nonumber\\
                        &\qquad\qquad \left. -\frac{3}{8} E \partial^2 \mathcal{X}_{-}  +\frac{5}{8} \mathcal{X}_{+} \partial^2 U
                        +\mathcal{X}_{+} \partial^2 H  - \frac{11}{8}
 \mathcal{X}_{-} \partial^2 E +\frac{1}{2} \mathcal{X}_{+} \partial T
 +\frac{1}{2} \mathcal{X}_{+} \partial \widetilde{H}\right.
\nonumber\\
&\qquad \qquad \left.
                        +\frac{1}{2} \mathcal{X}_{-} \partial \widetilde{E} +\frac{3}{2} \widetilde{E} \partial \mathcal{X}_{-} + \tau \widetilde{\mathcal{X}}_{+} +\frac{5}{6} \partial^3 \mathcal{X}_{+}
                    \right\}~,
\\
                    \mathcal{W}(z) \widetilde{\mathcal{X}}_{-}(0) &\sim \frac{3 \mathcal{X}_{-}}{2z^4}
                    +\frac{1}{z^3}\left\{
                        \frac{15}{8} U \mathcal{X}_{-} +\frac{3}{4} H \mathcal{X}_{-} +\frac{3}{4} F \mathcal{X}_{+} +\frac{3 \partial \mathcal{X}_{-}}{4}
                    \right\}
\nonumber\\
                    &\qquad + \frac{1}{z^2}\left\{
                        -\frac{15}{8} U \partial \mathcal{X}_{-} -\frac{3}{4}H \partial \mathcal{X}_{-} -\frac{3}{4}F \partial \mathcal{X}_{+} +2\mathcal{X}_{+} \partial F
                        +\frac{65}{16} \mathcal{X}_{-} \partial U 
\right.
\nonumber\\
&\qquad \qquad \left.+\frac{35}{8} \mathcal{X}_{-} \partial H
 +\frac{5}{32} U U \mathcal{X}_{-}-\frac{15}{8} U H \mathcal{X}_{-}
                        -\frac{15}{8} U F \mathcal{X}_{+} -\frac{19}{8} H H \mathcal{X}_{-}
\right.
\nonumber\\
                        &\qquad\qquad \left.  -\frac{19}{8}E F \mathcal{X}_{-} +2 \mathcal{X}_{-} T +\frac{\partial^2 \mathcal{X}_{-}}{16}
                    \right\}
\nonumber\\
                    &\qquad +\frac{1}{z} \left\{
                        \frac{1}{2} E \mathcal{X}_{-} \widetilde{F}
 -\frac{1}{2} F \mathcal{X}_{-} \widetilde{E} -\frac{1}{2} \partial H
 \partial \mathcal{X}_{-} -\frac{3}{4} \partial F \partial
 \mathcal{X}_{+} + \frac{5}{4}U \mathcal{X}_{-} \partial U
                         \right.
\nonumber\\
&\qquad \qquad \left.  -H \mathcal{X}_{-} \partial H -\frac{1}{2}E
 \mathcal{X}_{-} \partial F -\frac{1}{2}F \mathcal{X}_{-} \partial E +\frac{5}{8}U \partial^2 \mathcal{X}_{-}
                        -\frac{1}{8} F \partial^2 \mathcal{X}_{+}
\right. 
\nonumber\\
                        &\qquad\qquad  -\frac{1}{8} \mathcal{X}_{+} \partial^2 F +\frac{5}{8} \mathcal{X}_{-} \partial^2 U +\frac{1}{2} \mathcal{X}_{-} \partial T
                        -\frac{1}{4} \mathcal{X}_{-} \partial
 \widetilde{H}
\nonumber\\
                        &\qquad\qquad \left. +\frac{1}{2} \widetilde{F} \partial \mathcal{X}_{+} + \tau \widetilde{\mathcal{X}}_{-} +\frac{1}{6} \partial^3 \mathcal{X}_{-}
                    \right\}~,
\\
                    \mathcal{W}(z) \widetilde{\mathcal{Y}}_{+}(0) &\sim \frac{3 \mathcal{Y}_{+}}{2z^4}
                    +\frac{1}{z^3}\left\{
                        -\frac{15}{8} U \mathcal{Y}_{+} -\frac{3}{4} H \mathcal{Y}_{+} -\frac{3}{4}E \mathcal{Y}_{-} +\frac{3 \partial \mathcal{Y}_{+}}{4}
                    \right\}
\nonumber\\
                    &\qquad +\frac{1}{z^2} \left\{
                        \frac{15}{8} U \partial \mathcal{Y}_{+} +\frac{3}{4} H \partial \mathcal{Y}_{+} +\frac{3}{4} E \partial \mathcal{Y}_{-} -\frac{65}{16} \mathcal{Y}_{+} \partial U
                        +\frac{3}{8} \mathcal{Y}_{+} \partial H
  \right.
\nonumber\\
                        &\qquad\qquad \left. -2\mathcal{Y}_{-} \partial E +\frac{5}{32} U U \mathcal{Y}_{+}-\frac{15}{8} U H \mathcal{Y}_{+}
                        -\frac{15}{8} U E \mathcal{Y}_{-} -\frac{19}{8}H
 H \mathcal{Y}_{+} \right.
\nonumber\\
&\qquad\qquad\left. 
-\frac{19}{8} E F \mathcal{Y}_{+} +2 \mathcal{Y}_{+} T +\frac{5 \partial^2 \mathcal{Y}_{+}}{4}
                    \right\}
\nonumber\\
                    &\qquad +\frac{1}{z}\left\{
                        \frac{3}{2} E \mathcal{Y}_{+} \widetilde{F} -\frac{3}{2} F \mathcal{Y}_{+} \widetilde{E} +\frac{1}{2} \partial H \partial \mathcal{Y}_{+} +\frac{5}{4} \partial E \partial \mathcal{Y}_{-}
                        +\frac{5}{4} U \mathcal{Y}_{+} \partial U \right.
\nonumber\\
&\qquad \qquad\left.
 -\frac{1}{2} H \mathcal{Y}_{-} \partial E -\frac{1}{4} E \mathcal{Y}_{+} \partial F +\frac{1}{2} E \mathcal{Y}_{-} \partial H
                        +\frac{1}{4} F \mathcal{Y}_{+} \partial E
 -\frac{5}{8} U \partial^2 \mathcal{Y}_{+}
\right.
\nonumber\\
                        &\qquad\qquad  +\frac{3}{8} E \partial^2 \mathcal{Y}_{-} -\frac{5}{8} \mathcal{Y}_{+} \partial^2 U + \mathcal{Y}_{+} \partial^2 H
                        +\frac{11}{8} \mathcal{Y}_{-} \partial^2 E +\frac{1}{2} \mathcal{Y}_{+} \partial T
\nonumber\\
                        &\qquad\qquad \left.  -\frac{1}{2} \mathcal{Y}_{+} \partial \widetilde{H} +\frac{1}{2} \mathcal{Y}_{-} \partial \widetilde{E}
                        +\frac{3}{2} \widetilde{E} \partial \mathcal{Y}_{-} - \tau \widetilde{\mathcal{Y}}_{+} +\frac{5}{6} \partial^3 \mathcal{Y}_{+}
                    \right\}~,
\\
                    \mathcal{W}(z) \widetilde{\mathcal{Y}}_{-}(0) &\sim \frac{3\mathcal{Y}_{-}}{2z^4}
                    +\frac{1}{z^3} \left\{
                        -\frac{15}{8} U \mathcal{Y}_{-} +\frac{3}{4} H \mathcal{Y}_{-} -\frac{3}{4} F \mathcal{Y}_{+} +\frac{3 \partial \mathcal{Y}_{-}}{4}
                    \right\}
\nonumber\\
                    &\qquad +\frac{1}{z^2}\left\{
                        \frac{15}{8} U \partial \mathcal{Y}_{-} -\frac{3}{4}H \partial \mathcal{Y}_{-} +\frac{3}{4} F \partial \mathcal{Y}_{+} -2 \mathcal{Y}_{+} \partial F
                        -\frac{65}{16} \mathcal{Y}_{-} \partial U
 \right.
\nonumber\\
&\qquad \qquad \left.
+ \frac{35}{8} \mathcal{Y}_{-} \partial H +\frac{5}{32}U U \mathcal{Y}_{-} +\frac{15}{8} U H \mathcal{Y}_{-}
                        -\frac{15}{8}U F \mathcal{Y}_{+} -\frac{19}{8} H
 H \mathcal{Y}_{-}
\right.
\nonumber\\
                        &\qquad\qquad \left.  -\frac{19}{8} E F \mathcal{Y}_{-} +2 \mathcal{Y}_{-} T +\frac{\partial^2 \mathcal{Y}_{-}}{16}
                    \right\}\\
                    &\qquad +\frac{1}{z}\left\{
                        -\frac{1}{2} E \mathcal{Y}_{-} \widetilde{F} +\frac{1}{2} F \mathcal{Y}_{-} \widetilde{E} -\frac{1}{2} \partial H \partial \mathcal{Y}_{-} +\frac{3}{4} \partial F \partial \mathcal{Y}_{+}
                        +\frac{5}{4} U \mathcal{Y}_{-} \partial U 
\right.
\nonumber\\
&\qquad\qquad\left.
-H \mathcal{Y}_{-} \partial H -\frac{1}{2}E \mathcal{Y}_{-} \partial F
 -\frac{1}{2} F \mathcal{Y}_{-} \partial E -\frac{5}{8} U \partial^2
 \mathcal{Y}_{-} +\frac{1}{8} F \partial^2 \mathcal{Y}_{+} 
\right.
\nonumber\\
                        &\qquad \qquad
                        +\frac{1}{8}
 \mathcal{Y}_{+} \partial^2 F-\frac{5}{8} \mathcal{Y}_{-} \partial^2 U +\frac{1}{2} \mathcal{Y}_{-} \partial T +\frac{1}{4} \mathcal{Y}_{-} \partial \widetilde{H} +\frac{1}{2} \widetilde{F} \partial \mathcal{Y}_{+} \\
                        &\qquad\qquad \left. - \tau \widetilde{\mathcal{Y}}_{-}+\frac{1}{6} \partial^3 \mathcal{Y}_{-}
                    \right\}~.
\end{align}
The non-vanishing OPEs of $\widetilde{H}, \widetilde{E},\widetilde{F},
\mathcal{X}_\pm,\mathcal{Y}_\pm,\widetilde{\mathcal{X}}_\pm$ and
$\widetilde{\mathcal{Y}}_\pm$ with $\widetilde{H},
\widetilde{E},\widetilde{F}$ are written as
\begin{align}
 \widetilde{H}(z) \widetilde{H}(0) &\sim -\frac{2}{z^4} +\frac{4 H^2+2 E F -2 \partial H}{z^2}
                    +\frac{4 H \partial H +E \partial F + F \partial
 E}{z}~,
\\
\widetilde{H}(z) \widetilde{E}(0) &\sim \frac{2 E}{z^3} +\frac{H E +\frac{1}{2}\partial E}{z^2}
 +\frac{-6 H \partial E + E \partial H + 2H H E + 2 E E F + 3\partial^2
 E}{z}~,
\\
 \widetilde{H}(z) \widetilde{F}(0) &\sim -\frac{2 F}{z^3} +\frac{H F -\frac{1}{2}\partial F}{z^2}
 +\frac{-6 H \partial F +9 F \partial H -2H H F -2 E F^2 -5 \partial^2
 F}{z}~,
\\
 \widetilde{E}(z) \widetilde{E}(0) &\sim \frac{E^2}{2z^2} +\frac{E \partial E}{2z}~,
 \\
 \widetilde{F}(z) \widetilde{F}(0) &\sim \frac{F^2}{2z^2} +\frac{F \partial F}{2z}~,
\\
 \widetilde{E}(z) \widetilde{F}(0) &\sim -\frac{1}{z^4} +\frac{2H}{z^3}
 +\frac{H^2 +\frac{3}{2} E F -\frac{1}{2}\partial H}{z^2}
\nonumber\\
&\qquad 
 +\frac{1}{z} \left\{
                        -H \partial H -\frac{1}{2} E \partial F +2 F \partial E +2 H^3 +2 H E F +2 \partial^2 H
                    \right\}~,
\\
 \widetilde{H}(z) \mathcal{X}_{+}(0) &\sim \frac{\mathcal{X}_{+}}{z^2}
 +\frac{2 H \mathcal{X}_{+} -E \mathcal{X}_{-}}{z}~,
\\
 \widetilde{H}(z) \widetilde{\mathcal{X}}_{+}(0) &\sim
 \frac{\widetilde{\mathcal{X}}_{+}}{z^2} +\frac{2 H
 \widetilde{\mathcal{X}}_{+}- E \widetilde{\mathcal{X}}_{-}}{z}~,
\\
                    \widetilde{H}(z) \mathcal{X}_{-}(0) &\sim
 -\frac{\mathcal{X}_{-}}{z^2} +\frac{2 H \mathcal{X}_{-} + F
 \mathcal{X}_{+}}{z}~,
\\
 \widetilde{H}(z) \widetilde{\mathcal{X}}_{-}(0) &\sim
 -\frac{\widetilde{\mathcal{X}}_{-}}{z^2} +\frac{2 H
 \widetilde{\mathcal{X}}_{-}+ F \widetilde{\mathcal{X}}_{+}}{z}~,
\\
                    \widetilde{H}(z) \mathcal{Y}_{+}(0) &\sim
 -\frac{\mathcal{Y}_{+}}{z^2} +\frac{-2 H \mathcal{Y}_{+} -E
 \mathcal{Y}_{-}}{z}~,
\\
 \widetilde{H}(z) \widetilde{\mathcal{Y}}_{+}(0)
 &\sim -\frac{\widetilde{\mathcal{Y}}_{+}}{z^2} +\frac{-2 H
 \widetilde{\mathcal{Y}}_{+} - E \widetilde{\mathcal{Y}}_{-}}{z}~,
\\
                    \widetilde{H}(z) \mathcal{Y}_{-}(0) &\sim
 \frac{\mathcal{Y}_{-}}{z^2} +\frac{-2 H \mathcal{Y}_{-} + F
 \mathcal{Y}_{+}}{z}~,
\\
 \widetilde{H}(z) \widetilde{\mathcal{Y}}_{-}(0)
 &\sim \frac{\widetilde{\mathcal{Y}}_{-}}{z^2} +\frac{-2 H
 \widetilde{\mathcal{Y}}_{-} + F \widetilde{\mathcal{Y}}_{+}}{z}~,
\\
\widetilde{E}(z) \mathcal{X}_+(0) &\sim \frac{E\mathcal{X}_+}{2z}~,
\\
 \widetilde{E}(z) \widetilde{\mathcal{X}}_{+}(0) &\sim \frac{E
 \widetilde{\mathcal{X}}_{+}}{2z}~,
\\
\widetilde{E}(z) \mathcal{X}_-(0) &\sim -\frac{\mathcal{X}_+}{z^2} +
 \frac{-H\mathcal{X}_+ + \frac{3}{2}E\mathcal{X}_-}{z}~,
\\
 \widetilde{E}(z) \widetilde{\mathcal{X}}_{-}(0) &\sim
 -\frac{\widetilde{\mathcal{X}}_{+}}{z^2} +\frac{-H
 \widetilde{\mathcal{X}}_{+} +\frac{3}{2} E
 \widetilde{\mathcal{X}}_{-}}{z}~,
\\
\widetilde{E}(z) \mathcal{Y}_+(0) &\sim -\frac{E\mathcal{Y}_+}{2z}~,
\\
 \widetilde{E}(z) \widetilde{\mathcal{Y}}_{+}(0) &\sim -\frac{E
 \widetilde{\mathcal{Y}}_{+}}{2z}~,
\\
\widetilde{E}(z) \mathcal{Y}_-(0) &\sim -\frac{\mathcal{Y}_+}{z^2} +
 \frac{-H\mathcal{Y}_+ - \frac{3}{2}E\mathcal{Y}_-}{z}
\\
 \widetilde{E}(z) \widetilde{\mathcal{Y}}_{-}(0) &\sim
 -\frac{\widetilde{\mathcal{Y}}_{+}}{z^2} +\frac{-H
 \widetilde{\mathcal{Y}}_{+}-\frac{3}{2}E
 \widetilde{\mathcal{Y}}_{-}}{z}~,
\\
 \widetilde{F}(z) \mathcal{X}_{+}(0) &\sim
 -\frac{\mathcal{X}_{-}}{z^2} +\frac{H
 \mathcal{X}_{-} +\frac{3}{2}F
 \mathcal{X}_{+}}{z}~,
\\
 \widetilde{F}(z) \widetilde{\mathcal{X}}_{+}(0) &\sim
 -\frac{\widetilde{\mathcal{X}}_{-}}{z^2} +\frac{H
 \widetilde{\mathcal{X}}_{-} +\frac{3}{2}F
 \widetilde{\mathcal{X}}_{+}}{z}~,
\\
 \widetilde{F}(z)\mathcal{X}_{-}(0) &\sim \frac{F
 \mathcal{X}_{-}}{2z}~,
\\
 \widetilde{F}(z) \widetilde{\mathcal{X}}_{-}(0) &\sim \frac{F
 \widetilde{\mathcal{X}}_{-}}{2z}~,
\\
 \widetilde{F}(z) \mathcal{Y}_{+}(0) &\sim
 -\frac{\mathcal{Y}_{-}}{z^2} +\frac{H
 \mathcal{Y}_{-} -\frac{3}{2} F
 \mathcal{Y}_{+}}{z}~,
\\
 \widetilde{F}(z) \widetilde{\mathcal{Y}}_{+}(0) &\sim
 -\frac{\widetilde{\mathcal{Y}}_{-}}{z^2} +\frac{H
 \widetilde{\mathcal{Y}}_{-} -\frac{3}{2} F
 \widetilde{\mathcal{Y}}_{+}}{z}~,
\\
 \widetilde{F}(z) \mathcal{Y}_{-}(0) &\sim -\frac{F
 \mathcal{Y}_{-}}{2z}~,
\\
 \widetilde{F}(z) \widetilde{\mathcal{Y}}_{-}(0) &\sim -\frac{F
 \widetilde{\mathcal{Y}}_{-}}{2z}~.
\end{align}
Finally, the non-vanishing OPEs among
$\mathcal{X}_\pm,\mathcal{Y}_\pm,\widetilde{\mathcal{X}}_\pm$ and
$\widetilde{\mathcal{Y}}_\pm$ are expressed as follows:
\begin{align}
 \mathcal{X}_{+}(z) \mathcal{Y}_{+}(0) &\sim \frac{E}{z^2} +\frac{-\frac{5}{2}UE + \widetilde{E} +\frac{1}{2}\partial E}{z}~,\\
\mathcal{X}_-(z)\mathcal{Y}_-(0) &\sim \frac{F}{z^2}
 +\frac{-\frac{5}{2}UF + \widetilde{F}+ \frac{1}{2}F'}{z}~,
\\
 \mathcal{X}_{\pm}(z) \mathcal{Y}_{\mp}(0) &\sim \frac{1}{z^3}
 +\frac{-\frac{5}{2}U \mp H}{z^2} 
                    +\frac{
                        \frac{15}{8} U^2 \pm \frac{5}{2} U H
 +\frac{H^2}{2} +\frac{EF+FE}{4} -\frac{5\partial U}{4}\mp \frac{\partial H+\widetilde{H}}{2}-T}{z}~,
\\
 \mathcal{X}_{+}(z) \widetilde{\mathcal{Y}}_{+}(0) &\sim -\frac{E
 \tau}{z}~,
\\
 \mathcal{X}_{-}(z) \widetilde{\mathcal{Y}}_{-}(0)
 &\sim -\frac{F\tau}{z}~,
\\
 \mathcal{X}_{\pm}(z) \widetilde{\mathcal{Y}}_{\mp}(0)
 &\sim -\frac{\tau}{z^2} +\frac{\frac{5}{2}U \tau \pm  H \tau -3
 \mathcal{W} -\frac{1}{4}\partial \tau}{z}~,
\\
 \mathcal{Y}_{+}(z) \widetilde{\mathcal{X}}_{+}(0) &\sim \frac{E
 \tau}{z}~,
\\
 \mathcal{Y}_{-}(z) \widetilde{\mathcal{X}}_{-}(0) &\sim \frac{F
 \tau}{z}~,
\\
 \mathcal{Y}_{\pm }(z) \widetilde{\mathcal{X}}_{\mp}(0) &\sim
 -\frac{\tau}{z^2} +\frac{-\frac{5}{2} U \tau \pm  H \tau + 3 \mathcal{W}
 -\frac{1}{4}\partial \tau}~,
 \\
 \widetilde{\mathcal{X}}_{\pm}(z)
 \widetilde{\mathcal{X}}_{\pm}(0) &\sim \frac{\mathcal{X}_{\pm}
 \mathcal{X}_{\pm}}{2z^2} +\frac{\mathcal{X}_{\pm} \partial
 \mathcal{X}_{\pm}}{2z}~,
\\
 \widetilde{\mathcal{X}}_{\pm}(z)
 \widetilde{\mathcal{X}}_{\mp}(0) &\sim \frac{\mathcal{X}_{\pm}
 \mathcal{X}_{\mp}}{2z^2} +\frac{\mathcal{X}_{\mp} \partial
 \mathcal{X}_{\pm}}{2z}~,
\\
 \widetilde{\mathcal{Y}}_{\pm}(z)
 \widetilde{\mathcal{Y}}_{\pm}(0) &\sim \frac{\mathcal{Y}_{\pm}
 \mathcal{Y}_{\pm}}{2z^2} +\frac{\mathcal{Y}_{\pm} \partial
 \mathcal{Y}_{\pm}}{2z}~,
\\
                    \widetilde{\mathcal{Y}}_{\pm}(z)
 \widetilde{\mathcal{Y}}_{\mp}(0) &\sim \frac{\mathcal{Y}_{\pm}
 \mathcal{Y}_{\mp}}{2z^2} +\frac{\mathcal{Y}_{\mp} \partial
 \mathcal{Y}_{\pm}}{2z}~,
\\
 \widetilde{\mathcal{X}}_{+}(z) \widetilde{\mathcal{Y}}_{+}(0) &\sim
 \frac{E}{z^4} +\frac{-\frac{5}{2} U E + \widetilde{E} +
 \frac{1}{2}\partial E}{z^3} 
\nonumber\\
                    &\quad + \frac{1}{z^2} \left\{
                        -\frac{5}{2} U \widetilde{E} -\frac{3}{2} E T -\frac{5}{4} U \partial E -\frac{9}{2} H \partial E -\frac{5}{4} E \partial U
                        -\frac{3}{2} E \partial H \right.
\nonumber\\
                        &\qquad \quad \left. +\frac{5}{4}U U E+2 H H E+2 E E F +\frac{3\partial^2 E}{2}+\frac{\partial \widetilde{E}}{2}
                    \right\}
\nonumber\\
                    &\quad +\frac{1}{z} \left\{
                        -\frac{5}{4} U \partial \widetilde{E} -\frac{3}{4} E \partial T +\frac{5}{12} E \partial \widetilde{H} -\frac{3}{4} T \partial E
                        -\frac{4}{3} \widetilde{H} \partial E
 -\frac{5}{4} \widetilde{E} \partial U + \widetilde{E} \partial H
 \right.
\nonumber\\
                        &\qquad \quad +\frac{5}{4} U U \widetilde{E}+\frac{15}{4} U E T               
                        +\frac{5}{6} H E \widetilde{H} +\frac{5}{3} E F \widetilde{E} -\frac{5}{2} E \hat{W} -\frac{5}{8} \partial U \partial E
                        -3 \partial H \partial E 
\nonumber\\
                        &\qquad\quad  +\frac{5}{8} U U \partial E
 +\frac{45}{4} U H \partial E +\frac{5}{4} U E \partial U +\frac{15}{4} U E \partial H + H H \partial E
                        +2 H E \partial H
\nonumber\\
                        &\qquad\quad \left.  +E E \partial F +2 E F
 \partial E -\frac{15}{4} U \partial^2 E -\frac{5}{4}H \partial^2
 E-\frac{5}{12} E \partial^2 U +\frac{5}{4}E \partial^2 H \right.
\nonumber\\
&\qquad\quad \left.+\frac{25U U U E }{12} -5 U H H E -5 U E E F
                        -\frac{3 T \widetilde{E}}{2} +\frac{13\partial^3 E}{12} +\frac{7\partial^2 \widetilde{E}}{3}
                    \right\}~,
\\
 \widetilde{\mathcal{X}}_{-}(z) \widetilde{\mathcal{Y}}_{-}(0) &\sim
 \frac{F}{z^4} +\frac{-\frac{5}{2} U F +\widetilde{F}
 +\frac{1}{2}\partial F}{z^3} 
\nonumber\\
                    &\quad +\frac{1}{z^2} \left\{
                        -\frac{5}{2} U \widetilde{F}-\frac{3}{2} F T -\frac{5}{4} U \partial F +\frac{9}{2} H \partial F -\frac{5}{4} F \partial U
                        -\frac{13}{2} F \partial H  \right.
\nonumber\\
                        &\qquad\quad \left. +\frac{5}{4} U U F +2 H H F+2 E F F +\frac{7\partial^2 F}{2}+\frac{\partial \widetilde{F}}{2}
                    \right\}
\nonumber\\
                    &\quad +\frac{1}{z} \left\{
                        -\frac{5}{4} U \partial \widetilde{F} -\frac{3}{4} F \partial T +\frac{5}{12} F \partial \widetilde{H} -\frac{3}{4} T \partial F +\frac{4}{3} \widetilde{H} \partial F -\frac{5}{4} \widetilde{F} \partial U 
                        -\widetilde{F} \partial H \right.
\nonumber\\
                        &\qquad\quad +\frac{5}{4} U U \widetilde{F}+\frac{15}{4} U F T +\frac{5}{6} H F \widetilde{H} +\frac{5}{3} F F \widetilde{E} -\frac{5}{2} F \hat{W} 
                        -\frac{5}{8} \partial U \partial F - \partial H
 \partial F 
\nonumber \\
                        &\qquad\quad +\frac{5}{8} U U \partial F
 -\frac{45}{4} U H \partial F+\frac{5}{4} U F \partial U +\frac{65}{4} U
 F \partial H + H H \partial F +2 H F \partial H 
\nonumber \\
                        &\qquad\quad \left. +2 E F \partial F + F F
 \partial E -\frac{35}{4} U \partial^2 F +\frac{5}{4} H \partial^2
 F-\frac{5}{12} F \partial^2 U -\frac{5}{4} F \partial^2 H \right.
\nonumber\\
&\qquad \quad \left.+\frac{25}{12} U U U F -5 U H H F -5 U E F F -\frac{3}{2} T \widetilde{F} +\frac{13}{12} \partial^3 F+\frac{7\partial^2 \widetilde{F}}{3}
                    \right\}~,
\\
 \widetilde{\mathcal{X}}_{+}(z) \widetilde{\mathcal{Y}}_{-}(0) &\sim \frac{1}{z^5} +\frac{-\frac{5}{2}U -H}{z^4}
 +\frac{1}{z^3} \left\{
 \frac{5}{4} U^2 +\frac{5}{2} U H + H^2 + E F -\frac{5 \partial U}{4}-\frac{3\partial H}{2}-\frac{3 T}{2}-\frac{\widetilde{H}}{2}
                    \right\} 
\nonumber\\
                    &\quad +\frac{1}{z^2} \left\{
                        \frac{15}{4} U T +\frac{5}{4} U \widetilde{H} +\frac{3}{2} H T -\frac{1}{6} H \widetilde{H} 
                        -\frac{1}{3} E \widetilde{F} +\frac{5}{4} U
 \partial U +\frac{15}{4} U \partial H +\frac{5}{4} H \partial U +3 H
 \partial H \right.
\nonumber\\
                        &\quad +\frac{3}{4} E \partial F +\frac{1}{4} F \partial E +\frac{25}{12} U^3 -\frac{5}{4} U U H -\frac{5}{2} U H^2 
                        -\frac{5}{2} U E F -2 H^3 -2 H E F 
\nonumber\\
                        &\quad \left. -\frac{5\partial^2 U}{12}-\frac{\partial^2 H}{2}-\frac{3\partial T}{4}-\frac{\partial \widetilde{H}}{6}-\frac{5 \hat{W}}{2}
                    \right\}
\nonumber\\
                    &\quad +\frac{1}{z} \left\{
                        \frac{15}{8} U \partial T +\frac{5}{6} U \partial \widetilde{H} +\frac{3}{4}H \partial T +\frac{15}{8} T \partial U -2T \partial H
                        +\frac{5}{8} \widetilde{H} \partial U
 +\frac{7}{12} \widetilde{H} \partial H +\frac{3}{2} \widetilde{E}
 \partial F \right.
\nonumber\\
                        &\quad -2 \widetilde{F} \partial E -\frac{115}{16} U U T
                        -\frac{5}{8} U U \widetilde{H} -\frac{15}{4} U H T +\frac{5}{12} U H \widetilde{H} +\frac{5}{6} U F \widetilde{E} +\frac{11}{4} H H T 
                        -\frac{5}{6} H H \widetilde{H} 
\nonumber\\
                        &\quad +\frac{11}{4} E F T -\frac{5}{6} E F \widetilde{H} +\frac{25}{4} U \hat{W} +\frac{5}{2} H \hat{W} 
                        +\frac{55}{32} (\partial U)^2
 +\frac{15}{8} \partial U \partial H -\frac{97}{32} (\partial H)^2 -\frac{69}{32} \partial E \partial F 
\nonumber\\
                        &\quad +\frac{25}{8} U U \partial U
 -\frac{115}{16} U U \partial H -\frac{5}{4}U H \partial U
 -\frac{15}{2}U H \partial H -\frac{15}{8} U E \partial F -\frac{5}{8} U
 F \partial E -\frac{5}{4} H H \partial U
\nonumber \\                       
                        &\quad +\frac{21}{8} H H \partial H +\frac{7}{8} H E \partial F +\frac{7}{8} H F \partial E -\frac{5}{4} E F \partial U +\frac{7}{8} E F \partial H +\frac{35}{48} U \partial^2 U
                        +\frac{5}{4} U \partial^2 H 
\nonumber\\
                        &\quad +\frac{5}{12} H \partial^2 U+\frac{45}{16} H \partial^2 H -\frac{5}{64} E \partial^2 F +\frac{185}{64} F \partial^2 E
                        -\frac{2225}{384} U^4 -\frac{25}{12} U U U H 
\nonumber\\
                        &\quad +\frac{105}{16} U U H^2 +\frac{105}{16} U U E F +5 U H^3 +5 U H E F - T^2
                        +\frac{3}{4} T \widetilde{H} - \tau^2
 -\frac{15}{32} \widetilde{H}^2
\nonumber\\
                        &\quad \left. -\frac{15}{8} \widetilde{E} \widetilde{F} -\frac{35}{48} \partial^3 U+\frac{119}{48}\partial^3 H+\frac{3 \partial^2 T}{8}-\frac{43 \partial^2 \widetilde{H}}{24}-\frac{5\partial \hat{W}}{4}
                    \right\}~,
\\
                    \widetilde{\mathcal{X}}_{-}(z) \widetilde{\mathcal{Y}}_{+}(0) &\sim \frac{1}{z^5} +\frac{-\frac{5}{2}U + H}{z^4}
                    +\frac{1}{z^3} \left\{
                        \frac{5}{4} U^2 -\frac{5}{2} U H + H^2 + E F -\frac{5\partial U}{4}-\frac{\partial H}{2}-\frac{3T}{2}+\frac{\widetilde{H}}{2}
                    \right\}
\nonumber\\
                    &\quad +\frac{1}{z^2} \left\{
                        \frac{15}{4}  U T -\frac{5}{4} U \widetilde{H} -\frac{3}{2} H T -\frac{1}{6} H \widetilde{H} 
                        -\frac{1}{3} E \widetilde{F} +\frac{5}{4} U
 \partial U +\frac{5}{4} U \partial H -\frac{5}{4} H \partial U -H
 \partial H \right.
\nonumber\\
                        &\quad +\frac{1}{4} E \partial F
                        +\frac{3}{4} F \partial E +\frac{25}{12} U^3 +\frac{5}{4} U U H -\frac{5}{2} U H^2 
                        -\frac{5}{2} U E F +2 H^3 +2 H E F 
\nonumber\\
                        &\quad \left. -\frac{5\partial^2 U}{12}+\frac{\partial^2 H}{2}-\frac{3\partial T}{4}+\frac{\partial \widetilde{H}}{3}-\frac{5 \hat{W}}{2}
                    \right\}
\nonumber\\
                    &\quad +\frac{1}{z} \left\{
                        \frac{15}{8} U \partial T -\frac{5}{12} U \partial \widetilde{H}-\frac{3}{4}H \partial T +\frac{15}{8} T \partial U
                        -\frac{7}{2}T \partial H -\frac{5}{8}
 \widetilde{H} \partial U -\frac{13}{12} \widetilde{H} \partial H -2
 \widetilde{E} \partial F \right.
\nonumber\\
                        &\quad +\frac{3}{2} \widetilde{F} \partial E -\frac{115}{16} U U T +\frac{5}{8} U U \widetilde{H} +\frac{15}{4} U H T +\frac{5}{12} U H \widetilde{H} +\frac{5}{6} U F \widetilde{E} +\frac{11}{4} H H T
                        +\frac{5}{6} H H \widetilde{H} 
\nonumber\\
                        &\quad +\frac{11}{4} E F T +\frac{5}{6} E F \widetilde{H} +\frac{25}{4} U \hat{W}
                        -\frac{5}{2}H \hat{W} +\frac{55}{32} (\partial U)^2 +\frac{5}{8} \partial U \partial H -\frac{161}{32} (\partial H)^2
                        -\frac{69}{32} \partial E \partial F
\nonumber\\
                        &\quad +\frac{25}{8} U U \partial U -\frac{95}{16} U U \partial H +\frac{5}{4} U H \partial U 
                        +\frac{5}{2}U H \partial H -\frac{5}{8} U E
 \partial F -\frac{15}{8} U F \partial E -\frac{5}{4}H H \partial U 
\nonumber\\
                        &\quad +\frac{69}{8}H H \partial H
                        +\frac{23}{8} H E \partial F +\frac{23}{8}H F \partial E -\frac{5}{4} E F \partial U +\frac{23}{8} E F \partial H
                        +\frac{35}{48} U \partial^2 U -\frac{5}{4} U
 \partial^2 H 
\nonumber\\
                        &\quad -\frac{5}{12} H \partial^2 U +\frac{45}{16} H \partial^2 H -\frac{85}{64}E \partial^2 F +\frac{265}{64} F \partial^2 E
                        -\frac{2225}{384} U^4 +\frac{25}{12} U U U H
 +\frac{105}{16} U U H^2 
\nonumber\\
                        &\quad +\frac{105}{16} U U E F -5 U H^3 -5 U H E F - T^2 -\frac{3}{4} T \widetilde{H} - \tau^2 -\frac{15}{32}  \widetilde{H} \widetilde{H} -\frac{15}{8} \widetilde{E} \widetilde{F}
                        -\frac{35}{48} \partial^3 U 
\nonumber\\
                        &\quad \left. +\frac{191}{48}\partial^3 H+\frac{3 \partial^2 T}{8}+\frac{43 \partial^2 \widetilde{H}}{24}-\frac{5\partial \hat{W}}{4}
                    \right\}~,
\\
 \widetilde{\mathcal{X}}_{\pm}(z) \widetilde{\mathcal{Y}}_{\mp}(0) &\sim
 \frac{1}{z^5} +\frac{-\frac{5}{2}U \mp H}{z^4}
 +\frac{1}{z^3} \left\{
 \frac{5}{4} U^2 +\frac{5}{2} U H + H^2 + \frac{EF+FE\mp H'}{2} -\frac{5 \partial U}{4}-\frac{3 T}{2}\mp\frac{\widetilde{H}}{2}
                    \right\} 
\nonumber\\
                    &\quad +\frac{1}{z^2} \Bigg\{
-\frac{5}{2}UT \pm \frac{5}{4}U\widetilde{H}\pm
 \frac{3}{2}HT+\frac{1}{4}H\widetilde{H}
 +\frac{E\widetilde{F}+F\widetilde{E}}{4} -\frac{H\partial H}{4}\mp
 \frac{\partial \widetilde{H}}{4}-\frac{55}{16}U\partial U
\nonumber\\
&\qquad \quad \pm \frac{5}{4}\partial (UH) +\frac{\pm 2-1}{8}E\partial
 F + \frac{\mp 2 -1}{8}F\partial E 
\nonumber\\
&\qquad\quad +\frac{25}{32}U^3 \mp \frac{5}{4}UUH
 +\frac{5}{8}UH^2+\frac{5}{16}U(EF+FE) \mp 2H^3 \mp H(EF+FE)
\nonumber\\
&\qquad \quad +\frac{5}{8}\partial^2 U \mp \frac{1}{2}\partial^2H
 +\frac{1}{2}\partial T+\frac{5}{4}(\mathcal{X}_+\mathcal{Y}_- + \mathcal{X}_-\mathcal{Y}_+)
              \Bigg\}
\nonumber\\
 &\quad +\frac{1}{z} \Bigg\{
-\frac{35}{8}U\partial T\pm \frac{5}{8}U\partial \widetilde{H} \mp
 \frac{1}{2}H\partial T -\frac{5}{4}T\partial U \pm \frac{3}{4}T\partial
 H\pm \frac{5}{8}\widetilde{H}\partial U +\frac{3}{8}
 \widetilde{H}\partial H
\nonumber\\
&\qquad \quad + \frac{3\pm 19}{8}\widetilde{E}\partial F + \frac{3\mp
 19}{8}\widetilde{F}\partial E +\frac{135}{16}UUT \mp
 \frac{5}{8}UU\widetilde{H} \pm \frac{5}{2}UHT -
 \frac{5}{8}UH\widetilde{H} 
\nonumber\\
&\qquad \quad  -\frac{5}{8}U(F\widetilde{E} + E\widetilde{F}) +
 \frac{11}{4}HHT  \mp \frac{5}{4}HH\widetilde{H} +\frac{
11}{8}(EF+FE)T 
\nonumber\\
&\qquad \quad \mp \frac{5}{8}(EF+FE)\widetilde{H}
 -\frac{25}{8}U(\mathcal{X}_+\mathcal{Y}_- + \mathcal{X}_-\mathcal{Y}_+)
 \mp \frac{5}{4}H(\mathcal{X}_+\mathcal{Y}_-+
 \mathcal{X}_-\mathcal{Y}_+)
\nonumber\\
&\qquad \quad -\frac{5}{8}(\partial U)^2 \pm \frac{5}{8}(\partial
 U)(\partial H) -\frac{89}{32}(\partial H)^2 -\frac{89}{64}\Big((\partial
 E)(\partial F) + (\partial F)(\partial E)\Big) 
\nonumber\\
&\qquad \quad +\frac{825}{64}UU\partial U \mp \frac{5}{8} UU\partial H 
 \pm  \frac{55}{16}UH\partial U + \frac{15}{4}UH\partial H + \frac{15\mp
 5}{8}UE\partial F 
\nonumber\\
&\qquad \quad + \frac{15\pm 5}{8}UF\partial E + \frac{5}{16}HH\partial U
 \mp \frac{7}{4}HH\partial H - \frac{69 \pm  6}{16}HE\partial F +
 \frac{69\mp 6}{16}HF\partial E
\nonumber\\
&\qquad \quad +\frac{5}{32}(\partial U)(EF+FE) \mp \frac{1}{2}(\partial
 H)(EF+FE)-\frac{135}{32}U\partial^2 U \pm \frac{5}{4}U\partial^2 H
\nonumber\\
&\qquad \quad \mp \frac{5}{8}H\partial^2 U + \frac{173}{16}H\partial^2 H
 + \frac{139\pm 8}{64}E\partial^2F + \frac{139\mp 8}{64}F\partial^2E
 -\frac{325}{128}U^4
\nonumber\\
&\qquad \quad \mp\frac{25}{32}UUUH -\frac{5}{4}UUH^2 \pm
 \frac{15}{8}UH^3 + \frac{69}{8}H^4 -\frac{5}{8}UU(EF+FE)
\nonumber\\
&\qquad \quad \pm \frac{15}{16}UH(EF+FE) + \frac{69}{16}HH (EF+FE) - T^2
 \pm \frac{3}{4}T\widetilde{H} - \tau^2 -\frac{21}{8}\widetilde{H}^2
\nonumber\\
&\qquad \quad -\frac{15}{16}(\widetilde{E}\widetilde{F} +
 \widetilde{F}\widetilde{E}) + \frac{5}{6}\partial^3 U \mp
 \frac{5}{12}\partial^3 H + \partial^2T \mp
 \frac{37}{16}\partial^2\widetilde{H} +
 \frac{5}{8}\partial(\mathcal{X}_+\mathcal{Y}_- + \mathcal{X}_-\mathcal{Y}_+)
\Bigg\}~.
\end{align}
Note that there is a null operator of the form
\begin{align}
 (\partial H)^2 - 2 H(E\partial F - F\partial E) + 4 H\partial^2 H+
 \frac{1}{2}(E\partial^2 F + F\partial^2 E) + 4 H^4 +2
 HH(EF+FE)-\widetilde{H}^2~,
\label{eq:null}
\end{align}
which implies that the coefficient of $1/z$ on the RHS does not have a
unique exression; one can add \eqref{eq:null} with an arbitrary
coefficient to it.

\bibliography{VOA}
\bibliographystyle{utphys}

\end{document}